\documentclass[10pt,american,3p]{elsarticle}
\usepackage[T1]{fontenc}
\usepackage[latin9]{inputenc}
\usepackage{color}
\usepackage{textcomp}
\usepackage{mathrsfs}
\usepackage{amsmath}
\usepackage{amssymb}
\usepackage{fixltx2e}
\usepackage{mathdots}
\usepackage{graphicx}
\usepackage{esint}
\PassOptionsToPackage{version=3}{mhchem}
\usepackage{mhchem}

\makeatletter

\newcommand{\noun}[1]{\textsc{#1}}
\providecommand{\tabularnewline}{\\}

\RequirePackage{lineno}
\usepackage{pslatex}
\usepackage{graphicx}
\usepackage{amsbsy}
\journal{Journal of Computational Physics}
\usepackage{fontenc}
\usepackage{mathrsfs}
\renewcommand{\boldsymbol}[1]{\pmb{#1}} 
\biboptions{sort&compress}

\AtBeginDocument{
  
}

\makeatother

\usepackage{babel}
\begin{document}
\begin{frontmatter}

\title{On anisotropy function in crystal growth simulations using Lattice Boltzmann equation}


\author[CEA]{\noun{Amina Younsi}\fnref{fn1}}

\ead{amina.younsi@c-s.fr}

\author[CEA]{\noun{Alain Cartalade}\corref{cor1}}

\ead{alain.cartalade@cea.fr}

\address[CEA]{Den -- DM2S, STMF, LMSF, CEA, Université de Paris-Saclay, F-91191, Gif-sur-Yvette, France.}

\fntext[fn1]{Present address: CS, 22 avenue Galilée, F-92350, Le Plessis Robinson, France.}

\cortext[cor1]{Corresponding author. Tel.:+33 (0)1 69 08 40 67}

\begin{abstract}

In this paper, we present the ability of the Lattice Boltzmann (LB)
equation, usually applied to simulate fluid flows, to simulate various
shapes of crystals. Crystal growth is modeled with a phase-field model
for a pure substance, numerically solved with a LB method in 2D and
3D. This study focuses on the anisotropy function that is responsible
for the anisotropic surface tension between the solid phase and the
liquid phase. The anisotropy function involves the unit normal vectors
of the interface, defined by gradients of phase-field. Those gradients
have to be consistent with the underlying lattice of the LB method
in order to avoid unwanted effects of numerical anisotropy. Isotropy
of the solution is obtained when the directional derivatives method,
specific for each lattice, is applied for computing the gradient terms.
With the central finite differences method, the phase-field does not
match with its rotation and the solution is not any more isotropic.
Next, the method is applied to simulate simultaneous growth of several
crystals, each of them being defined by its own anisotropy function.
Finally, various shapes of 3D crystals are simulated with standard
and non standard anisotropy functions which favor growth in $\left\langle 100\right\rangle $-,
$\left\langle 110\right\rangle $- and $\left\langle 111\right\rangle $-directions.

\end{abstract}

\begin{keyword}

Lattice Boltzmann method, phase-field model, crystal growth, anisotropy
function, directional derivatives method.

\end{keyword}

\end{frontmatter}

\section{\label{sec:Introduction}Introduction}

Simulation of crystal growth \citep{Boettinger_etal_AnnRevMatRes2002,SingerLoginova_RepProgPhys2008,Steinbach_Review_MSMSE2009,Provatas-Elder_Book_2010}
is a problem of phase change between a solid phase and a liquid phase
separated by an interface. Modeling of crystal growth necessitates
to follow that interface which is characterized by its surface energy
and its kinetic mobility. Those two properties are two anisotropic
functions depending on the underlying structure of the crystal.

The phase-field method has become, in recent years, one of the most
popular methods for simulating crystal growth and microstructure evolution
in materials \citep{Boettinger_etal_AnnRevMatRes2002,SingerLoginova_RepProgPhys2008,Provatas-Elder_Book_2010}.
In this approach, the geometry of domains and interfaces is described
by one or several scalar functions, the phase fields, that take constant
values within each domain and vary smoothly but rapidly through the
interfaces. The evolution equations for the phase fields, which give
the interface dynamics without the need for an explicit front-tracking
algorithm, are nonlinear partial differential equations (PDEs) that
can be obtained from the principles of out-of-equilibrium thermodynamics.
Therefore, they also naturally incorporate thermodynamic boundary
conditions at the interfaces, such as the Gibbs-Thomson condition.
Moreover, it is straightforward to introduce interfacial anisotropy
in phase-field models, which makes it possible to perform accurate
simulations of dendritic growth.

In phase-field models, the interfacial energy and the mobility are
decomposed into a product of a constant value with a function depending
locally on the normal vector of the interface \citep{Kobayashi_PhysD1993,Wang_etal_PhysD1993,Wheeler-Murray-Schaefer_PhysD1993,Warren-Boettinger_ActaMetallMater1995,Kim-Kim-Suzuki_PRE1998,Kim-Kim-Suzuki_PRE1999,Karma-Rappel_PRE1996,Karma-Rappel_PRE1998,Anderson_etal_Convection_PhysD2000,Bragard_etal_InterfSci2002,Echebarria_etal_PRE2004,Ramirez_etal_BinaryAlloy_PRE2004,Medvedev-Kassner_LBMCrystGrowthFlows_PRE2005,Plapp_DirectionalSolidification_JCG2007,Plapp_PRE2011}.
In the rest of this paper, this function will be called <<anisotropy
function>> and will be noted $a_{s}(\mathbf{n})$, where $\mathbf{n}$
is the unit normal vector of the interface. This function is responsible
for the characteristic shape of the crystals, it is involved in the
phase-field equation. Let us remind that in the <<sharp interface>>
formulations \citep{Zhao_etal_IJNMF2005}, the anisotropy function
appears in the Gibbs-Thomson condition which gives the interface temperature
as a function of the melting temperature, corrected by the curvature
and the kinetic mobility of the interface. In the literature, for
most of 3D crystal growth simulations, the anisotropy function is
chosen such as the branches of crystal grow along the main axes of
the coordinate system, i.e. in the $x$-, $y$- and $z$-directions
\citep{Karma-Rappel_PRL1996,Karma-Rappel_PRE1998,Chen_etal_2Dvs3DMorpho_IJHMT2009,Li_etal_OpSplitting_JCG2011,Bollada_et_JCP2015}.
This particular direction of growth is called the $\left\langle 100\right\rangle $-direction
in 3D. Thus, the crystal presents generally a dendritic shape with
six tips, two of them being directed along the $x$-axis, two other
directed along the $y$-axis and two last ones directed along the
$z$-axis.

Phase-field models are composed of highly non-linear PDEs that necessitate
to take a special care when discretizing those PDEs, particularly
when evaluating the laplacian term in the phase-field equation \citep{Karma-Rappel_PRE1998,Nestler_etal_JCP2005}.
Indeed, when a standard finite difference method is used, involving
only the first nearest neighbors, a numerical error can occur on the
crystal shape: the growth of a sphere has not the spherical shape
during the simulation. Such an error is called <<grid anisotropy>>
and this problem is fixed by using a finite difference method using
eighteen nearest neighbors in 3D \citep{Karma-Rappel_PRE1998,Bragard_etal_InterfSci2002}.

The Lattice Boltzmann (LB) equation is a popular numerical method
for simulating fluid flows modeled by Navier-Stokes equations \citep{Chen-Doolen_AnnRevFlMech1998,Succi_LBM_Book_2001,Guo-Shu_BookLBM2013}.
The method has achieved many successes in problems involving complex
fluids, such as two-phase flows \citep{Kendon_etal_SpinodalDecomp_JFM2001,Zheng_etal_LargeDensityRatio_JCP2006,Liu-Valocchi_etal_AdvWR2014},
flows interacting with a magnetic field \citep{Dellar_JCP2002,Lin_CrystalMagnetism_CF2014}
and even those that are influenced by crystal growth and solidification
\citep{Medvedev-Kassner_LBMCrystGrowthFlows_PRE2005,Medvedev_etal_PRE2006,Chatterjee-Chakraborty_PhysLettA2006,Rojas-Takaki-Ohno_PF-LBM_JCP2015}.
In this method, the main quantity is a distribution function that
is moved and performs a collision at each node of a lattice. Macroscopic
variables such as density and velocity are calculated at each time
step by updating moments of zeroth and first order of this distribution
function. 

Many works exist in the literature that combine the lattice Boltzmann
method with models of solidification or crystal growth \citep{Medvedev-Kassner_LBMCrystGrowthFlows_PRE2005,Rasin-Miller-Succi_PhaseField-CrystGrowth_PRE2005,Chatterjee-Chakraborty_PhysLettA2006,Medvedev_etal_PRE2006,Huber_etal_IJHFF2008,Sun_etal_ActaMater2009,Lin_CrystalMagnetism_CF2014}.
However, in those papers, the LBE is often used to simulate fluid
flow, whereas the model of phase change is simulated with another
numerical method (e.g. finite difference with an explicit time scheme
\citep{Rasin-Miller-Succi_PhaseField-CrystGrowth_PRE2005}). In \citep{Miller-Succi_JSP2002,Miller_etal_BinaryAlloy_PhysA2006},
the LB equation was applied to simulate a phase-field model, but the
phase-field equation does not correspond to the model solved in this
present work (\citep{Karma-Rappel_PRE1998}). The main difference
is that the anisotropy function $a_{s}(\mathbf{n})$ does not explicitly
appear in the models. We propose here to take into account explicitly
that anisotropy function by modifying first the standard lattice Boltzmann
equation and second the equilibrium distribution function. We will
introduce the directional derivatives method to compute accurately
that function. On that basis, various anisotropy functions $a_{s}(\mathbf{n})$
will be implemented and simulated in the framework of LB method.

In other examples where the LBE is applied to simulate solidification
in presence of interfacial anisotropy, the model used to track the
interface between the solid and the liquid is not based on the phase-field
theory. For instance in \citep{Sun_etal_ActaMater2009}, the Gibbs-Thomson
condition at the interface is explicitly solved in the numerical procedure,
which corresponds to a <<sharp interface>> method. In \citep{Jiaung_etal_NHT-B_2001,Chatterjee-Chakraborty_PhysLettA2006,Huber_etal_IJHFF2008}
the model is based on the <<enthalpy-porosity>> approach, an alternative
model of solid/liquid phase transition for a pure substance \citep{Voller_etal_IJNME1987,Brent_etal_NHT1988}.
Finally another class of LB method exists to simulate crystal growth.
The method is based on a reaction-diffusion model \citep{Kang_etal_GRL2004}
by applying a first-order kinetic-reaction as boundary conditions
at the fluid-solid interface. This method is extended to study dissolution
and precipitation in \citep{Chen_etal_Dissolution-Precipi_IJHMT2014,Min_etal_Reaction-Diffu_IJHMT2016}.
In that approach, the interface is not explicitly followed by using
a PDE and no anisotropy function is involved in the model.

Recently, the lattice Boltzmann schemes were applied on a phase-field
model, with or without anti-trapping current, in order to simulate
crystal growth of a binary mixture \citep{Younsi_etal_ProcIHTC15-2014,Cartalade_etal_CAMWA2015}.
In those references, numerical methods are well suited for simulating
crystal growth with low and moderate Lewis numbers. In more general
problems involving high Lewis numbers, the LB numerical schemes have
to be modified to take into account a diffusion coefficient which
can be ten thousands times smaller than the thermal diffusivity. Nevertheless,
the main advantage of this new algorithm lies in its formulation that
is identical to the one used in fluid flow. A monolithic code that
uses the same distribution functions and same libraries can be used
to simulate fluid flow and solidification process. Thank to the same
stages of collision and displacement, which can be applied for each
equation, a seamless integration with fluid flow simulations could
be performed in the future.

In continuation of reference \citep{Cartalade_etal_CAMWA2015}, crystal
growth is discussed here in terms of anisotropy function $a_{s}(\mathbf{n})$,
when the lattice Boltzmann schemes are used for simulations. For simplicity,
we focus the presentation on the solidification of a pure substance
with the model presented in \citep{Karma-Rappel_PRE1998}. Let us
emphasize that the methods presented in this paper are also suited
for dilute binary mixtures because the anisotropy function is involved
only in the phase-field equation. In the LB framework, the grid anisotropy
effects on the crystal shape are avoided by using for phase-field
equation the lattices D3Q15 or D3Q19 in 3D or D2Q9 in 2D \citep{Cartalade_etal_CAMWA2015}.
However, a numerical error may remain if $a_{s}(\mathbf{n})$ is calculated
by using the standard method of central finite differences. Indeed,
that function necessitates to calculate the interface normal vector
$\mathbf{n}$, defined by the gradient of phase-field. Those derivatives
must be consistent with the moving directions of underlying lattice
in order to avoid additional numerical errors. In this paper, those
unwanted numerical errors will be called <<lattice anisotropy>>.
The lattice anisotropy occurs especially when the growth of crystal
occurs in different directions from the standard $\left\langle 100\right\rangle $
one, e.g. $\left\langle 110\right\rangle $ or $\left\langle 111\right\rangle $.

We show in this work, that the lattice anisotropy can be decreased
by using the directional derivatives method \citep{Lee-Liu_DropImpact_LBM_JCP2010}
for calculating the gradients. Such derivatives are calculated along
each direction of propagation on the lattice. The number of directional
derivatives is equal to the number of displacement vectors in the
LB method. The three gradient components are obtained by calculating
their moment of first order. With a standard method of central finite
differences, the solution is not isotropic and the phase-field does
not match with its rotation. The directional derivatives method has
already demonstrated its performance for hydrodynamics problem in
order to reduce parasitic currents for two-phase flow problem \citep{Lee-Fischer_PRE2006,Lee_Parasitic_CAMWA2009}.
Here, we study the impact of the gradient calculation for crystal
growth simulations. To the best of authors' knowledge, the effects
of directional derivatives in lattice Boltzmann scheme were not studied
for such problems. The entire approach demonstrates the ability of
the LB equation for simulating various shapes of crystal in 2D and
3D.

This paper is organized as follows. Section \ref{sec:Mathematical-Model}
will present the mathematical model for crystal growth based on the
phase-field theory. Next, Section \ref{sec:Lattice-Boltzmann-schemes}
will present the lattice Boltzmann methods that we used to simulate
the phase-field model. Section \ref{sec:Anisotropic-functions} will
remind the main formulations of anisotropy function $a_{s}(\mathbf{n})$,
which can be defined either with the local angle $\varphi$ (between
$\mathbf{n}$ and the $x$-axis) in 2D, or by using the components
of normal vector $\mathbf{n}=(n_{x},\, n_{y},\, n_{z})^{T}$ in 3D.
A comparison between the directional derivatives method and the finite
differences method will be presented in this section. That section
will also present simulations on simultaneous growth of several crystals,
each of them being defined by its own anisotropy function. Finally,
section \ref{sec:Simulations} will present 3D simulations of various
crystals shapes obtained with standard and non standard anisotropy
functions which favor the growth in directions $\left\langle 100\right\rangle $,
$\left\langle 110\right\rangle $ and $\left\langle 111\right\rangle $.

\section{\label{sec:Mathematical-Model}Phase-field model for a pure substance}

In this work, we focus on the solidification of a pure substance.
For such a problem, the main physical process is the heat diffusion
in solid and liquid. The heat fluxes are modeled by the Fourier's
law and we assume for simplicity that properties such as specific
heat $C_{p}$ and thermal diffusivity $\kappa$ are constant and identical
in each phase. The solidification can be modeled by heat equation
applied in the bulk phases with two additional equations at the interface.
The first one is the energy conservation and the second one is the
Gibbs-Thomson condition. During the solidification, the velocity of
the interface multiplied by the latent heat $L$, released into the
liquid, is balanced with the difference between the heat fluxes in
the solid and the liquid. For the second interfacial boundary condition,
the Gibbs-Thomson condition gives the interfacial temperature as a
function of the melting temperature $T_{m}$ corrected by the curvature
and the mobility of the interface. The mathematical equations of that
<<sharp interface>> model can be found in many references such as
\citep{Karma-Rappel_PRE1998,Plapp-Karma_JCP2000,Bragard_etal_InterfSci2002,SingerLoginova_RepProgPhys2008}.

In phase-field theory, the interfacial boundary conditions are replaced
by the phase-field equation. The interface position is given by a
continuous function $\phi\equiv\phi(\mathbf{x},\, t)$, the phase-field
(dimensionless), defined over the whole computational domain. The
interface is now a diffuse one with an interface thickness noted $W_{0}$.
In our work we assume that $\phi=-1$ and $\phi=+1$ correspond respectively
to the liquid phase and the solid phase. The model is thermodynamically
consistent, i.e. it is derived by postulating a free energy functional
depending upon a double-well potential plus a gradient term \citep{Karma-Rappel_PRL1996,Karma-Rappel_PRE1996,Karma-Rappel_PRE1998}.
The model is composed of two PDEs, the first one for the phase-field
$\phi$ and the second one for the normalized temperature $u\equiv u(\mathbf{x},\, t)$
(dimensionless) defined by $u(\mathbf{x},\, t)=C_{p}(T(\mathbf{x},\, t)-T_{m})/L$:

\begin{subequations}

\begin{align}
\tau(\mathbf{n})\frac{\partial\phi}{\partial t} & =W_{0}^{2}\boldsymbol{\nabla}\cdot(a_{s}^{2}(\mathbf{n})\boldsymbol{\nabla}\phi)+W_{0}^{2}\boldsymbol{\nabla}\cdot\boldsymbol{\mathcal{N}}+(\phi-\phi^{3})-\lambda u(1-\phi^{2})^{2},\label{eq:PhaseField_KarmaRappel}\\
\frac{\partial u}{\partial t} & =\kappa\boldsymbol{\nabla}^{2}u+\frac{1}{2}\frac{\partial\phi}{\partial t},\label{eq:Temp_Ramirez}
\end{align}
where $\boldsymbol{\mathcal{N}}\equiv\boldsymbol{\mathcal{N}}(\mathbf{x},\, t)$
is defined by:

\begin{equation}
\boldsymbol{\mathcal{N}}(\mathbf{x},\, t)=\bigl|\boldsymbol{\nabla}\phi\bigr|^{2}a_{s}(\mathbf{n})\left(\frac{\partial a_{s}(\mathbf{n})}{\partial(\partial_{x}\phi)},\,\frac{\partial a_{s}(\mathbf{n})}{\partial(\partial_{y}\phi)},\,\frac{\partial a_{s}(\mathbf{n})}{\partial(\partial_{z}\phi)}\right)^{T}.\label{eq:TermesAnisotropes}
\end{equation}
The anisotropy function $a_{s}(\mathbf{n})$ (dimensionless) for a
growing direction $\left\langle 100\right\rangle $ is:

\begin{equation}
a_{s}(\mathbf{n})=1-3\varepsilon_{s}+4\varepsilon_{s}\sum_{\alpha=x,y,z}n_{\alpha}^{4}.\label{eq:As_Function_Classical}
\end{equation}

In Eq. (\ref{eq:PhaseField_KarmaRappel}), $\phi$ is the phase-field,
$W_{0}$ is the interface thickness, $\lambda$ is the coupling coefficient
with the normalized temperature $u(\mathbf{x},\, t)$. The normal
vector $\mathbf{n}\equiv\mathbf{n}(\mathbf{x},\, t)$ is defined such
as:

\begin{equation}
\mathbf{n}(\mathbf{x},\, t)=-\frac{\boldsymbol{\nabla}\phi}{\bigl|\boldsymbol{\nabla}\phi\bigr|},\label{eq:Def_Normal_Vector}
\end{equation}
directed from the solid to the liquid. The coefficient $\tau(\mathbf{n})$
is the kinetic coefficient of the interface, it is defined as $\tau(\mathbf{n})=\tau_{0}a_{s}^{2}(\mathbf{n})$
where $\tau_{0}$ is the kinetic characteristic time. Let us notice
that each term of Eq. (\ref{eq:PhaseField_KarmaRappel}) is dimensionless.
The physical dimensions of $W_{0}$, $\boldsymbol{\mathcal{N}}$,
$\tau_{0}$ and $\lambda$ are respectively $[W_{0}]\equiv[\mathscr{L}]$,
$[\boldsymbol{\mathcal{N}}]\equiv[\mathscr{L}]^{-1}$, $[\tau_{0}]\equiv[\mathcal{T}]$
and $[\lambda]\equiv[-]$, where $[\mathscr{L}]$ indicates the length
dimension and $[\mathcal{T}]$ indicates the time dimension. In Eq.
(\ref{eq:Temp_Ramirez}), the physical dimension of $\kappa$ is $[\mathscr{L}]^{2}/[\mathcal{T}]$.

\end{subequations}

In phase-field models, the capillary effects can be simulated by defining
the interface thickness as a function depending on the normal vector
of the interface $\mathbf{n}$. In other words, the definition $W(\mathbf{n})=W_{0}a_{s}(\mathbf{n})$
where $a_{s}(\mathbf{n})$ is the anisotropy function of the interfacial
energy allows to simulate correctly the growth of crystal with anisotropic
shapes. In the model, the second term in the right-hand side of Eq.
(\ref{eq:PhaseField_KarmaRappel}) is responsible for such a dendritic
structure. The term $(\phi-\phi^{3})$ is the derivative of the double-well
potential and the last term $\lambda u(1-\phi^{2})^{2}$ is the coupling
term with the normalized temperature, interpolated in the diffuse
interface. Heat equation (Eq. (\ref{eq:Temp_Ramirez})) is a standard
diffusive equation with a source term involving the time derivative
of phase-field: $(1/2)\partial\phi/\partial t$. The meaning of this
term can be understood by multiplying Eq. (\ref{eq:Temp_Ramirez})
by the latent heat $L$ which is involved in the definition of $u$.
That term models the release of latent heat during the solidification
process.

The phase-field equation Eq. (\ref{eq:PhaseField_KarmaRappel}) involves
three parameters $W_{0}$, $\tau_{0}$ and $\lambda$. Matched asymptotic
expansions provide relationships between those parameters with the
capillary length $d_{0}$ and the interface kinetic coefficient $\beta$
of the sharp-interface model \citep{Karma-Rappel_PRE1998}:

\begin{subequations}

\begin{align}
d_{0} & =a_{1}\frac{W_{0}}{\lambda},\label{eq:Capillary-Length}\\
\beta & =a_{1}\left(\frac{\tau_{0}}{W_{0}\lambda}-a_{2}\frac{W_{0}}{\kappa}\right).\label{eq:Kinetic-Coeff}
\end{align}
Coefficients $a_{1}$ and $a_{2}$ are two numbers of order unity:
$a_{1}=5\sqrt{2}/8$, and $a_{2}\approx0.6267$. These relations make
it possible to choose phase-field parameters for prescribed values
of the capillary length (surface energy) and the interface mobility
(interface kinetic coefficient). Eqs. (\ref{eq:Capillary-Length})
and (\ref{eq:Kinetic-Coeff}) will be applied in Section \ref{sec:Anisotropic-functions}
for simulating a crystal which grows without kinetic coefficient ($\beta=0$).

Note that the interface width $W_{0}$ is a parameter that can be
freely chosen in this formulation; the asymptotic analysis remains
valid as long as $W_{0}$ remains much smaller than any length scale
present in the sharp-interface solution of the considered problem
(for example, a dendrite tip radius in the case of dendritic growth).
More details about the dependence of parameters with \textbf{$\mathbf{n}$}
and the link of the phase-field model with the sharp interface model
can be found in \citep{Bragard_etal_InterfSci2002}. That model of
crystal growth was applied in many references \citep{Nestler_etal_JCP2005,Bragard_etal_InterfSci2002,Lin-Chen-Lan_JCP2011,Li_etal_OpSplitting_JCG2011}
and serves as basis for model of dilute binary mixture \citep{Ramirez_etal_BinaryAlloy_PRE2004,Bollada_et_JCP2015}
and model of coupling with fluid flows \citep{Lu-Beckermann-Ramirez_JCG2005,Medvedev-Kassner_LBMCrystGrowthFlows_PRE2005,Chen_etal_2Dvs3DMorpho_IJHMT2009}.

\end{subequations}

The characteristic shape of the crystal is obtained by defining appropriately
the $a_{s}(\mathbf{n})$ function. If the growing direction $\left\langle 100\right\rangle $
is considered, this function is defined by Eq. (\ref{eq:As_Function_Classical}).
In this relationship, $\varepsilon_{s}$ is the strength of anisotropy
and $n_{\alpha}$ ($\alpha=x,\, y,\, z$) is the $\alpha$-component
of $\mathbf{n}$. The choice of function (\ref{eq:As_Function_Classical})
yields to a dendritic shape which presents four tips in 2D and six
tips in 3D \citep{Karma-Rappel_PRE1996,Karma-Rappel_PRL1996}. In
this paper, we focus on the way to simulate crystal growth with various
anisotropy functions $a_{s}(\mathbf{n})$ by using the LB method.

For all simulations in this paper, the computational domain is a square
in 2D or a cube in 3D, and zero fluxes are imposed on all boundaries
for both equations. For phase-field equation, a nucleus is initialized
as a diffuse sphere:

\begin{equation}
\phi(\mathbf{x},\,0)=\tanh\left[\frac{R_{s}-d_{s}}{\sqrt{2W{}_{0}}}\right],\label{eq:Initial_Condition}
\end{equation}
where $R_{s}$ is the radius, $d_{s}$ is the distance defined by
$d_{s}=[(x-x_{s})^{2}+(y-y_{s})^{2}+(z-z_{s})^{2}]^{1/2}$ and $\mathbf{x}_{s}=(x_{s},\, y_{s},\, z_{s})^{T}$
is the position of its center. With this initial condition, $\phi=+1$
inside the sphere (solid) and $\phi=-1$ outside (liquid). The initial
temperature is considered as a constant on the whole domain and fixed
below the melting temperature: $T<T_{m}$. The initial condition of
the normalized temperature is set such as:

\begin{equation}
u(\mathbf{x},\,0)=u_{0}<0.\label{eq:Intial-Cond_u}
\end{equation}

In the following simulations, the undercooling defined by $\Delta=-u_{0}$
will be specified.

\section{\label{sec:Lattice-Boltzmann-schemes}Lattice Boltzmann schemes}

LB methods are usually applied to simulate fluid flows. Here, the
standard LB equation with the classical equilibrium distribution function
(e.d.f.) have to be modified for simulating Eqs. (\ref{eq:PhaseField_KarmaRappel})--(\ref{eq:Def_Normal_Vector}).
Indeed, fluid flows model involves one scalar equation (mass balance)
plus one vectorial equation (momentum) whereas the phase-field model
is composed of two scalar equations which are coupled and non-linear.
LB methods for phase-field model of crystal growth are already presented
in details in reference \citep{Cartalade_etal_CAMWA2015}. They were
applied on a model of dilute binary mixture with anti-trapping current
which is an extension of a pure substance model. Following that reference,
LB schemes are described here for heat equation and phase-field equation.
The heat equation is a standard diffusion equation with a particular
source term. For a pedagogical presentation, we start the description
of algorithms with that equation, because the classical LB method
for a diffusive equation is applied (e.g. \citep{Dawson_etal_JCP1993}).
Modifications that have to be done for the phase-field equation will
be presented and commented in subsection \ref{sub:Phase-Field-equation}.

\subsection{\label{sub:Heat-equation}Heat equation}

The LB method works on a distribution function $f_{i}(\mathbf{x},\, t)$
depending on position $\mathbf{x}$ and time $t$. The index $i$
identifies the moving directions on a lattice: $i=0,\,...,\, N_{pop}$
where $N_{pop}$ is the total number of directions. The lattice choice
(space dimension and number of moving directions) depends on the physical
problem to be simulated. The lattices used in this work are presented
in subsection \ref{sub:Lattices}. The heat equation (\ref{eq:Temp_Ramirez})
is a diffusion equation with a source term involving the time derivative
of $\phi$. For that equation, the standard LB equation with the Bhatnagar-Gross-Krook
(BGK) approximation for the collision term is applied:

\begin{subequations}

\begin{equation}
f_{i}(\mathbf{x}+\mathbf{e}_{i}\delta x,\, t+\delta t)=f_{i}(\mathbf{x},\, t)-\frac{1}{\eta_{u}}\left[f_{i}(\mathbf{x},\, t)-f_{i}^{(0)}(\mathbf{x},\, t)\right]+w_{i}Q_{u}(\mathbf{x},\, t)\delta t.\label{eq:LBE_Eq_Temp}
\end{equation}
Note that the LB equation (Eq. (\ref{eq:LBE_Eq_Temp})) is an evolution
equation for $f_{i}(\mathbf{x},\, t)$ which is already discretized
in space, time, and moving directions. The space-step is noted $\delta x$
by assuming $\delta x=\delta y=\delta z$ and the time-step is noted
$\delta t$. Vectors of displacement on the lattice are noted $\mathbf{e}_{i}$
and $w_{i}$ are weights. The quantities $N_{pop}$, $\mathbf{e}_{i}$
and $w_{i}$ are lattice-dependent and will be specified in table
\ref{tab:Values-Lattices} of subsection \ref{sub:Lattices}. In Eq.
(\ref{eq:LBE_Eq_Temp}), the BGK collision term is the second term
in the right-hand side. It relaxes the distribution function $f_{i}(\mathbf{x},\, t)$
toward an equilibrium distribution function (e.d.f.) $f_{i}^{(0)}(\mathbf{x},\, t)$
with a relaxation rate $\eta_{u}$. In such an equation, $f_{i}(\mathbf{x},\, t)$
can be regarded as an intermediate function introduced to calculate
the normalized temperature $u(\mathbf{x},\, t)$ and updated at each
time step by:

\begin{equation}
u(\mathbf{x},\, t)=\sum_{i=0}^{N_{pop}}f_{i}(\mathbf{x},\, t).\label{eq:Temperature_LB}
\end{equation}
Eq. (\ref{eq:Temperature_LB}) is called the moment of zeroth order
of the distribution function $f_{i}$. Moments of first and second
order are defined respectively by $\sum_{i=0}^{N_{pop}}f_{i}\mathbf{e}_{i}$
(vector) and $\sum_{i=0}^{N_{pop}}f_{i}\mathbf{e}_{i}\mathbf{e}_{i}$
(tensor of second order). In Eq. (\ref{eq:LBE_Eq_Temp}), the equilibrium
distribution function $f_{i}^{(0)}(\mathbf{x},\, t)$ and the source
term $Q_{u}(\mathbf{x},\, t)$ are given by:

\begin{align}
f_{i}^{(0)}(\mathbf{x},\, t) & =w_{i}u(\mathbf{x},\, t),\label{eq:EqFunction_Tempr}\\
Q_{u}(\mathbf{x},\, t) & =\frac{1}{2}\frac{\partial\phi}{\partial t}.\label{eq:Source_Tempr}
\end{align}

By carrying out the Chapman-Enskog's expansions of Eq. (\ref{eq:LBE_Eq_Temp})
with Eq. (\ref{eq:EqFunction_Tempr}), the first- and second-order
moments of the equilibrium distribution function are equal to $\sum_{i}f_{i}^{(0)}\mathbf{e}_{i}=\mathbf{0}$
and $\sum_{i}f_{i}^{(0)}\mathbf{e}_{i}\mathbf{e}_{i}=u(\mathbf{x},\, t)e^{2}\overline{\overline{\mathbf{I}}}$,
where $\overline{\overline{\mathbf{I}}}$ is the identity tensor of
rank 2. An additional lattice-dependent coefficient, $e^{2}$, arises
from the second-order moment of $f_{i}^{(0)}$. Values of $e^{2}$
are given in table \ref{tab:Values-Lattices} for four lattices used
in this work. The thermal diffusivity $\kappa$ is related to the
relaxation rate of collision $\eta_{u}$ by:

\begin{equation}
\kappa=e^{2}\left(\eta_{u}-\frac{1}{2}\right)\frac{\delta x^{2}}{\delta t}.\label{eq:ThermalDiffusivity}
\end{equation}
The index $u$ in $Q_{u}$ and $\eta_{u}$ indicates that both quantities
are relative to the heat equation. Note that the relaxation rate $\eta_{u}$
is a constant because the thermal diffusivity $\kappa$ was assumed
constant in the model. Being given $\kappa$, $\delta x$ and $\delta t$,
the relaxation parameter is initialized before the time loop and kept
constant during the simulation. In a more general case, $\kappa$
can be a function depending on space and time. In that case, the relationship
(\ref{eq:ThermalDiffusivity}) must be inverted and the relaxation
parameter has to be updated at each time step. Let us notice that
the factor $\delta x^{2}/\delta t$ in the right-hand side is homogeneous
to $[\mathscr{L}]^{2}/[\mathcal{T}]$, which is consistent with the
physical dimension of $\kappa$.

The principle of the LB scheme is the following. Once the normalized
temperature $u$ is known with the initial condition, the equilibrium
distribution function $f_{i}^{(0)}$ is computed by using Eq. (\ref{eq:EqFunction_Tempr}).
The collision stage (right-hand side of Eq. (\ref{eq:LBE_Eq_Temp}))
is next calculated and yields an intermediate distribution function
that will be streamed in each direction (left-hand side of Eq. (\ref{eq:LBE_Eq_Temp})).
Finally after updating the boundary conditions, the new temperature
is calculated by using Eq. (\ref{eq:Temperature_LB}) and the algorithm
is iterated in time.

Notice that the collision term is local and the scheme is fully explicit,
i.e., all terms in the right-hand side of Eq. (\ref{eq:LBE_Eq_Temp})
are defined at position $\mathbf{x}$ and time $t$. Also note that
the source term $Q_{u}$ involves the time derivative of the phase
field which is approximated here by an explicit Euler scheme: $\partial\phi/\partial t\cong(\phi(\mathbf{x},\, t)-\phi(\mathbf{x},\, t-\delta t))/\delta t$.
In practice, the heat equation must be solved after solving the phase-field
equation. At the first time-step, the derivative is obtained by the
difference between the phase-field at the first time-step and its
initial condition.

Finally, this scheme can be easily extended to simulate the Advection-Diffusion
Equation (ADE) (e.g. \citep{Walsh-Saar_WRR2010}): $\partial u/\partial t=\kappa\boldsymbol{\nabla}^{2}u-\boldsymbol{\nabla}\cdot(u\mathbf{v})$,
where $\mathbf{v}$ is the advective velocity, by modifying the equilibrium
distribution function such as $f_{i}^{(0)\, ADE}=w_{i}u\left[1+e^{-2}\mathbf{e}_{i}\cdot\mathbf{v}\delta t/\delta x\right]$.
Moments of zeroth-, first- and second-order of $f_{i}^{(0)\, ADE}$
are respectively $u$, $\mathbf{v}u\delta t/\delta x$ and $e^{2}u\overline{\overline{\mathbf{I}}}$.
In next subsection, the LB method for the phase-field equation is
presented with an analogy with this ADE.

\end{subequations}

\subsection{\label{sub:Phase-Field-equation}Phase-Field equation}

Phase-field equation (Eq. (\ref{eq:PhaseField_KarmaRappel})) looks
like an ADE with two differences. The first one is the presence of
an additional factor $\tau(\mathbf{n})=\tau_{0}a_{s}^{2}(\mathbf{n})$
in front of the time derivative $\partial\phi/\partial t$ in the
left-hand side. The second difference is the presence of the divergence
term $W_{0}^{2}\boldsymbol{\nabla}\cdot\boldsymbol{\mathcal{N}}$
in the right-hand side, which is not strictly speaking one <<advective>>
term. In order to handle those two terms, the standard LB scheme has
to be modified. First, a new distribution function $g_{i}(\mathbf{x},\, t)$
is introduced. The LBE for the phase-field equation works on that
function. The first difference necessitates a modification of the
evolution equation for $g_{i}$. The second difference requires a
modification of the equilibrium distribution function $g_{i}^{(0)}(\mathbf{x},\, t)$.

The numerical scheme is:

\begin{subequations}

\begin{equation}
a_{s}^{2}(\mathbf{n})g_{i}(\mathbf{x}+\mathbf{e}_{i}\delta x,\, t+\delta t)=g_{i}(\mathbf{x},\, t)-(1-a_{s}^{2}(\mathbf{n}))g_{i}(\mathbf{x}+\mathbf{e}_{i}\delta x,\, t)-\frac{1}{\eta_{\phi}(\mathbf{x},\, t)}\left[g_{i}(\mathbf{x},\, t)-g_{i}^{(0)}(\mathbf{x},\, t)\right]+w_{i}Q_{\phi}(\mathbf{x},\, t)\frac{\delta t}{\tau_{0}},\label{eq:LBE_Eq_Phase}
\end{equation}

\noindent \begin{flushleft}
with the equilibrium distribution function $g_{i}^{(0)}(\mathbf{x},\, t)$
defined by:
\par\end{flushleft}

\begin{equation}
g_{i}^{(0)}(\mathbf{x},\, t)=w_{i}\left(\phi(\mathbf{x},\, t)-\frac{1}{e^{2}}\mathbf{e}_{i}\cdot\boldsymbol{\mathcal{N}}(\mathbf{x},\, t)\frac{\delta t}{\delta x}\frac{W_{0}^{2}}{\tau_{0}}\right).\label{eq:Feq_Phase}
\end{equation}

\end{subequations}

In Eq. (\ref{eq:LBE_Eq_Phase}), the anisotropy function $a_{s}(\mathbf{n})$
has to be specified. It is defined by Eq. (\ref{eq:As_Function_Classical})
for simulating one crystal that grows in a preferential direction
$\left\langle 100\right\rangle $. In order to simulate other preferential
directions of growth, a modification of that function is necessary.
Analytic expressions of various anisotropy functions will be specified
in Section \ref{sub:Formulations-anisotropy}. Whatever the relationship
used, calculation of $a_{s}(\mathbf{n})$ requires the computation
of $\mathbf{n}$ which is defined by the gradient of phase-field $\boldsymbol{\nabla}\phi$.
The calculation of that gradient will be presented in Section \ref{sub:Derivees-directionnelles}.
In Eq. (\ref{eq:Feq_Phase}), the vector $\boldsymbol{\mathcal{N}}(\mathbf{x},\, t)$
is defined by Eq. (\ref{eq:TermesAnisotropes}). Once the anisotropy
function $a_{s}(\mathbf{n})$ is set, the components of $\boldsymbol{\mathcal{N}}$
are calculated by deriving $a_{s}(\mathbf{n})$ with respect to $\partial_{x}\phi$,
$\partial_{y}\phi$ and $\partial_{z}\phi$. In Eqs. (\ref{eq:LBE_Eq_Phase})--(\ref{eq:Feq_Phase}),
functions $a_{s}(\mathbf{n})$ and $\boldsymbol{\mathcal{N}}(\mathbf{x},\, t)$
are treated explicitly in time.

The lattice Boltzmann scheme for the phase-field equation differs
\textcolor{black}{from} the standard LB method for ADE on two points.
The first difference is the presence in Eq. (\ref{eq:LBE_Eq_Phase})
of \emph{(i)} a factor $a_{s}^{2}(\mathbf{n})$ in front of $g_{i}(\mathbf{x}+\mathbf{e}_{i}\delta x,\, t+\delta t)$
in the left-hand side and \emph{(ii)} an additional term $(1-a_{s}^{2}(\mathbf{n}))g_{i}(\mathbf{x}+\mathbf{e}_{i}\delta x,\, t)$
in the right-hand side. The latter term is non-local in space, i.e.,
it is involved in the collision step at time $t$ and needs the knowledge
of $g_{i}$ at \textcolor{black}{the neighboring} nodes $\mathbf{x}+\mathbf{e}_{i}\delta x$.
Those two terms appear to handle the factor $a_{s}^{2}(\mathbf{n})$
in front of the time derivative $\partial\phi/\partial t$ in Eq.
(\ref{eq:PhaseField_KarmaRappel}). It is straightforward to prove
it by carrying out the Taylor's expansion of Eq. (\ref{eq:LBE_Eq_Phase}).

\begin{subequations}

The second difference with the LB algorithm for ADE, is the definition
of the equilibrium distribution function $g_{i}^{(0)}$ (Eq. (\ref{eq:Feq_Phase})).
Moments of zeroth-, first- and second-order of this function are respectively:

\begin{align}
\sum_{i=0}^{N_{pop}}g_{i}^{(0)}(\mathbf{x},\, t) & =\phi(\mathbf{x},\, t)\label{eq:M0_g0}\\
\sum_{i=0}^{N_{pop}}g_{i}^{(0)}(\mathbf{x},\, t)\mathbf{e}_{i} & =-\boldsymbol{\mathcal{N}}(\mathbf{x},\, t)\frac{\delta t}{\delta x}\frac{W_{0}^{2}}{\tau_{0}}\label{eq:M1_g0}\\
\sum_{i=0}^{N_{pop}}g_{i}^{(0)}(\mathbf{x},\, t)\mathbf{e}_{i}\mathbf{e}_{i} & =\phi(\mathbf{x},\, t)e^{2}\overline{\overline{\mathbf{I}}}\label{eq:M2_g0}
\end{align}
where $\overline{\overline{\mathbf{I}}}$ is still the identity tensor
of rank 2. The main difference with the e.d.f. of ADE appears in the
moment of first order (Eq. (\ref{eq:M1_g0})). In that term, the scalar
field $\phi(\mathbf{x},\, t)$ is not involved in the right-hand side
because it does not appear in the divergence term of Eq. (\ref{eq:PhaseField_KarmaRappel}).
The absence of phase field $\phi(\mathbf{x},\, t)$ in the divergence
term, explains its presence in the first term inside the brackets
(\ref{eq:Feq_Phase}), and not before the brackets. Moreover, note
the sign change in front of the scalar product $\mathbf{e}_{i}\cdot\boldsymbol{\mathcal{N}}$,
which corresponds to the sign change of $+\boldsymbol{\nabla}\cdot\boldsymbol{\mathcal{N}}$
compared to $-\boldsymbol{\nabla}\cdot(\mathbf{v}u)$ for the ADE.
A dimensional analysis shows that Eqs. (\ref{eq:LBE_Eq_Phase}) and
(\ref{eq:Feq_Phase}) are both dimensionless. In the e.d.f., the second
term in the brackets is dimensionless since $[\boldsymbol{\mathcal{N}}]\equiv[\mathscr{L}]^{-1}$,
$[W_{0}]\equiv[\mathscr{L}]$ and $[\tau_{0}]\equiv[\mathcal{T}]$.

\end{subequations}

\begin{subequations}

The algorithm works in the same way than the previous one for the
temperature field $u(\mathbf{x},\, t)$. After the collision stage
and the streaming step, the phase-field $\phi(\mathbf{x},\, t)$ is
calculated by the moment of zeroth order:

\begin{equation}
\phi(\mathbf{x},\, t)=\sum_{i=0}^{N_{pop}}g_{i}(\mathbf{x},\, t).\label{eq:Def_M0_Phi}
\end{equation}
The scalar function $Q_{\phi}(\mathbf{x},\, t)$ is the source term
of the phase-field equation (\ref{eq:PhaseField_KarmaRappel}) which
is defined by:

\begin{align}
Q_{\phi}(\mathbf{x},\, t) & =\left[\phi-\lambda u(1-\phi^{2})\right](1-\phi^{2}).\label{eq:LBE_SourceTerm}
\end{align}

In Eq. (\ref{eq:PhaseField_KarmaRappel}), coefficient $a_{s}^{2}(\mathbf{n})$
in the first term of the right-hand side plays a similar role as a
<<diffusion>> coefficient depending on position and time (through
$\mathbf{n}$ that depends on $\phi$). The relaxation rate $\eta_{\phi}(\mathbf{x},\, t)$
is a function of position and time and must be updated at each time
step by the relationship:

\begin{equation}
\eta_{\phi}(\mathbf{x},\, t)=\frac{1}{e^{2}}a_{s}^{2}(\mathbf{n})\frac{W_{0}^{2}}{\tau_{0}}\frac{\delta t}{\delta x^{2}}+\frac{1}{2}.\label{eq:LBE_RelaxationTime}
\end{equation}

\end{subequations}

In Eq. (\ref{eq:LBE_RelaxationTime}), we can check that the first
term of the right-hand side is dimensionless: $W_{0}^{2}/\tau_{0}$
is dimensionally balanced with $\delta t/\delta x^{2}$. The origin
of factor $W_{0}^{2}/\tau_{0}$ inside $g_{i}^{(0)}$ (Eq. (\ref{eq:Feq_Phase}))
and $\eta_{\phi}$ (Eq. (\ref{eq:LBE_RelaxationTime})), can be understood
by dividing each term of Eq. (\ref{eq:PhaseField_KarmaRappel}) by
$\tau_{0}$. The first term in the right-hand side becomes $\boldsymbol{\nabla}\cdot\left[(a_{s}^{2}(\mathbf{n})W_{0}^{2}/\tau_{0})\boldsymbol{\nabla}\phi\right]$,
the second one becomes $\boldsymbol{\nabla}\cdot(\boldsymbol{\mathcal{N}}W_{0}^{2}/\tau_{0})$
and the last one is $Q_{\phi}/\tau_{0}$. Each term $(a_{s}^{2}(\mathbf{n})W_{0}^{2}/\tau_{0})$,
$(\boldsymbol{\mathcal{N}}W_{0}^{2}/\tau_{0})$ and $Q_{\phi}/\tau_{0}$
is respectively involved in Eqs. (\ref{eq:LBE_RelaxationTime}), (\ref{eq:Feq_Phase})
and (\ref{eq:LBE_Eq_Phase}). More rigorously, after the asymptotic
expansions of Eq. (\ref{eq:LBE_Eq_Phase}), the moments equation for
$g_{i}^{(0)}$ is compared to Eq. (\ref{eq:PhaseField_KarmaRappel})
divided by $\tau_{0}$. The identification of each term yields to
the relationships (\ref{eq:M0_g0})--(\ref{eq:M2_g0}) and (\ref{eq:LBE_RelaxationTime}).
The Chapman-Enskog's expansions of that numerical scheme can be found
in Appendices of reference \citep{Cartalade_etal_CAMWA2015}.

The LB scheme for phase-field equation uses the same BGK collision
rule than temperature equation. Other approximations of the collision
term exist in the literature: Multiple-Relaxation Time (MRT) \citep{dHumieres_etal_PhilTranRoySoc2002}
and Two-Relaxation Time (TRT) \citep{Ginzburg_AdWR2005a}. With the
MRT approximation the collisions are carried out in the moments space.
For ADE, tensorial diffusion coefficients can be taken into account
for problems that involve diffusion coefficients varying with direction
\citep{Yoshida-Nagaoka_JCP2010}. Moreover, with the MRT and TRT approximations,
it is possible to simulate problems with higher Peclet numbers than
those reached with the BGK collision. High Peclet number and anisotropic
diffusion coefficient are not studied here.

\subsection{\label{sub:Lattices}Lattices}

\begin{figure}
\begin{centering}
\begin{tabular}{ccc}
{\small (a) D2Q5} &  & {\small (b) D2Q9}\tabularnewline
\noalign{\vskip2mm}
\includegraphics[scale=0.4]{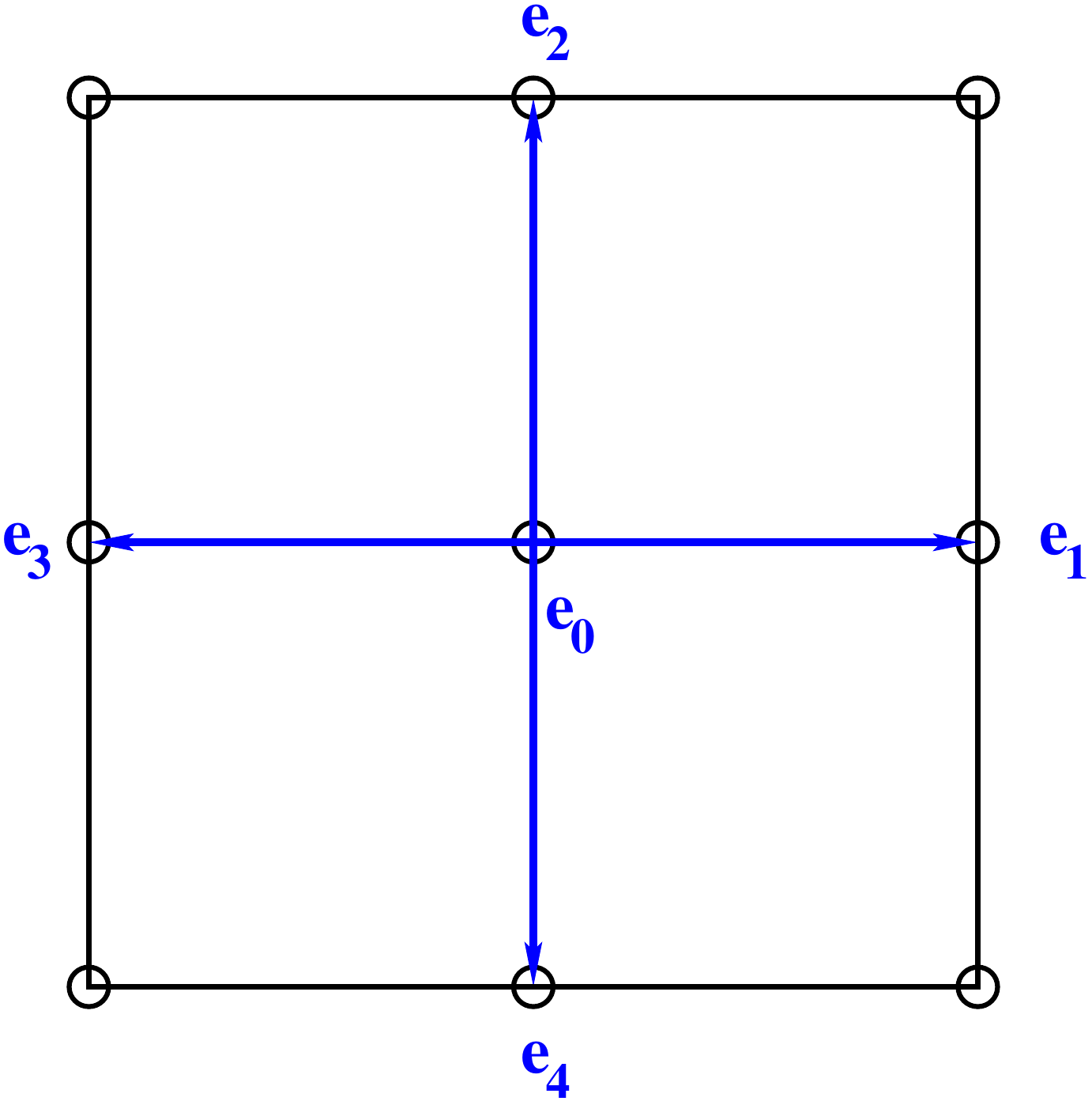} & $\qquad$ & \includegraphics[scale=0.4]{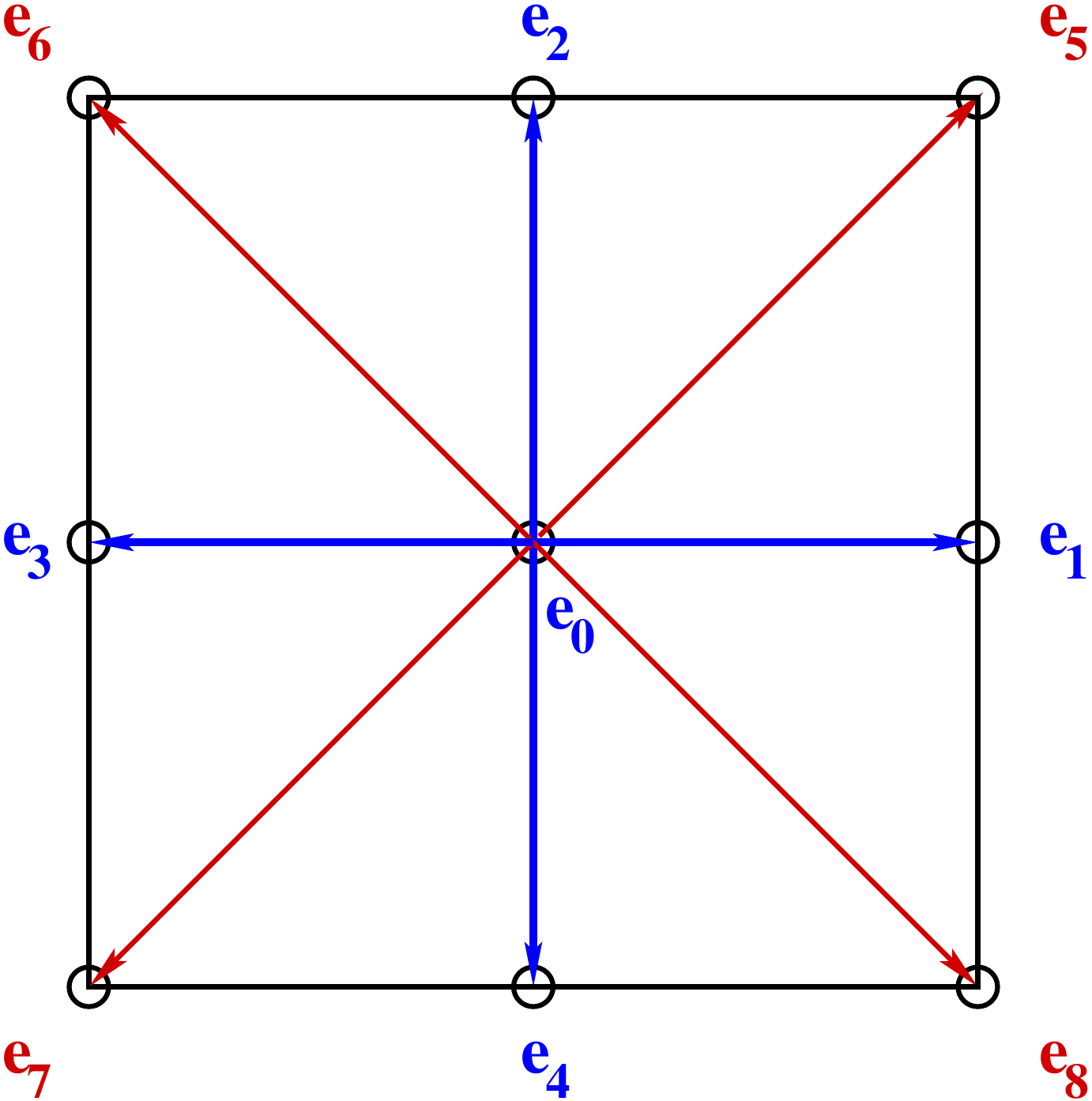}\tabularnewline
 &  & \tabularnewline
 &  & \tabularnewline
{\small (c) D3Q7} &  & {\small (d) D3Q15}\tabularnewline
\noalign{\vskip2mm}
\includegraphics[scale=0.32]{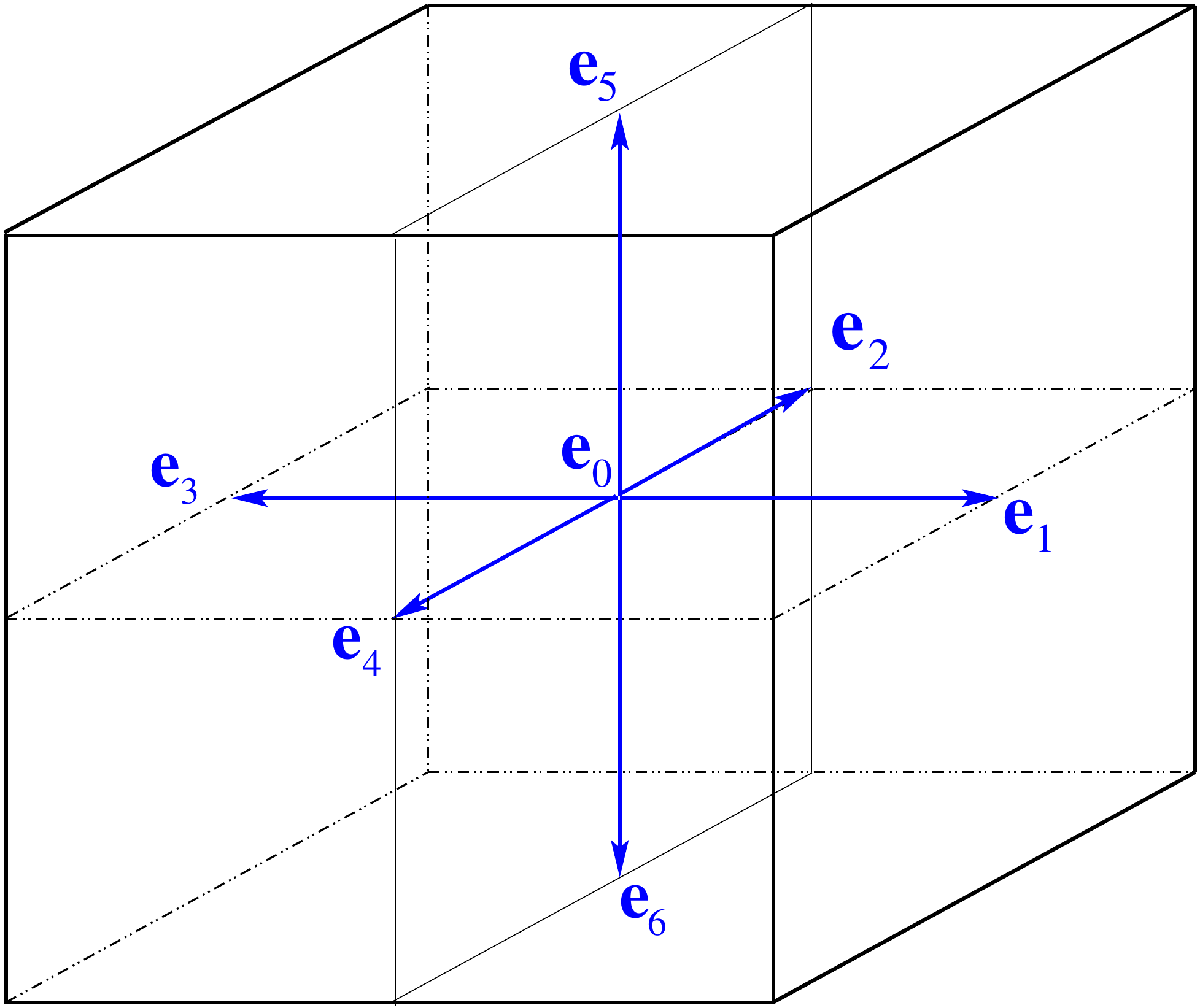} &  & \includegraphics[scale=0.32]{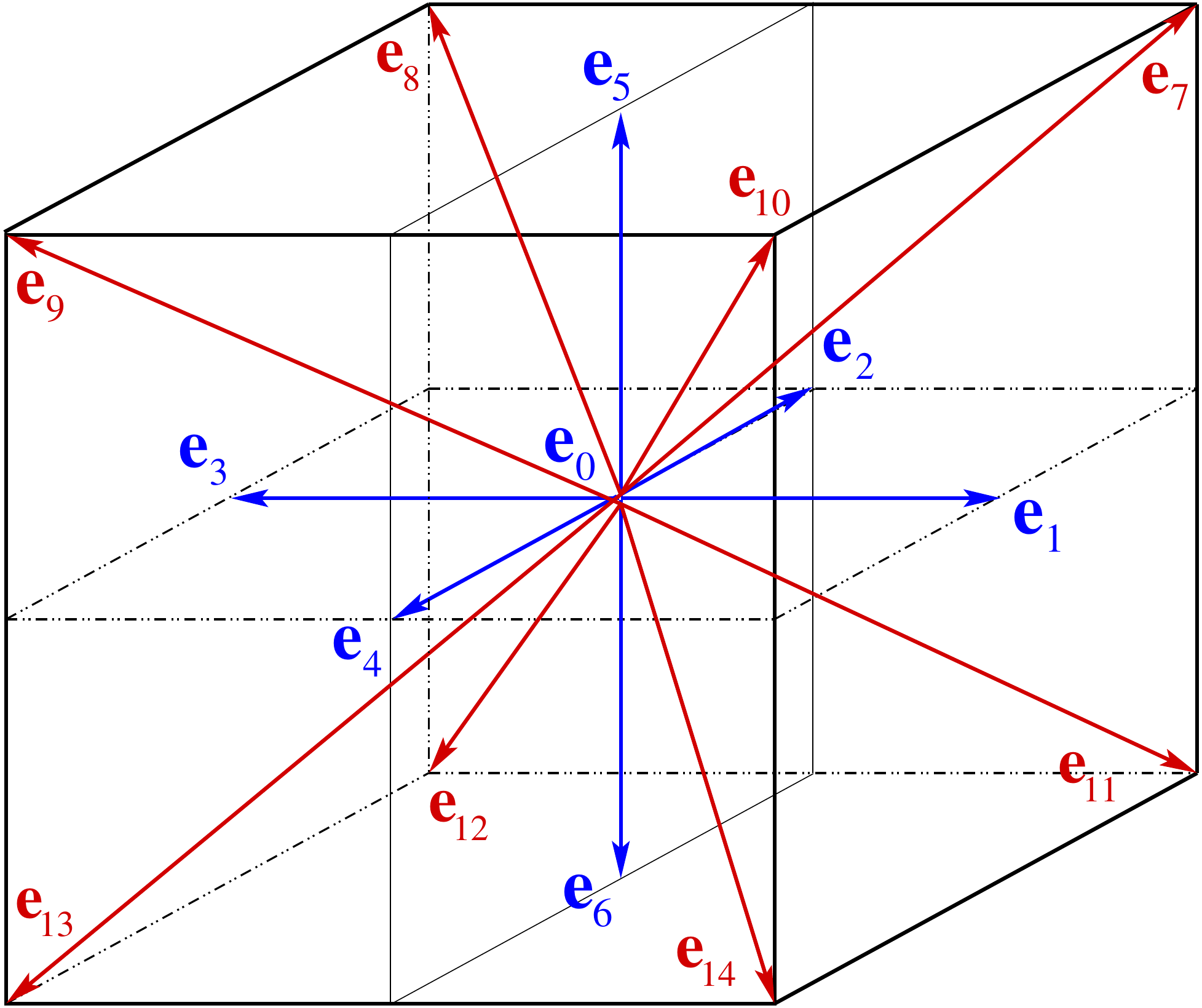}\tabularnewline
\end{tabular}
\par\end{centering}

\caption{\label{fig:Lattices}(a) Two-dimensional simulations: lattices (a)
D2Q5 for temperature equation and (b) D2Q9 for phase-field equation.
Three-dimensional simulations: (c) D3Q7 for temperature equation and
(d) D3Q15 for phase-field equation.}
\end{figure}

Several lattices exist in the LB method, they differ by the space
dimension and the number of moving directions. In this work, 2D and
3D simulations are carried out with four lattices: D2Q5, D2Q9, D3Q7
and D3Q15 (see Fig. \ref{fig:Lattices}). The lattices D2Q5 and D2Q9
are applied for two-dimensional simulations (D2) and uses respectively
five (Q5) and nine (Q9) directions of displacement. The lattice D2Q5
is applied for temperature equation and D2Q9 for phase-field equation.
For three-dimensional simulations, the temperature $u$ is computed
on a lattice D3Q7, whereas the phase-field $\phi$ is computed on
a lattice D3Q15. The first one is defined by seven directions of displacement
(Q7) and the second one is defined by fifteen directions (Q15). The
displacement vectors $\mathbf{e}_{i}$ are defined below for each
lattice.

\begin{table}
\begin{centering}
{\footnotesize }%
\begin{tabular}{cc}
\hline 
\textbf{Lattices} & \textbf{Weights}\tabularnewline
\hline 
\noalign{\vskip1mm}
\begin{tabular}{ccc}
2D & $N_{pop}$ & $e^{2}$\tabularnewline
D2Q5 & $4$ & $1/3$\tabularnewline
D2Q9 & $8$ & $1/3$\tabularnewline
\hline 
\noalign{\vskip1mm}
3D & $N_{pop}$ & $e^{2}$\tabularnewline
D3Q7 & $6$ & $1/4$\tabularnewline
D3Q15 & $14$ & $1/3$\tabularnewline
\end{tabular} & %
\begin{tabular}{ccc}
$w_{0}$ & $w_{1,...,4}$ & $w_{5,...,8}$\tabularnewline
$1/3$ & $1/6$ & $-$\tabularnewline
$4/9$ & $1/9$ & $1/36$\tabularnewline
\hline 
\noalign{\vskip1mm}
$w_{0}$ & $w_{1,...,6}$ & $w_{7,...,14}$\tabularnewline
$1/4$ & $1/8$ & $-$\tabularnewline
$2/9$ & $1/9$ & $1/72$\tabularnewline
\end{tabular}\tabularnewline
\hline 
\end{tabular}
\par\end{centering}{\footnotesize \par}

~

\caption{\label{tab:Values-Lattices}Values of $N_{pop}$, $w_{i}$, and $e^{2}$
for lattices D2Q5, D2Q9, D3Q7 and D3Q15.}
\end{table}

Although the phase-field equation (Eq. (\ref{eq:PhaseField_KarmaRappel}))
is a scalar one, the simplest lattices such as D2Q5 and D3Q7 are avoided.
Indeed, some unwanted effects of grid anisotropy can occur by using
them. For instance, the choice $\varepsilon_{s}=0$ in Eq. (\ref{eq:As_Function_Classical})
does not yield to a solution with a spherical shape (see \citep{Cartalade_etal_CAMWA2015}).
Let us mention that the same problem occurs when the laplacian term
of phase-field equation is discretized with a finite difference method
which uses only the first four (in 2D) or six (in 3D) neighbors \citep{Karma-Rappel_PRE1998}.
That problem is fixed by using eighteen neighboring nodes in 3D simulations
\citep{Bragard_etal_InterfSci2002}. At last, the lattices D3Q19 and
D3Q27 are not applied because the same solution as D3Q15 is obtained
by using them. Those two lattices use more moving directions and require
more memory.

The two-dimensional lattices D2Q5 and D2Q9 are defined as follows.
The number of moving directions are respectively $N_{pop}=4$ and
$N_{pop}=8$. For D2Q5, vectors $\mathbf{e}_{i}$ are defined by $\mathbf{e}_{0}=(0,\,0)^{T}$,
$\mathbf{e}_{1}=(1,\,0)^{T}$, $\mathbf{e}_{2}=(0,\,1)^{T}$, $\mathbf{e}_{3}=(-1,\,0)^{T}$
and $\mathbf{e}_{4}=(0,\,-1)^{T}$ (Fig. \ref{fig:Lattices}a). For
D2Q9 the first five vectors are the same (blue vectors on Fig. \ref{fig:Lattices}b)
and the last four vectors correspond to the four diagonals of the
square (red vectors on Fig. \ref{fig:Lattices}b): $\mathbf{e}_{5}=(1,\,1)^{T}$,
$\mathbf{e}_{6}=(-1,\,1)^{T}$, $\mathbf{e}_{7}=(-1,\,-1)^{T}$, $\mathbf{e}_{8}=(1,\,-1)^{T}$.

For D3Q7, the number of moving directions is $N_{pop}=6$. Vectors
$\mathbf{e}_{i}$ are defined such as $\mathbf{e}_{0}=(0,\,0,\,0)^{T}$,
$\mathbf{e}_{1}=(1,\,0,\,0)^{T}$, $\mathbf{e}_{2}=(0,\,1,\,0)^{T}$,~$\mathbf{e}_{3}=(-1,\,0,\,0)^{T}$,
$\mathbf{e}_{4}=(0,\,-1,\,0)^{T}$, $\mathbf{e}_{5}=(0,\,0,\,1)^{T}$
and $\mathbf{e}_{6}=(0,\,0,\,-1)^{T}$ (Fig. \ref{fig:Lattices}c).
For D3Q15, the number of moving directions is $N_{pop}=14$. The first
seven vectors $\mathbf{e}_{i}$ are the same as D3Q7 (blue vectors
on Fig. \ref{fig:Lattices}d). The eight other directions correspond
to the eight diagonals of the cube (red vectors on Fig. \ref{fig:Lattices}d)
and are defined by: $\mathbf{e}_{7}=(1,\,1,\,1)^{T}$, $\mathbf{e}_{8}=(-1,\,1,\,1)^{T}$,
$\mathbf{e}_{9}=(-1,\,-1,\,1)^{T}$, $\mathbf{e}_{10}=(1,\,-1,\,1)^{T}$,
$\mathbf{e}_{11}=(1,\,1,\,-1)^{T}$, $\mathbf{e}_{12}=(-1,\,1,\,-1)^{T}$,
$\mathbf{e}_{13}=(-1,\,-1,\,-1)^{T}$, $\mathbf{e}_{14}=(1,\,-1,\,-1)^{T}$
.

Finally, values of $e^{2}$ and $w_{i}$ for those four lattices are
given in Table \ref{tab:Values-Lattices}. All quantities are calculated
at nodes and the weights $w_{i}$ are defined such as $\sum_{i=1,...,N_{pop}}w_{i}=1$.
As usual in LB method, the condition $\delta x=\delta y=\delta z$
is assumed: meshes are squares (in 2D) or cubes (in 3D). In this work,
for all boundaries of computational domain, the standard bounce-back
method is applied for both distribution functions $f_{i}$ and $g_{i}$:

\begin{subequations}

\begin{align}
f_{i}(\mathbf{x},\, t) & =f_{i'}(\mathbf{x},\, t),\label{eq:Bounce-Back}\\
g_{i}(\mathbf{x},\, t) & =g_{i'}(\mathbf{x},\, t),
\end{align}
where $i'$ is the opposite direction of $i$. For instance, for temperature
equation, after the streaming step, components $f_{1}$, $f_{5}$
and $f_{8}$ are unknown at the left boundary. The bounce-back method
consists to update each component $f_{i}$ with the value of $f_{i'}$
which moves in opposite direction i.e. $f_{1}=f_{3}$, $f_{5}=f_{7}$
and $f_{8}=f_{6}$.

\end{subequations}

LB methods are composed of Eqs. (\ref{eq:LBE_Eq_Temp})--(\ref{eq:ThermalDiffusivity})
for temperature equation and Eqs. (\ref{eq:LBE_Eq_Phase}), (\ref{eq:Feq_Phase})
and (\ref{eq:Def_M0_Phi})--(\ref{eq:LBE_RelaxationTime}) for phase-field
equation. Numerical implementation of those schemes is validated in
\citep{Cartalade_etal_CAMWA2015} by comparing the tip velocity evolution
with a finite difference method described in reference \citep{Plapp-Karma_JCP2000}.
The comparisons were carried out on a D2Q9 lattice with an anisotropy
function defined by Eq. (\ref{eq:As_Function_Classical}) for two
undercoolings $\Delta=0.3$ and $\Delta=0.55$. The methodology was
applied to simulate hydrodynamics effects on crystal growth in \citep{Cartalade_etal_MSpro2014}
and crystal growth of dilute binary mixture in \citep{Younsi_etal_ProcIHTC15-2014}.

\subsection{Validations}

\begin{figure}

\begin{centering}
\includegraphics[angle=-90,scale=0.4]{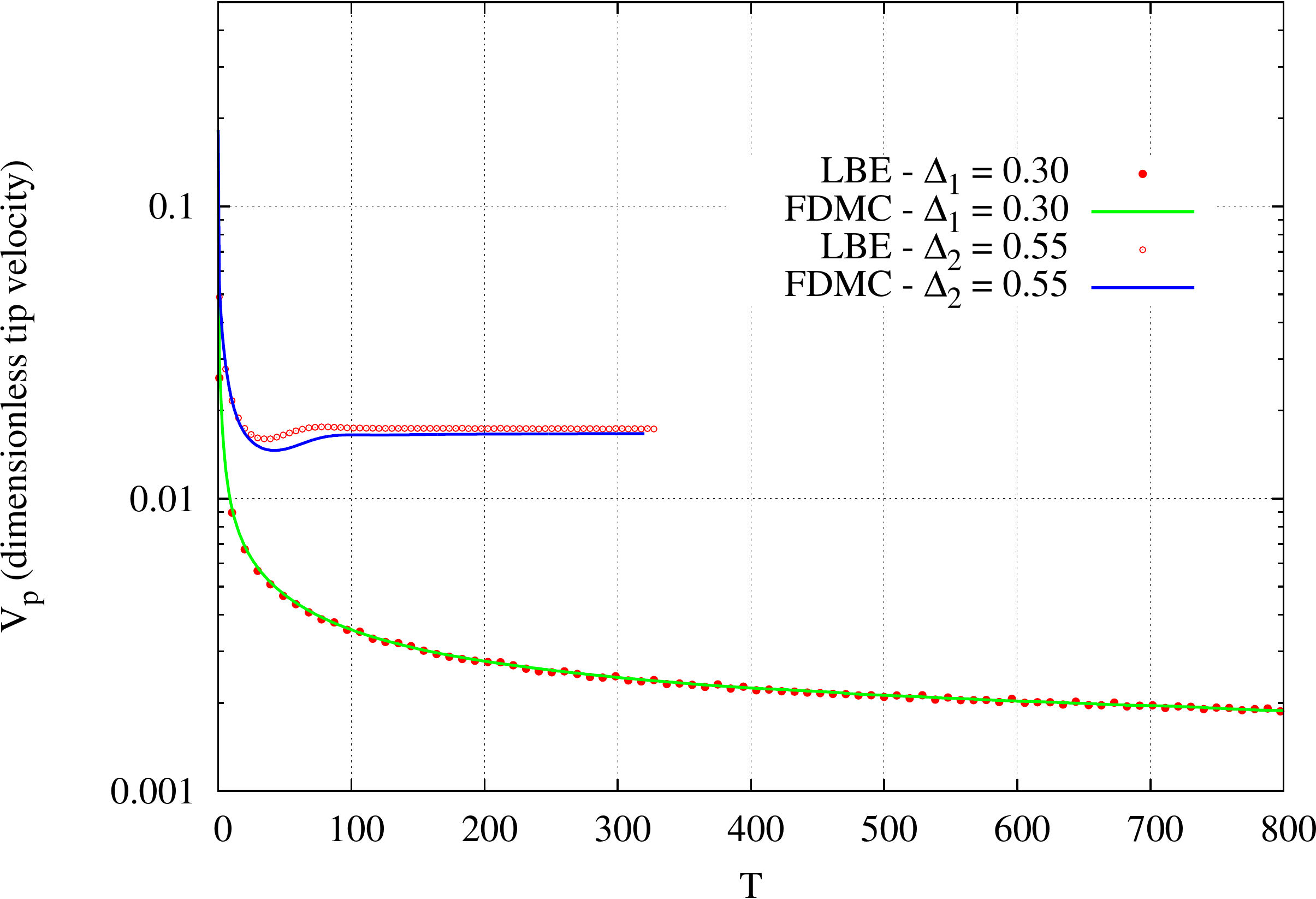}
\par\end{centering}

~

\caption{\label{fig:Validation-LBM}Dimensionless tip velocity $V_{p}$ as
a function of time for FDMC (solid lines) and LBE (red dots) for two
undercoolings $\Delta_{1}=0.30$ and $\Delta_{2}=0.55$ (from \citep{Cartalade_etal_CAMWA2015}).
Parameters are $\kappa=4$, $\lambda^{\star}=6.3826$, $d_{0}=0.1385$
and $\varepsilon_{s}=0.05$.}

\end{figure}

That LB algorithm was validated with another numerical code by comparing
the tip velocity $V_{p}$ and the dendrite shape ($\phi=0$) for two
undercoolings: $\Delta_{1}=0.30$ and $\Delta_{2}=0.55$. We used
for the comparison a 2D numerical code based on a Finite Difference
method for phase-field equation and a Monte-Carlo algorithm for temperature
\citep{Plapp-Karma_JCP2000}. For LB schemes, we used the lattices
D2Q9 for Eq. (\ref{eq:PhaseField_KarmaRappel}) and D2Q5 for Eq. (\ref{eq:Temp_Ramirez}).
On Fig. \ref{fig:Validation-LBM}, results of first code are labeled
by <<FDMC>> and results for LB schemes are labeled by <<LBE>>.

The domain is a square discretized with meshes of size $\delta x$.
The initial seed is a diffuse circle of radius $R_{s}=10\delta x$
which is set at the origin of the computational domain. The problem
is symmetrical with respect to the $x$-axis and $y$-axis. The interface
thickness $W_{0}$ and the characteristic time $\tau_{0}$ are set
to $W_{0}=\tau_{0}=1$. The space step is chosen such as $\delta x/W_{0}=0.4$
\citep{Karma-Rappel_PRE1998}, the time step is $\delta t=0.008$
and the lengths of the system depend on the undercooling $\Delta=-u_{0}$.
A smaller undercooling necessitates a bigger mesh because of the larger
diffusive length. The time to reach the stationary velocity is also
more important. For $\Delta_{1}=0.30$ and $\Delta_{2}=0.55$, we
have used respectively a mesh of $1000^{2}$ nodes and $500^{2}$
nodes. The capillary length $d_{0}$ and the kinetic coefficient $\beta$
are respectively given by Eq. (\ref{eq:Capillary-Length}) and Eq.
(\ref{eq:Kinetic-Coeff}) with $a_{1}\approx0.8839$ and $a_{2}\approx0.6267$.
In the benchmark, we have chosen the parameter $\lambda$ such as
$\beta=0$, i.e. $\lambda^{\star}=\kappa\tau_{0}/a_{2}W_{0}^{2}$.
With $W_{0}=\tau_{0}=1$, the coefficient $\lambda^{\star}$ is equal
to $\lambda^{\star}=1.59566\kappa$. For a thermal diffusivity equals
to $\kappa=4$, the coefficients are $\lambda^{\star}=6.3826$ and
$d_{0}=0.1385$.

Results of such a benchmark are presented on Fig. \ref{fig:Validation-LBM}
for $\varepsilon_{s}=0.05$. In the comparisons, the solid lines represent
the results for FDMC method and the red dots represent the results
for LB schemes. The tip velocity $V_{p}$ is dimensionless by using
the factor $d_{0}/\kappa$ ($V_{p}=\tilde{V}_{p}d_{0}/\kappa$), the
position $x$ is also dimensionless by using the space-step ($x=\tilde{x}/\delta x$)
and the time $T$ is the time $t$ divided by $\tau_{0}$ ($T=t/\tau_{0}$).
That benchmark validates the LB schemes presented in Section \ref{sec:Lattice-Boltzmann-schemes}.
More details about relative errors of those results are commented
in subsection 4.1 of reference \citep{Cartalade_etal_CAMWA2015}.
The superposition of dendrite shapes $\phi=0$ can be found in this
same reference.

\section{\label{sec:Anisotropic-functions}Anisotropy functions}

This section focuses on the $a_{s}(\mathbf{n})$ function responsible
for the characteristic shapes of crystals. In LB schemes for phase-field
model, the $a_{s}$ function appears \emph{(i)} in the phase-field
equation (Eq. (\ref{eq:LBE_Eq_Phase})), \emph{(ii)} in the vectorial
function $\boldsymbol{\mathcal{N}}(\mathbf{x},\, t)$ of e.d.f. (Eq.
(\ref{eq:Feq_Phase})) and \emph{(iii)} in the relaxation rate $\eta_{\phi}$
(Eq. (\ref{eq:LBE_RelaxationTime})). This section investigates the
anisotropy function for which the growth of branches are not directed
along the main axes of the coordinate system, i.e. we focus on the
anisotropy function for a preferential growth other than $\left\langle 10\right\rangle $
in 2D and $\left\langle 100\right\rangle $ in 3D. First, main formulations
of the $a_{s}$ function are reminded and next, effects of gradient
calculation on the crystal shape are presented.

\subsection{\label{sub:Formulations-anisotropy}Spherical or Cartesian formulations}

Two main formulations are used in the literature to simulate dendritic
crystals. In 2D, the standard formulations are based on the angle
$\varphi\equiv\varphi(\mathbf{x},\, t)$ (e.g. \citep{Kobayashi_PhysD1993})
between the normal vector of the interface and the $x$-axis. In phase-field
models, this angle is expressed in terms of derivatives of the phase-field
$\phi$. Definition of normal vector is given by $\mathbf{n}=-\boldsymbol{\nabla}\phi/\bigl|\boldsymbol{\nabla}\phi\bigr|$
and its components are $n_{x}=-\partial_{x}\phi/\bigl|\boldsymbol{\nabla}\phi\bigr|$
and $n_{y}=-\partial_{y}\phi/\bigl|\boldsymbol{\nabla}\phi\bigr|$.
The angle $\varphi$ is defined by: $\tan\varphi=\sin\varphi/\cos\varphi=n_{y}/n_{x}=\partial_{y}\phi/\partial_{x}\phi$
i.e. it can be obtained by the relationship $\varphi=\tan^{-1}(\partial_{y}\phi/\partial_{x}\phi)$.
Therefore, if $\partial_{x}\phi\neq0$, the angle $\varphi$ is calculated
from the phase-field $\phi$. The standard form of $a_{s}(\varphi)$
in 2D is:

\begin{equation}
a_{s}(\varphi)=1+\varepsilon_{s}\cos\left[q\left(\varphi(\mathbf{x},\, t)-\varphi_{0}\right)\right],\label{eq:As_Theta}
\end{equation}
where $\varphi_{0}$ is a reference angle. Both coefficients $\varepsilon_{s}>0$
and $q$ are respectively the strength and mode of anisotropy. For
instance, the choice $q=4$ and $\varphi_{0}=0$ leads to the development
of crystal with four tips directed along the main axes $x$ and $y$.
Indeed, the growth is maximal when $\cos(4\varphi)$ is equal to one,
i.e. for multiple integers of $\varphi=\pi/2$. Other similar 2D formulations
exist in the literature. For instance in \citep{Zhao_etal_IJNMF2005}:
$a_{s}(\varphi)=1+\varepsilon_{s}\left[8/3\sin^{4}(q/2(\varphi-\varphi_{0}))-1\right]$.
The power four of the sine function yields an asymmetry between the
maximal and minimal values of $a_{s}$. Also note the modification
of $a_{s}(\varphi)$ for simulating crystals that are faceted \citep{Debierre_etal_PRE2003,Miura_WulfShape_JCG2013}.
Those shapes are not studied in this work.

A formulation of $a_{s}(\varphi)$, defined by an angle $\varphi,$
is equivalent to $a_{s}(\mathbf{n})$ defined by a normal vector $\mathbf{n}$.
By expanding the particular case $a_{s}(\varphi)=1+\varepsilon_{s}\cos(4\varphi)$
in powers of $\cos\varphi$ and $\sin\varphi$ (with $\cos(4\varphi)=\cos2(2\varphi)$),
and next by identifying $n_{x}=\cos\varphi$ and $n_{y}=\sin\varphi$,
the 2D relationship $a_{s}(\mathbf{n})=1-3\varepsilon_{s}+4\varepsilon_{s}(n_{x}^{4}+n_{y}^{4})$
is obtained \citep{Karma-Rappel_PRE1998}. More generally, the function
$\cos(q\varphi)$ can be derived from a formalism involving spherical
harmonics. On a spherical surface, any scalar function can be decomposed
on a basis of spherical functions depending on two angles $\varphi$
and $\theta$: the spherical harmonics. The distance $r$ is the third
coordinate to define the position on a sphere. For instance $\cos(4\varphi)$
corresponds to the real part of the spherical harmonic $Y_{4,4}(\theta,\,\varphi)$
in the $XY$-plane (i.e. with $\theta=\pi/2$): $\mbox{Re}[Y_{4,4}(\pi/2,\,\varphi)]\propto\cos(4\varphi)$.
The symbol $\propto$ means that the spherical harmonic $Y_{4,4}$
is proportional to that function with a normalization factor.

An application of such a formalism is given here. The $a_{s}(\mathbf{n})$
function that is derived will be used in Section \ref{sec:Simulations},
for comparisons between a 2D code which uses a $\varphi$-formulation
and a 3D code which uses a $\mathbf{n}$-formulation. That benchmark
will be carried out in order to validate numerical implementation
of the directional derivatives method. If we want to derive a $a_{s}(\mathbf{n})$
function that is equivalent to $a_{s}(\varphi)=1+\varepsilon_{s}\cos(6\varphi)$,
the real part of $Y_{6,6}(\theta,\,\varphi)$ is considered: $\mbox{Re(}Y_{6,6}(\theta,\,\varphi))\propto\cos(6\varphi)\sin^{6}\theta$.
After developing $\cos(6\varphi)$ and using the variable change from
spherical to Cartesian coordinates, $\sin\theta\cos\varphi=x/r$,
$\sin\theta\sin\varphi=y/r$ and $\cos\theta=z/r$, the calculation
leads to $\mbox{Re}(Y_{6,6}(\mathbf{x}))\propto(x^{6}-15x^{4}y^{2}+15x^{2}y^{4}-y^{6})/r^{6}$.
After identifying $n_{x}=x/r$, $n_{y}=y/r$ and $n_{z}=z/r$, the
anisotropy function is:

\begin{equation}
a_{s}(\mathbf{n})=1+\varepsilon_{s}(n_{x}^{6}-15n_{x}^{4}n_{y}^{2}+15n_{x}^{2}n_{y}^{4}-n_{y}^{6}).\label{eq:As_Y66}
\end{equation}
Eq. (\ref{eq:As_Y66}) favors crystal growth in six main growing directions
in the $XY$-plane. That function will be used in Section \ref{sub:Validation_2D-3D}.
The same procedure can be followed for other characteristic shapes
after identification of spherical harmonics suited to the problem.

In 3D, any anisotropy function $a_{s}(\mathbf{n})$ can be expressed
as a linear combination of cubic harmonics. The cubic harmonics are
linear combinations of real spherical harmonics with an appropriate
cubic symmetry \citep{Fehlner-Vosko_CJP1976,Mueller-Priestley_PR1966,Puff_PhyStaSol1970}.
For instance, function $Q(\mathbf{n})=n_{x}^{4}+n_{y}^{4}+n_{z}^{4}$
of Eq. (\ref{eq:As_Function_Classical}), which favors the growth
in $\left\langle 100\right\rangle $ preferential direction, can be
derived from the cubic harmonic $K_{4,1}(\theta,\,\varphi)\propto\left[5\cos^{4}\theta-3+5\sin^{4}\theta(\cos^{4}\varphi+\sin^{4}\varphi)\right]$.
After the variable change in Cartesian coordinates, that function
becomes $K_{4,1}(\theta,\,\varphi)\propto\left[5(n_{x}^{4}+n_{y}^{4}+n_{z}^{4})-3\right]$.
By considering an additional cubic harmonic $K_{6,1}$, the $a_{s}(\mathbf{n})$
function can be generalized for other growing directions. For instance,
the $\left\langle 110\right\rangle $-direction can be simulated by
\citep{Hoyt-Asta-Karma_MatSciEng2003}:

\begin{align}
a_{s}(\mathbf{n}) & =1+\varepsilon_{s}\left(\sum_{\alpha=x,y,z}n_{\alpha}^{4}-\frac{3}{5}\right)+\gamma\left(3\sum_{\alpha=x,y,z}n_{\alpha}^{4}+66n_{x}^{2}n_{y}^{2}n_{z}^{2}-\frac{17}{7}\right),\label{eq:As_MatSci}
\end{align}
where $\varepsilon_{s}$ is the anisotropy strength in the $\left\langle 100\right\rangle $-directions
and $\gamma$ is the anisotropy strength in the $\left\langle 110\right\rangle $-directions.
The first term of the right-hand side is the cubic harmonic $K_{0,0}=1$,
the second one is $K_{4,1}\propto[5Q(\mathbf{n})-3]$ and the last
one is the cubic harmonic $K_{6,1}\propto[462S(\mathbf{n})+21Q(\mathbf{n})-17]$
with $S(\mathbf{n})=n_{x}^{2}n_{y}^{2}n_{z}^{2}$. We can refer to
\citep{Fehlner-Vosko_CJP1976,Putzai-etal_WulfShape_JCP2014} for other
cubic harmonics of higher order, defined in terms of functions $Q(\mathbf{n})$
and $S(\mathbf{n})$.

Let us mention that several experiments and simulations, involving
the anisotropy function (Eq. (\ref{eq:As_MatSci})), were carried
out in the literature to study the <<dendrite orientation transition>>
from $\left\langle 100\right\rangle $ to $\left\langle 110\right\rangle $
\citep{Karma_Orientation_Nature2006,Dantzig-Rappaz_MMTA2013}. In
those references, phase-field simulations are performed by varying
systematically the anisotropy parameters $\varepsilon_{s}$ and $\gamma$
to explore the role of these parameters on the resulting microstructures.
Simulations are carried out for pure substance and binary mixture
for equiaxed crystals and directional solidification.

\begin{figure*}
\begin{centering}
\begin{tabular}{ccccc}
{\small (a) $Q(\mathbf{n})=n_{x}^{4}+n_{y}^{4}+n_{z}^{4}$} &  & {\small (b) $S(\mathbf{n})=n_{x}^{2}n_{y}^{2}n_{z}^{2}$} &  & {\small (c) $3Q(\mathbf{n})+66S(\mathbf{n})$}\tabularnewline
 &  &  &  & \tabularnewline
\includegraphics[scale=0.24]{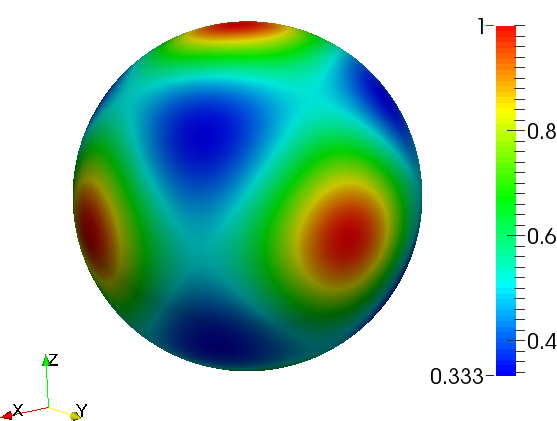} &  & \includegraphics[scale=0.24]{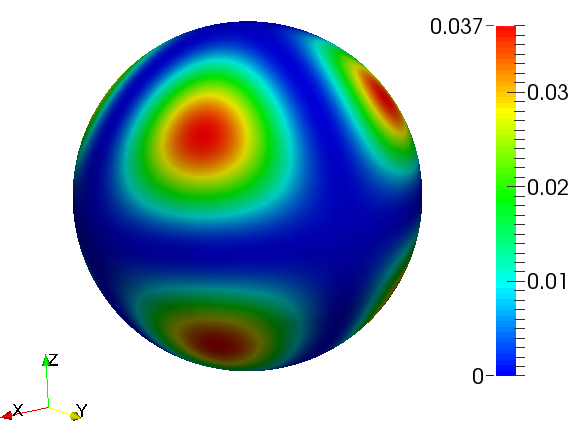} &  & \includegraphics[scale=0.24]{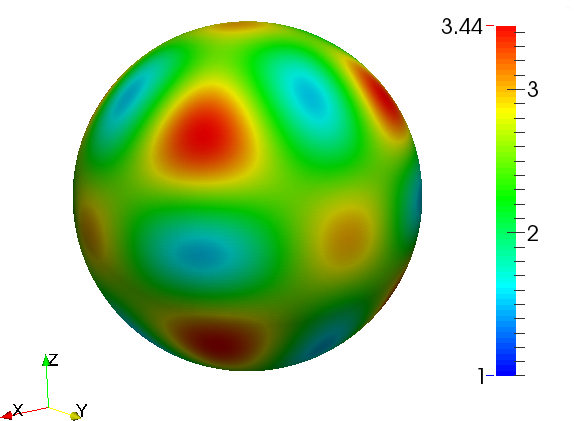}\tabularnewline
 &  &  &  & \tabularnewline
{\small (d) $a_{s}(\mathbf{n})$ with $\varepsilon_{s}=0.02$, $\gamma=-0.02$} &  & {\small (e) $a_{s}(\mathbf{n})$ with $\varepsilon_{s}=0$, $\gamma=-0.02$} &  & {\small (f) $a_{s}(\mathbf{n})$ with $\varepsilon_{s}=0.05$, $\gamma=0.02$}\tabularnewline
 &  &  &  & \tabularnewline
\includegraphics[scale=0.24]{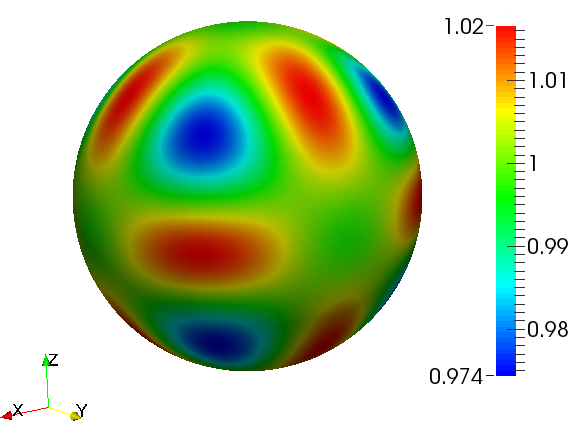} &  & \includegraphics[scale=0.24]{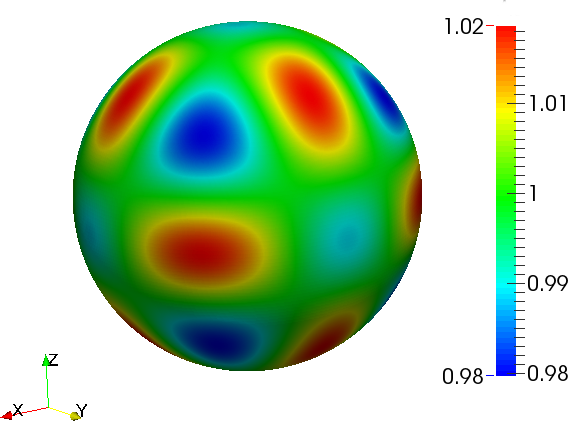} &  & \includegraphics[scale=0.24]{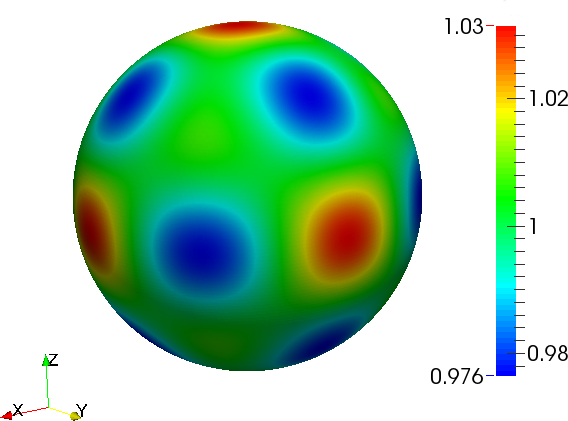}\tabularnewline
\end{tabular}
\par\end{centering}

\caption{\label{fig:Representations_As}Graphical representation of each term
of Eq. (\ref{eq:As_MatSci}).}
\end{figure*}

In order to see the influence of each term of Eq. (\ref{eq:As_MatSci}),
a graphical representation is plotted on a spherical surface. First,
the phase-field $\phi(\mathbf{x},\,0)$ is initialized (by using Eq.
(\ref{eq:Initial_Condition})) inside a cubic domain composed of $200^{3}$
nodes. The initial sphere is set at the center of the domain $\mathbf{x}_{s}=(100,\,100,\,100)^{T}$
with a radius equals to $R_{s}=50$ lattice units (l.u.). Next, the
components $n_{x}$, $n_{y}$ and $n_{z}$ are derived by calculating
the gradient of $\phi$. Finally, each term of Eq. (\ref{eq:As_MatSci})
is calculated and post-processed on a spherical surface of radius
$R_{s}$ and centered at $\mathbf{x}_{s}$. Results are presented
in Figs. \ref{fig:Representations_As} for various values of $\varepsilon_{s}$
and $\gamma$. On those figures, the red and blue areas correspond
to the zones for which the growth is respectively maximal and minimal.
All figures are plotted for a same orientation of the coordinate system
to facilitate the comparison between functions. As expected, function
$Q(\mathbf{n})$ favors growth in the $\left\langle 100\right\rangle $-direction
(Fig. \ref{fig:Representations_As}a); $S(\mathbf{n})$ favors growth
in the $\left\langle 111\right\rangle $-direction (Fig. \ref{fig:Representations_As}b),
and $a_{s}(\mathbf{n})$ defined by Eq. (\ref{eq:As_MatSci}) with
$\varepsilon_{s}=0$ and $\gamma=-0.02$ favors growth in the $\left\langle 110\right\rangle $-direction
(Fig. \ref{fig:Representations_As}e). Simulations of crystal growth
with those three anisotropy functions will be carried out in Section
\ref{sec:Simulations}.

\subsection{\label{sub:Derivees-directionnelles}Effect of directional derivatives
of gradients in LB schemes}

For each formulation of $a_{s}$, it is necessary to compute the gradient
of $\phi$ for deriving the components of $\mathbf{n}$. It is also
necessary to compute the derivatives of $a_{s}(\mathbf{n})$ with
respect to $\partial_{x}\phi$, $\partial_{y}\phi$ and $\partial_{z}\phi$
involved in the vector $\boldsymbol{\mathcal{N}}(\mathbf{x},\, t)$.
Two approaches were compared in this work. In the first one, the gradient
is calculated by using the classical formula of central finite differences,
i.e. in 2D: $\partial_{x}\phi\simeq(\phi_{j+1,\, k}-\phi_{j-1,\, k})/2\delta x$
and $\partial_{y}\phi\simeq(\phi_{j,\, k+1}-\phi_{j,\, k-1})/2\delta x$,
where $j$ and $k$ are indexes of coordinates $x$ and $y$.

In the second approach, the method based on the directional derivatives
is applied. The method has already demonstrated its performance for
hydrodynamics problem in order to reduce parasitic currents for two-phase
flow problem \citep{Lee-Fischer_PRE2006,Lee_Parasitic_CAMWA2009,Lee-Liu_DropImpact_LBM_JCP2010}.
The directional derivative is the derivative along each moving direction
on the lattice. Taylor's expansion at second-order of a differentiable
scalar function $\phi(\mathbf{x})$ at $\mathbf{x}+\mathbf{e}_{i}\delta x$
and $\mathbf{x}-\mathbf{e}_{i}\delta x$ yields the following approximation
of directional derivatives:

\begin{subequations}

\begin{equation}
\mathbf{e}_{i}\cdot\boldsymbol{\nabla}\phi\bigr|_{\mathbf{x}}=\frac{1}{2\delta x}\left[\phi(\mathbf{x}+\mathbf{e}_{i}\delta x)-\phi(\mathbf{x}-\mathbf{e}_{i}\delta x)\right].\label{eq:Def_DeriveeDirectionnelle}
\end{equation}
The number of directional derivatives is equal to the number of moving
direction $\mathbf{e}_{i}$ on the lattice i.e. $N_{pop}$. The gradient
is obtained by:

\begin{equation}
\boldsymbol{\nabla}\phi\bigr|_{\mathbf{x}}=\frac{1}{e^{2}}\sum_{i=0}^{N_{pop}}w_{i}\mathbf{e}_{i}\left(\mathbf{e}_{i}\cdot\boldsymbol{\nabla}\phi\bigr|_{\mathbf{x}}\right).\label{eq:Grad_DF-Centree}
\end{equation}

The three components of gradient $\partial_{x}\phi$, $\partial_{y}\phi$
and $\partial_{z}\phi$ are obtained by calculating each directional
derivative with the relationship (\ref{eq:Def_DeriveeDirectionnelle})
and next, by calculating the moment of first order with Eq. (\ref{eq:Grad_DF-Centree}).
In Eq. (\ref{eq:Grad_DF-Centree}), each direction is weighted by
coefficients $w_{i}$ and the sum is normalized by factor $1/e^{2}$.
For gradient computation, the number and directions of directional
derivatives are consistent with the number and moving directions of
distribution functions. The directional derivatives method is consistent
with the lattice that are used for simulations. This is the main advantage
of the method. Indeed, on lattices D2Q9 and D3Q15, the distribution
functions $f_{i}$ and $g_{i}$ can move in diagonal directions. With
this method, each derivative along the direction of propagation is
used to calculate the gradient. Thus, the derivatives contain the
contributions of diagonal directions of lattices D2Q9 and D3Q15. Those
relationships can also be used for other lattices, such as D3Q19 and
D3Q27. Note that the second-order central difference is used in Eq.
(\ref{eq:Def_DeriveeDirectionnelle}) for approximating the directional
derivatives. Other approximations exist \citep{Lee-Liu_DropImpact_LBM_JCP2010}
such as the first-order and second-order upwind schemes (or biased
differences) respectively defined by $\mathbf{e}_{i}\cdot\boldsymbol{\nabla}^{up_{1}}\phi\bigr|_{\mathbf{x}}=[\phi(\mathbf{x}+\mathbf{e}_{i}\delta x)-\phi(\mathbf{x})]$/$\delta x$
and $\mathbf{e}_{i}\cdot\boldsymbol{\nabla}^{up_{2}}\phi\bigr|_{\mathbf{x}}=\left[-\phi(\mathbf{x}+2\mathbf{e}_{i}\delta x)+4\phi(\mathbf{x}+\mathbf{e}_{i}\delta x)-3\phi(\mathbf{x})\right]/(2\delta x)$.
If necessary, the second derivative of $\phi$ can also be obtained
by $(\mathbf{e}_{i}\cdot\boldsymbol{\nabla})^{2}\phi\bigr|_{\mathbf{x}}=\left[\phi(\mathbf{x}+\mathbf{e}_{i}\delta x)-2\phi(\mathbf{x})+\phi(\mathbf{x}-\mathbf{e}_{i}\delta x)\right]/\delta x^{2}$.
Here, the central difference approximation Eq. (\ref{eq:Def_DeriveeDirectionnelle})
is applied for all simulations.

\end{subequations}

We compare the impact of both methods on the solution $\phi$. We
simulate Eqs. (\ref{eq:PhaseField_KarmaRappel})--(\ref{eq:TermesAnisotropes})
in 2D with $a_{s}(\varphi)$ defined by Eq. (\ref{eq:As_Theta}) for
$q=6$ and $\varphi_{0}=0$. Components of gradient are calculated
by using first \emph{(i)} the classical formula of central finite
difference ($FD$) and second \emph{(ii)} the directional derivatives
($DD$) given by Eqs. (\ref{eq:Def_DeriveeDirectionnelle})--(\ref{eq:Grad_DF-Centree}).
For simulations, the mesh is composed of 800$\times$800 nodes. The
parameters are $\delta x=\delta y=0.01$, $\delta t=1.5\times10^{-5}$,
$\tau_{0}=1.5625\times10^{-4}$, $W_{0}=0.0125$, $\kappa=1$, the
undercooling is $\Delta=0.30$, $\lambda=10$, $\varepsilon_{s}=0.05$.
The seed is initialized at the center of the domain $\mathbf{x}_{c}=(400,\,400)$,
and the radius is $R_{s}=10$ lattice unit (l.u.).

\begin{figure*}
\begin{centering}
\begin{tabular}{ccc}
{\small (a) $\phi=0$ with standard finite differences} & $\qquad$ & {\small (b) $\phi=0$ with directional derivatives method}\tabularnewline
\includegraphics[angle=-90,scale=0.36]{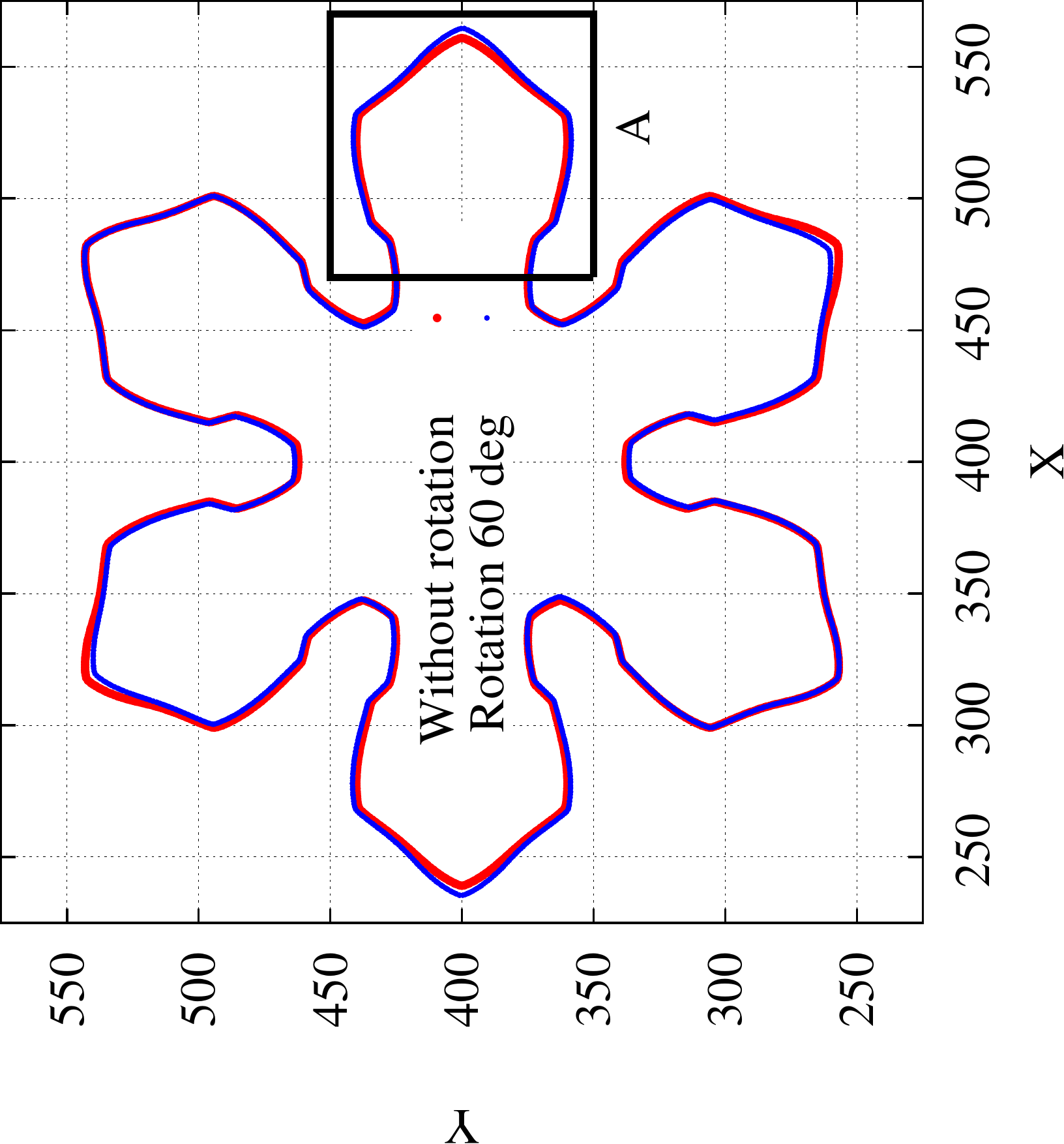} &  & \includegraphics[angle=-90,scale=0.36]{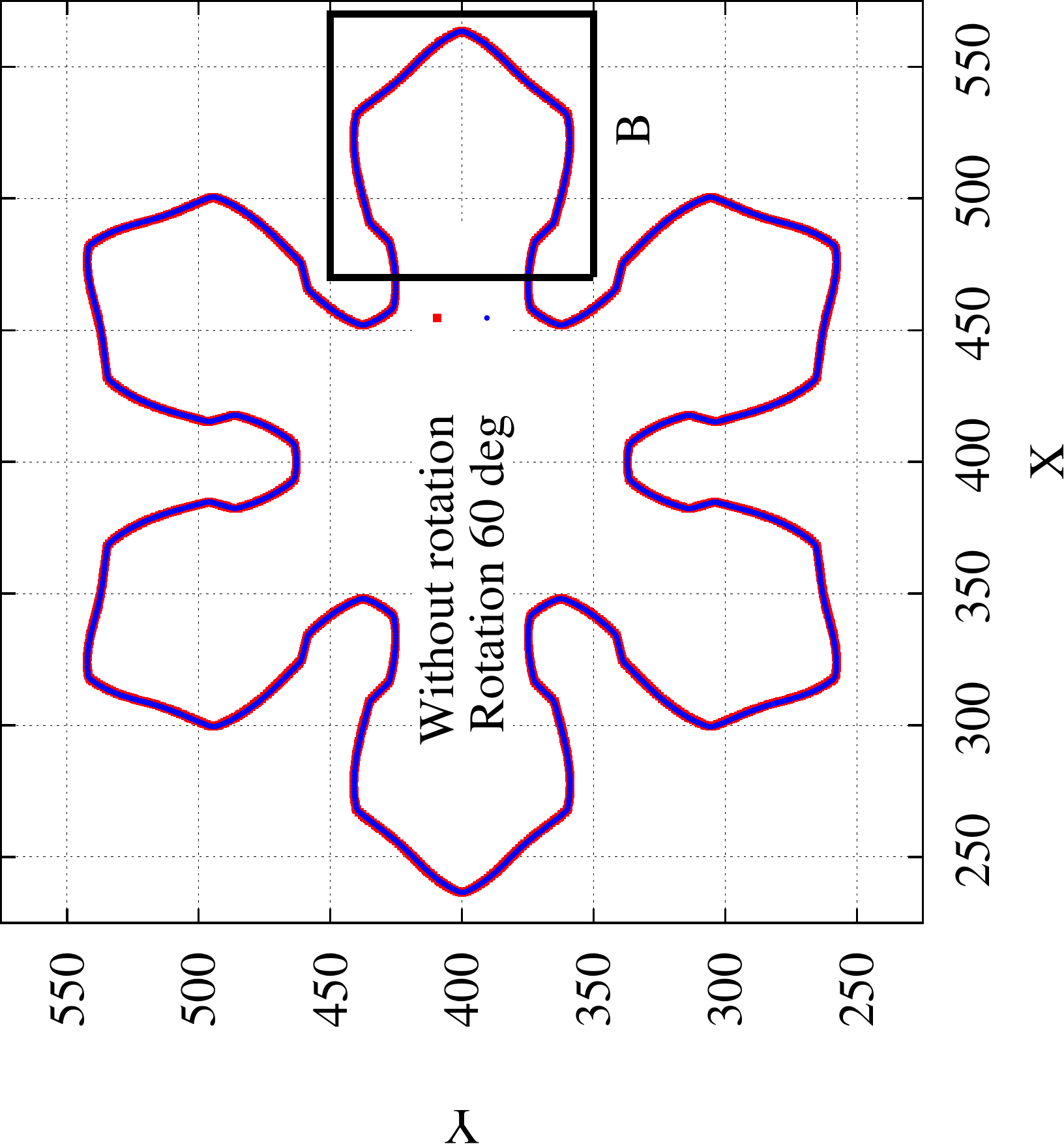}\tabularnewline
 &  & \tabularnewline
\includegraphics[angle=-90,scale=0.27]{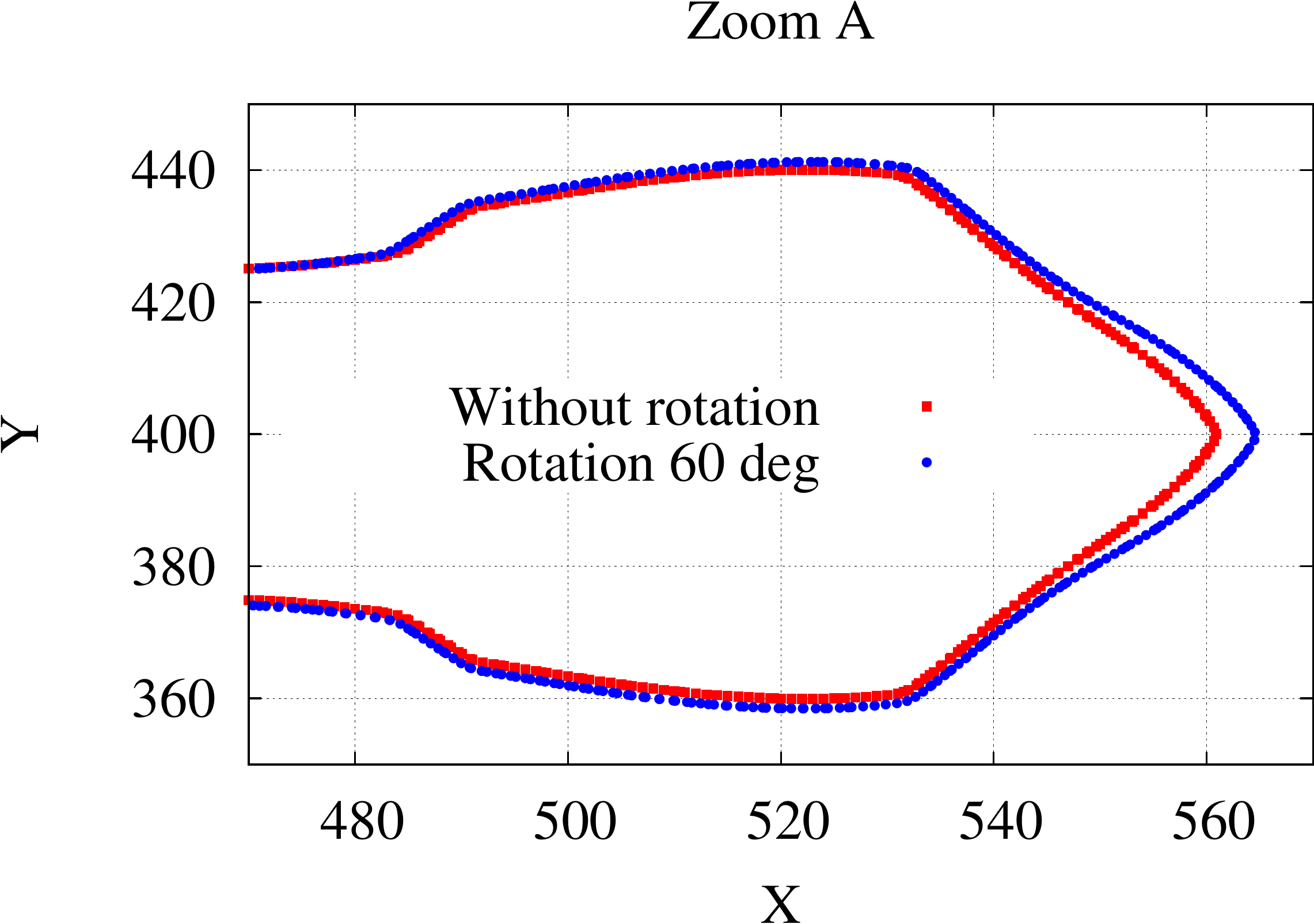} &  & \includegraphics[angle=-90,scale=0.27]{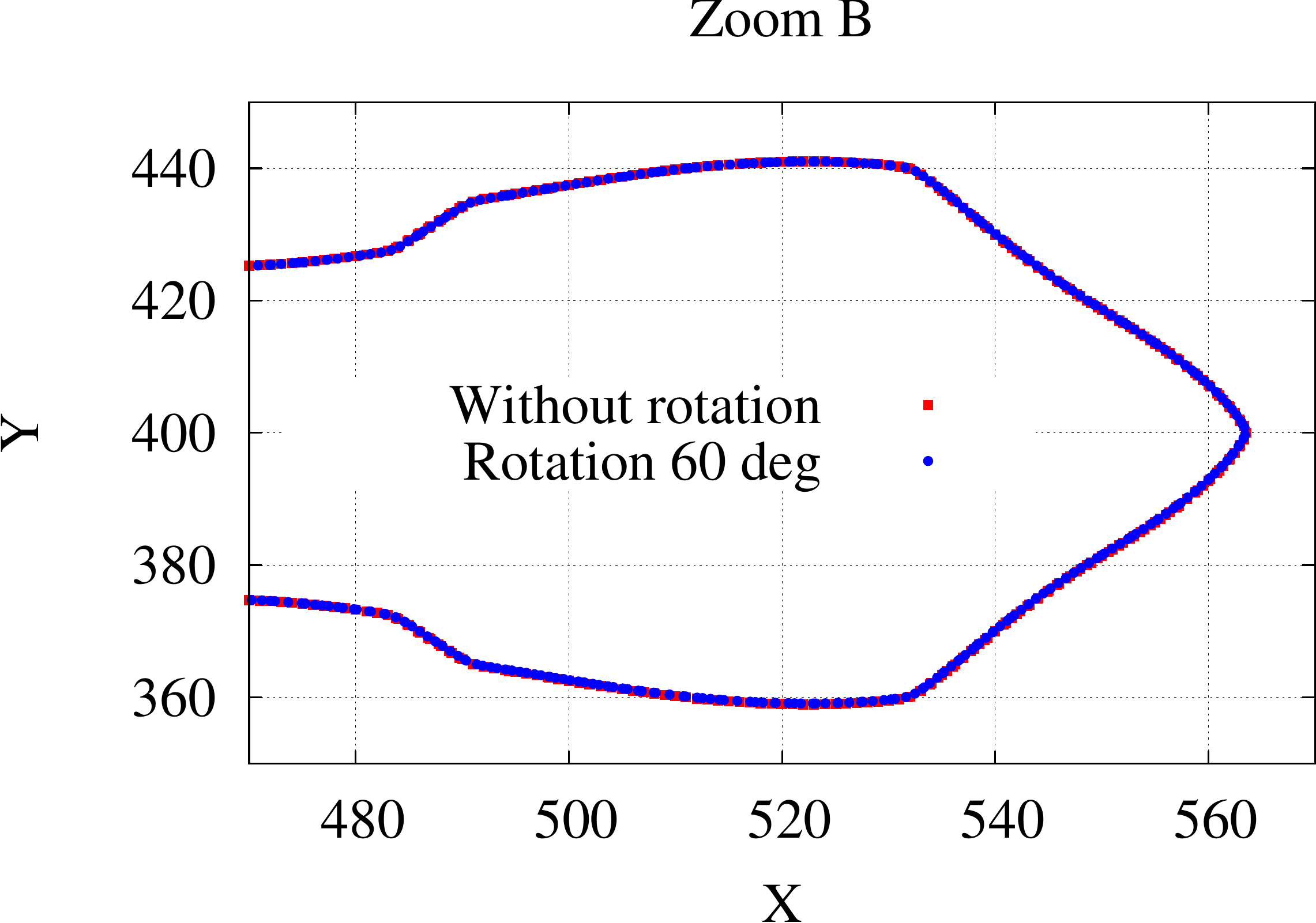}\tabularnewline
\end{tabular}
\par\end{centering}

~

\caption{\label{fig:Effet-Deriv-Direc}Effect of using a central finite difference
method (a), or a directional derivatives method (b) to calculate the
gradients of $\phi$ on a D2Q9 lattice. Without directional derivatives,
the phase-field $\phi=0$ (red) at $t=10^{5}\delta t$ does not match
with its rotation of 60\textdegree{} (blue). The rotation matches
perfectly when the directional derivatives method is applied.}
\end{figure*}

\begin{figure}

\begin{centering}
\begin{tabular}{ccc}
{\small (a) Difference $\phi^{rot}(\mathbf{x})-\phi(\mathbf{x})$
with $FD$} &  & {\small (b) Difference $\phi^{rot}(\mathbf{x})-\phi(\mathbf{x})$
with $DD$}\tabularnewline
\noalign{\vskip3mm}
\includegraphics[scale=0.23]{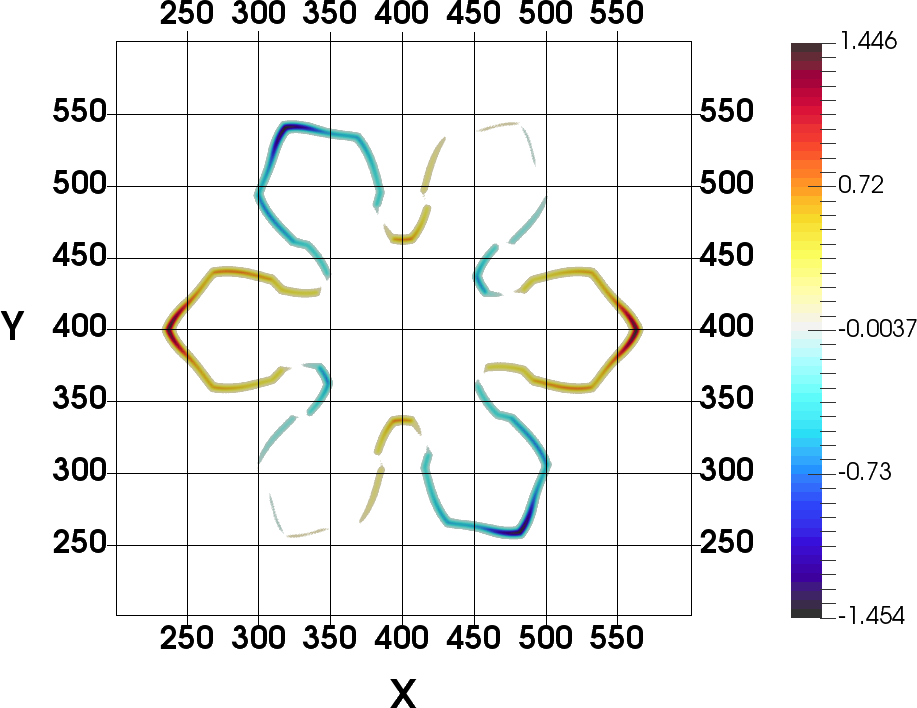} &  & \includegraphics[scale=0.23]{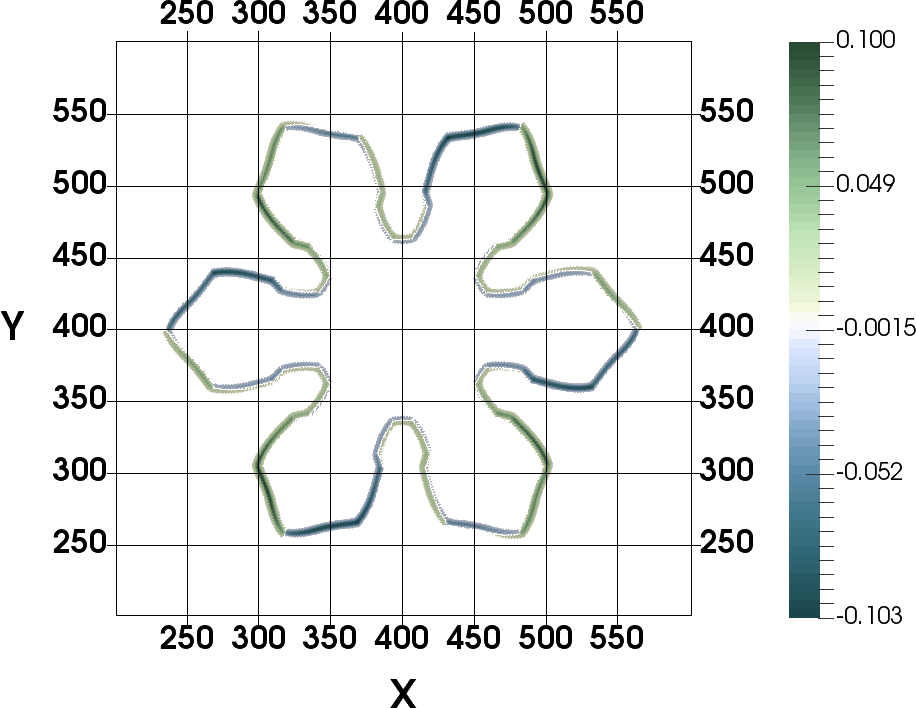}\tabularnewline
\end{tabular}
\par\end{centering}

\caption{\label{fig:Differences-Fields}Differences between $\phi^{rot}(\mathbf{x})-\phi(\mathbf{x})$
(a) with Finite Difference and (b) with Directional Derivatives. $\phi^{rot}(\mathbf{x})$
is obtained from $\phi(\mathbf{x})$ after a rotation of 60\textdegree{}.}

\end{figure}

For both methods, we present on Fig. \ref{fig:Effet-Deriv-Direc}
the solution $\phi=0$ at $t=10^{5}\delta t$ (red curve) and the
same solution with a rotation of 60\textdegree{} (blue curve). When
the gradients are calculated with a finite differences method, the
phase-field $\phi=0$ does not match perfectly with its rotation of
60\textdegree{} (Fig. \ref{fig:Effet-Deriv-Direc}a). The phase-field
is not any more isotropic and the numerical solution presents some
lattice anisotropy effects. Those effects can be seen more precisely
on Fig. \ref{fig:Differences-Fields}a which plots the difference
$\phi^{rot}(\mathbf{x})-\phi(\mathbf{x})$ with the $FD$ method.
In two main directions defined by $\mathbf{n}_{a}=(1,\,0)$ and $\mathbf{n}_{b}=(1/2,\,-\sqrt{3}/2)$
the differences are maximal at the tips (respectively $1.446$ and
$-1.454$), whereas in the third direction $\mathbf{n}_{c}=(1/2,\,+\sqrt{3}/2)$
the difference is lower than $0.01$. When the gradients are calculated
with the $DD$ method, the phase-field $\phi=0$ is much more isotropic
(Fig. \ref{fig:Effet-Deriv-Direc}b) as confirmed by the differences
$\phi^{rot}(\mathbf{x})-\phi(\mathbf{x})$ plotted on Fig.\ref{fig:Differences-Fields}b.
The differences are now more uniformly distributed inside the domain
and the maximal and minimal values are divided by a factor $14$ (respectively
$0.1$ and $-0.103$). The lattice anisotropy is reduced.

\begin{figure*}
\begin{centering}
\begin{tabular}{ccc}
{\small (a) Profiles with finite differences} & $\qquad$ & {\small (b) Profiles with directional derivatives}\tabularnewline
\includegraphics[angle=-90,scale=0.3]{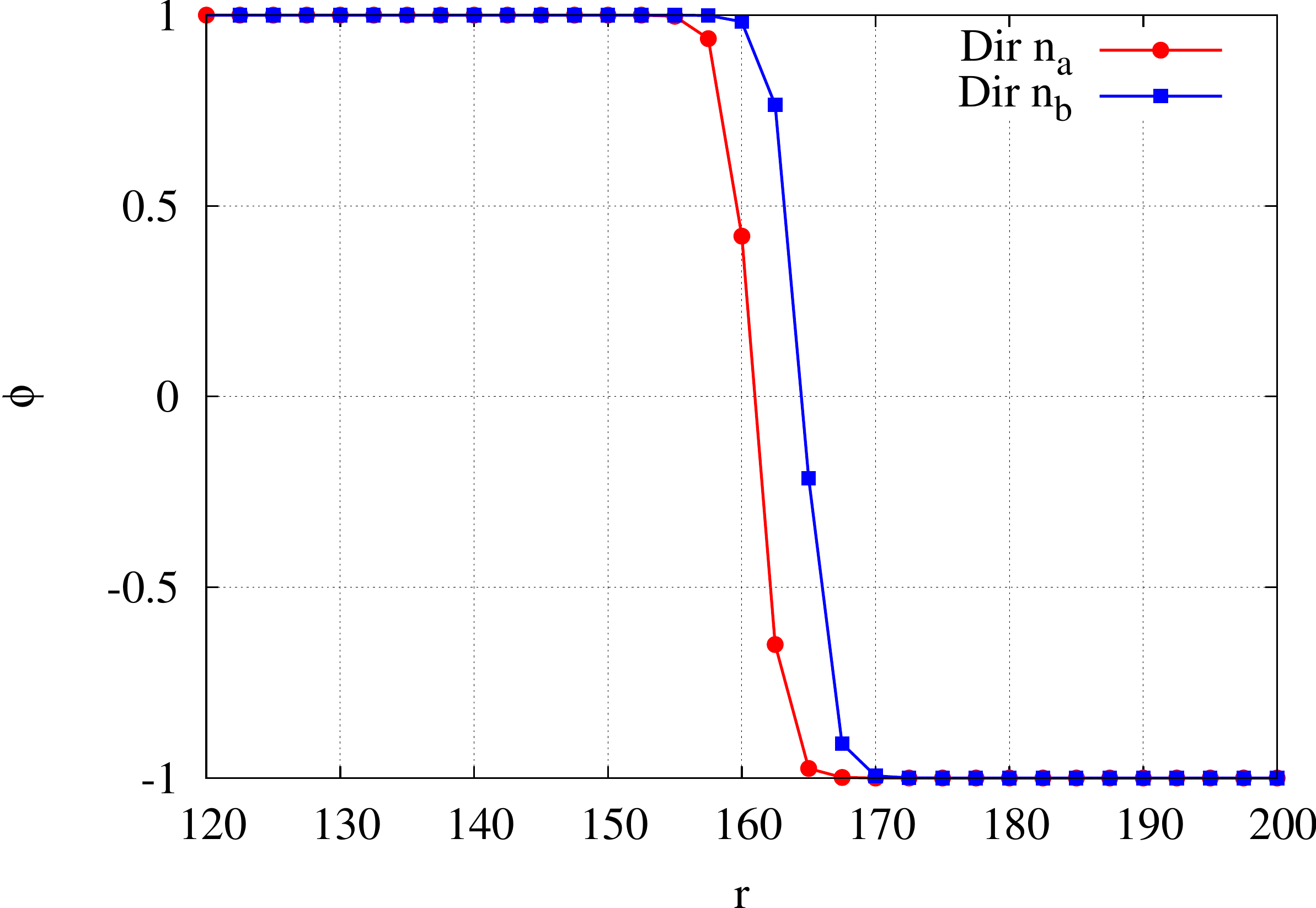} &  & \includegraphics[angle=-90,scale=0.3]{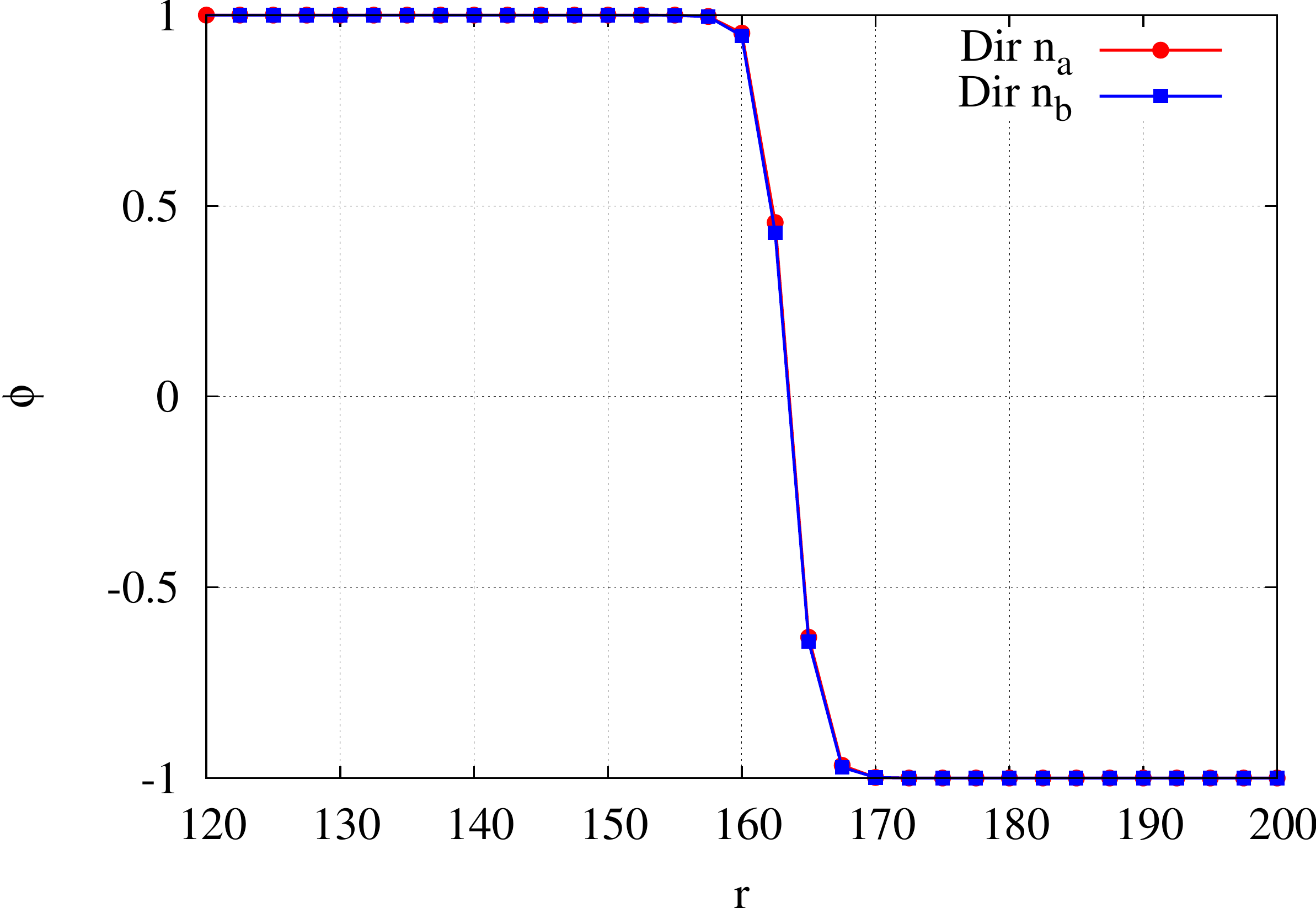}\tabularnewline
\end{tabular}
\par\end{centering}

~

\caption{\label{fig:Effet-Profiles}$\phi$-profiles along the directions $\mathbf{n}_{a}=(1,\,0)$
and $\mathbf{n}_{b}=(1/2,\,-\sqrt{3}/2)$. (a) With a finite difference
method the relative error ($\ell^{2}$-norm) is $1.71\times10^{-1}$
and (b) its value is equal to $3.07\times10^{-3}$ with a directional
derivatives method.}
\end{figure*}

In order to quantify the errors, the $\ell^{2}$ relative error norm
is calculated with two $\phi$-profiles plotted along two directions
$\mathbf{n}_{a}$ and $\mathbf{n}_{b}$ defined above. The $\ell^{2}$
relative error norm is defined by:

\begin{equation}
\mbox{Err}_{\ell^{2}}=\sqrt{\frac{\sum_{j}^{N_{j}}(\phi_{a,j}-\phi_{b,j})^{2}}{\sum_{j}^{N_{j}}\phi_{a,j}^{2}}},\label{eq:Error_L2}
\end{equation}
where $\phi_{a}$ and $\phi_{b}$ are the phase-fields collected along
the directions $\mathbf{n}_{a}$ and $\mathbf{n}_{b}$ respectively.
The profiles of $\phi_{a}$ and $\phi_{b}$ are presented on Figs.
\ref{fig:Effet-Profiles}a,b for both methods of gradient computation.
The phase-field $\phi_{a}$ sampled along the direction $\mathbf{n}_{a}$
is taken as a reference for the error calculation. $N_{j}$ is the
total number of values and $j$ is the index. The $\ell^{2}$ relative
error for the finite difference method is $\mbox{Err}{}_{\ell^{2}}^{FD}=1.71\times10^{-1}$
and the relative error for the directional derivatives method is $\mbox{Err}{}_{\ell^{2}}^{DD}=3.04\times10^{-3}$.
The error is decreased by a factor $56$ by using the $DD$ method.
This problem of lattice anisotropy is decreased when the relationships
(\ref{eq:Def_DeriveeDirectionnelle}) and (\ref{eq:Grad_DF-Centree})
are applied to calculate the gradients (Fig. \ref{fig:Effet-Deriv-Direc}b
and Fig. \ref{fig:Effet-Profiles}b).

\begin{figure*}
\begin{centering}
\begin{tabular}{ccc}
{\small (a) $a_{s}(\varphi)$ with finite differences method} & $\quad$ & {\small (b) $a_{s}(\varphi)$ with directional derivatives method}\tabularnewline
\noalign{\vskip2mm}
\includegraphics[scale=0.24]{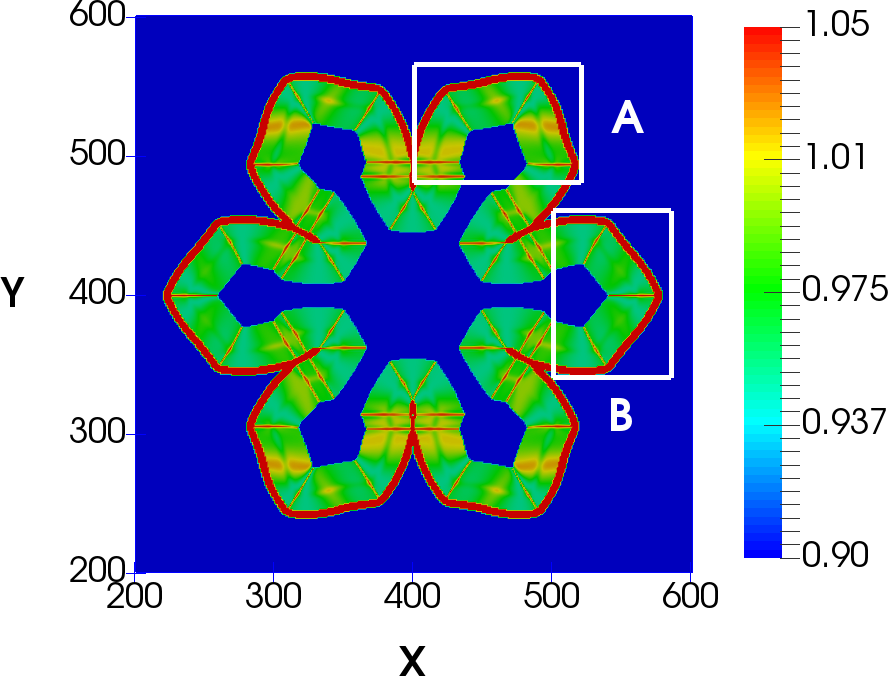} &  & \includegraphics[scale=0.24]{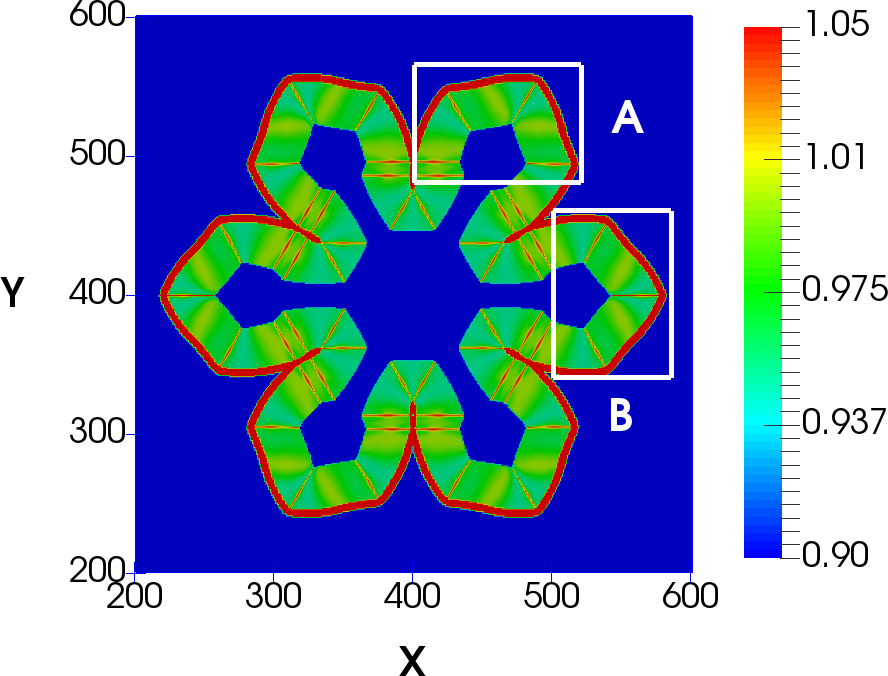}\tabularnewline
\end{tabular}
\par\end{centering}

~

\caption{\label{fig:Effect_on_As}Anisotropy function $a_{s}(\varphi)$ at
$t=10^{5}\delta t$. (a) With a finite differences method the values
of the $a_{s}$ function are different inside boxes A and B, whereas
(b) with a directional derivatives method, they are identical.}
\end{figure*}

The origin of those differences can be understood when the $a_{s}$
function is plotted as a function of position (Fig. \ref{fig:Effect_on_As}).
At $t=10^{5}\delta t$, with the finite differences method, the values
of $a_{s}$ depend on the direction of growth, the function is not
isotropic by rotation. In other words, on Fig. \ref{fig:Effect_on_As}a,
the first pattern inside the box B, which corresponds to a direction
of growth $\mathbf{n}_{a}$, is different of the second one inside
the box A corresponding to a direction $\mathbf{n}_{c}$. The pattern
of $a_{s}$ is not periodic. On the contrary, with the directional
derivatives method, the patterns in boxes A and B are identical (Fig.
\ref{fig:Effect_on_As}b). By considering the diagonals of the lattice
in the calculation of gradient, $a_{s}$ is more accurate and fully
symmetric by rotation. Let us mention that those differences do not
appear when $q=4$, even for gradients computed by central finite
difference, because the growth occurs in the main direction of the
coordinate system: $x$- and $y$-axis.

\begin{figure*}
\begin{centering}
\begin{tabular}{ccc}
{\small (a) Anisotropy function $a_{s}(\varphi)$ } &  & {\small (b) Temperature field $u$ and evolution of }$\phi=0$\tabularnewline
\noalign{\vskip2mm}
\includegraphics[scale=0.22]{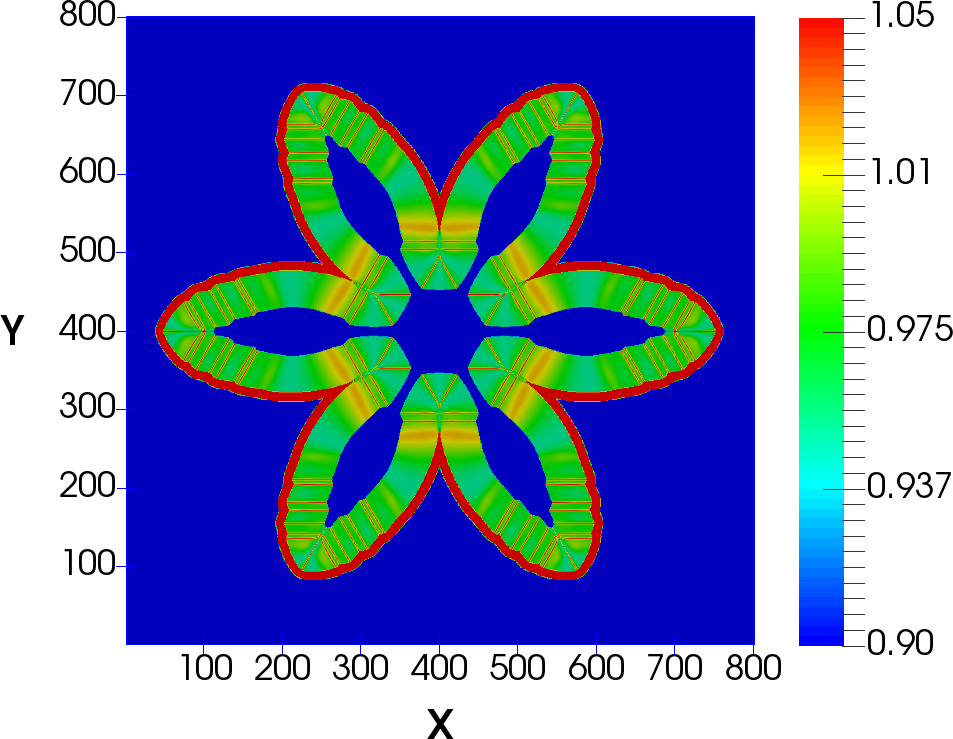} &  & \includegraphics[scale=0.22]{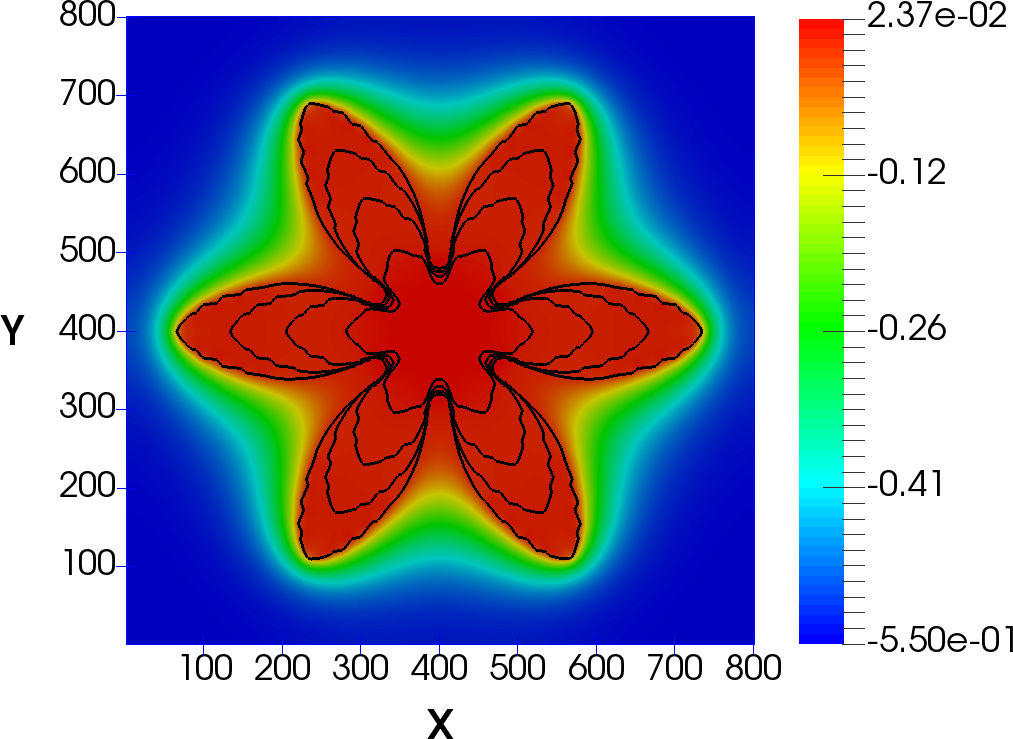}\tabularnewline
\end{tabular}
\par\end{centering}

~

\caption{\label{fig:Champs-6branches}Simulation for $\Delta=0.55$ and $q=6$,
results at $t=4\times10^{4}\delta t$. (a) Anisotropy function $a_{s}(\varphi)$
with directional derivatives. (b) Dimensionless temperature $u$ and
phase-field $\phi=0$ (black lines) at $t=1,\,2,\,3,\,4\times10^{4}\delta t$.}
\end{figure*}

Finally, the whole approach (LB+DD) is applied to simulate again the
crystal growth with $q=6$ and $\varphi_{0}=0$, by modifying the
undercooling and the interfacial kinetic. In this simulation, we choose
the parameter $\lambda$ such as the interfacial kinetic (Eq. (\ref{eq:Kinetic-Coeff}))
is canceled. The condition $\beta=0$ is satisfied for one particular
value of $\lambda$: $\lambda^{\star}=\kappa\tau_{0}/a_{2}W_{0}^{2}$.
By considering $W_{0}=1$ and $\tau_{0}=1$, the coefficient $\lambda^{\star}$
is equal to $\lambda^{\star}=\kappa/a_{2}=1.59566\kappa$. For a thermal
diffusivity equals to $\kappa=4$, we obtain $\lambda^{\star}=6.3826$
and the capillary length is $d_{0}=0.1385$. The mesh is composed
of 800$\times$800 nodes. Parameters are $\delta x=\delta y=0.4$,
$\delta t=0.008$, the undercooling is $\Delta=0.55$ and the anisotropic
strength is $\varepsilon_{s}=0.05$. The radius of the initial seed
is $R_{s}=10$ lattice unit (l.u.). Results are presented on Fig.
\ref{fig:Champs-6branches} at $t=4\times10^{4}\delta t$. The anisotropy
function is plotted on Fig. \ref{fig:Champs-6branches}a. On that
figure, we can observe that the pattern is identical in each branch
of the crystal. The $a_{s}$ function is symmetric by rotation of
60\textdegree{}. On Fig. \ref{fig:Champs-6branches}b, the temperature
field is plotted and the evolution of $\phi$ is superimposed at four
different times (black lines). We can check that the solution of $\phi$
is isotropic, the phase-field matches perfectly with its rotation
of 60\textdegree{}.

\subsection{Simultaneous growth of crystals with different anisotropy functions}

Now, we are interested in the simultaneous growth of several crystals
with different numbers of branches and different angle of reference
$\varphi_{0}$. The difficulty does not lie in the growth of several
crystals since their number is naturally taken into account in the
phase-field model, more precisely in the initial condition of $\phi$.
For instance, for simulating the growth of three crystals, three initial
grains are set with the relationship (\ref{eq:Initial_Condition}).
Each of them will grow progressively during the simulation but they
will have the same anisotropy function. Here, we consider that each
crystal can be defined by its own anisotropy function $a_{s}(\mathbf{n})$.

For that purpose, we consider that this function depends on a new
index $I\equiv I(\mathbf{x},\, t)$:

\begin{equation}
a_{s}^{(I)}(\mathbf{x},\, t)=1+\varepsilon_{s}^{(I)}\cos\left[q^{(I)}\left(\varphi(\mathbf{x},\, t)-\varphi_{0}^{(I)}\right)\right],\label{eq:As_Multi}
\end{equation}
where $I$ is a field indicating the crystal number. This new function
depends on position and time and varies from $1$ to $N_{I}$ where
$N_{I}$ is the number of crystals: it is equal to 1 for the first
crystal; 2 for the second one; 3 for the third one, and so on ...
Its value is zero everywhere else. For example in 2D, if we choose
$N_{I}=3$ crystals, with $q^{(1)}=4$, $q^{(2)}=5$ and $q^{(3)}=6$
respectively, the evolution of $I(\mathbf{x},\, t)$ has to be updated
at each time step. After initialization of $\phi(\mathbf{x},\, t)$
and $I(\mathbf{x},\, t)$, the nodes surrounding each crystal are
selected with a criterion based on the variation of $\phi$: $\bigl|\boldsymbol{\nabla}\phi\bigr|>\xi$
where $\xi$ is a small numerical value. This criterion is used to
identify all nodes that have a zero value and located near the diffuse
zone of each crystal. Then the index of those nodes takes the value
of the nearest node having an index different to zero.

\begin{figure*}
\begin{centering}
\begin{tabular}{cc}
{\small (a) Temperature field $u$ and evolution of }$\phi=0$ & {\small (b) Anisotropy function $a_{s}^{(I)}(\mathbf{x},\, t)$ at
$t=10^{5}\delta t$}\tabularnewline
\noalign{\vskip2mm}
\includegraphics[scale=0.21]{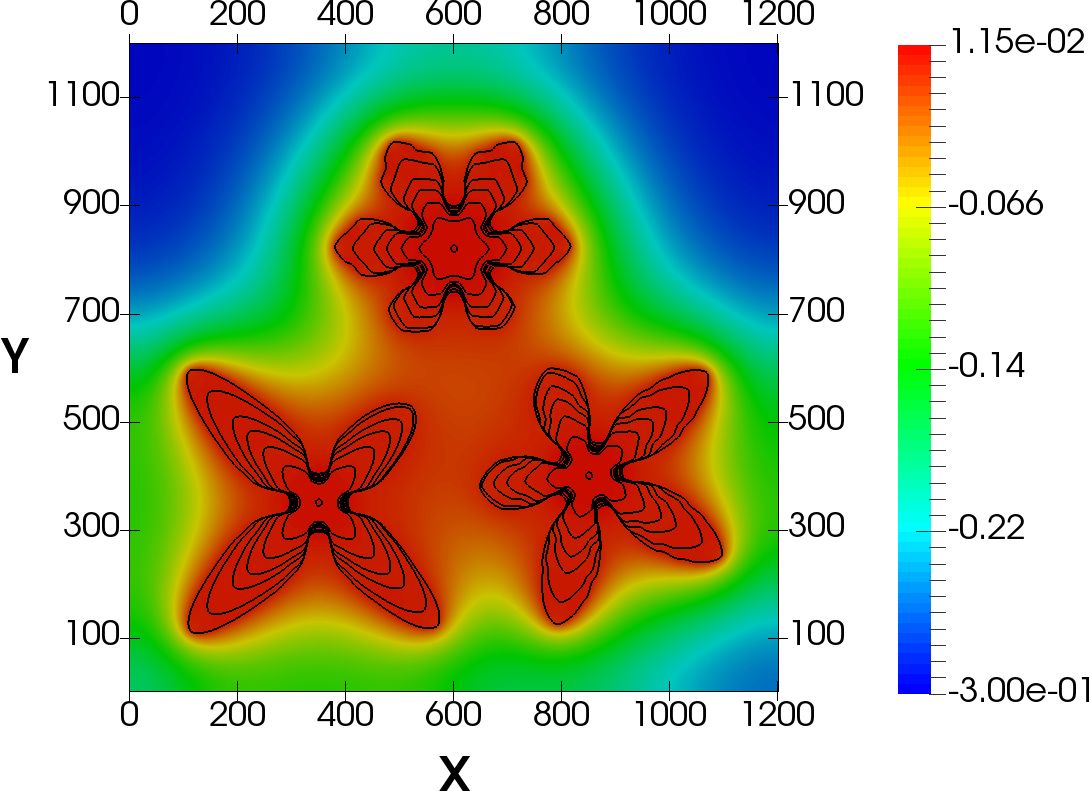} & \includegraphics[scale=0.21]{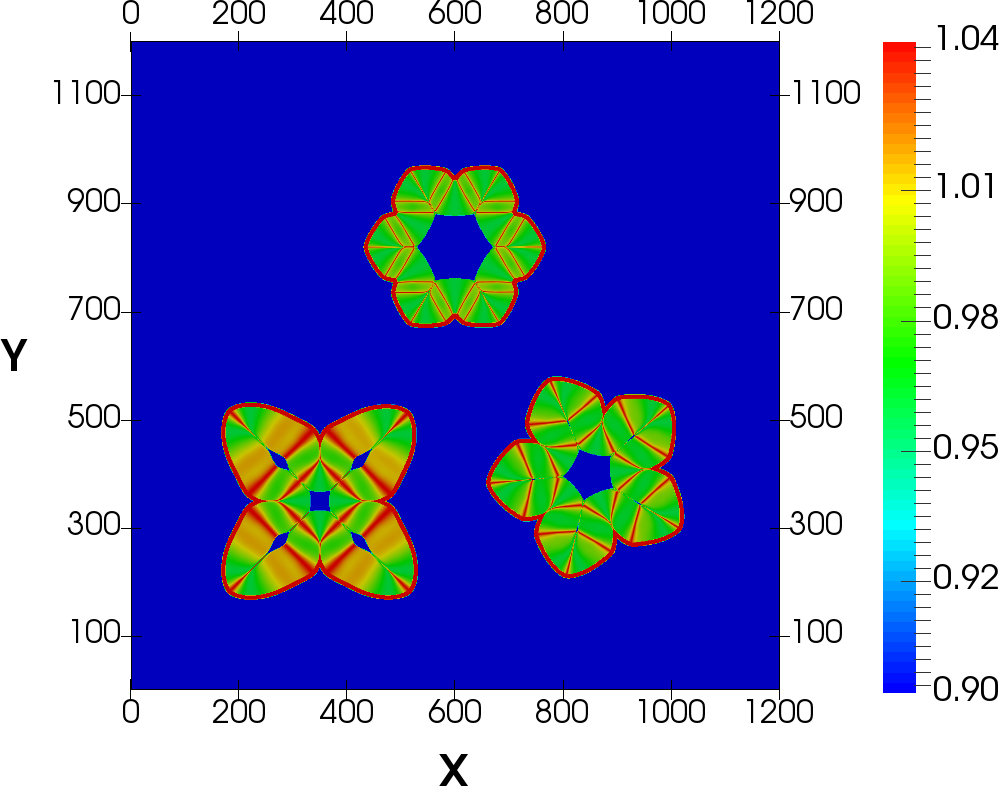}\tabularnewline
\end{tabular}
\par\end{centering}

\caption{\label{fig:Croissance-simul_multicristaux}Simultaneous growth of
three crystals with 4, 5 and 6 tips. (a) phase-field $\phi=0$ at
$t=0,\,2.5,\,5,\,7.5,\,10,\,15,\,20\times10^{4}\delta t$ with temperature
field at $t=20\times10^{4}\delta t$. (b) Anisotropy function $a_{s}^{(I)}(\mathbf{x},\, t)$
at $t=10^{5}\delta t$.}
\end{figure*}

Three seeds are initialized in a computational domain composed of
1200$\times$1200 nodes. The radius is set at $R_{s}=8$ l.u for each
of them and positions are $\mathbf{x}^{(1)}=(350,\,350)$, $\mathbf{x}^{(2)}=(850,\,400)$
and $\mathbf{x}^{(3)}=(600,\,820)$. An angle $\varphi_{0}^{(1)}=45$\textdegree{}
is set for the first crystal and $\varphi_{0}^{(2)}=5$\textdegree{}
for the second one. Parameters are $\delta x=5\times10^{-3}$, $\delta t=5\times10^{-6}$,
$W_{0}=0.012$, $\tau_{0}=10^{-4}$, $\lambda=10$, $\varepsilon_{s}^{1,\,2,\,3}=0.04$,
$\kappa=0.7$ and $\Delta=0.3$. The small numerical value for the
criterion is $\xi=10^{-7}$. Results are presented on Fig. \ref{fig:Croissance-simul_multicristaux}a.
Black lines represent the time evolution of $\phi=0$, and the temperature
field is plotted at the end of simulation. Crystals have respectively
four, five and six tips. On the figure, for short times ($<10^{5}\delta t$),
the branch size is identical for each crystal. At $t=10^{5}\delta t$
the pattern of each branch for each crystal is identical as confirmed
by Fig. \ref{fig:Croissance-simul_multicristaux}b. On the other hand,
for greater times, some branches are smaller because of the influence
of other crystals. The growth of those branches is limited by the
temperature field that is uniform and higher in the interaction zone
of crystals. In this area, the latent heat released during the solidification
is evacuated less rapidly in the system and the speed of growth decreases.

\section{\label{sec:Simulations}3D simulations}

In this section, we present 3D simulations based on anisotropy functions
defined in section \ref{sec:Anisotropic-functions}. Let us remind
that several experiments and simulations were carried out in the literature
to study the <<dendrite orientation transition>> from $\left\langle 100\right\rangle $
to $\left\langle 110\right\rangle $ for a pure substance and binary
mixtures \citep{Karma_Orientation_Nature2006,Dantzig-Rappaz_MMTA2013}.
Here, those simulations are performed to demonstrate the flexibility
of the LB method and its ability to simulate various crystal shapes.
Once the anisotropy function $a_{s}(\mathbf{n})$ is modified and
its derivatives with respect to $\partial_{\alpha}\phi$ ($\alpha=x,\, y,\, z$)
calculated, implementation of 3D schemes is straightforward by modifying
the lattices. In subsection \ref{sub:Validation_2D-3D}, the directional
derivatives method implemented in the 3D code is checked with the
2D code validated in \citep{Cartalade_etal_CAMWA2015} with a finite
difference code. In subsection \ref{sub:Sensitivity-of-parameters},
3D simulations with three different anisotropy functions will be presented.

\subsection{\label{sub:Validation_2D-3D}Comparisons between 2D and 3D codes}

In order to check numerical implementation of directional derivatives
method in the 3D code, a comparison is carried out with the 2D one.
The validation is based on a comparison of Eq. (\ref{eq:As_Y66}),
established in subsection \ref{sub:Formulations-anisotropy}, that
is equivalent to the 2D function $a_{s}(\varphi)=1+\varepsilon_{s}\cos(6\varphi)$.
Simulations of crystal growth are performed with a mesh composed of
$301{}^{2}$ nodes in 2D and $301\times301\times4$ nodes in 3D. Only
four nodes are used for the third dimension $z$ because Eq. (\ref{eq:As_Y66})
favors the growth of six tips in the $XY$-plane.

For both codes the parameters are set as follows. The space step is
equal to $\delta x=0.01$ and the time step is chosen such as $\delta t=1.5\times10^{-5}$.
The initial condition is a diffuse sphere initialized at the domain
center with a radius equals to $R_{s}=6$ lattice unit (l.u.). All
boundary conditions are zero-flux types. The initial undercooling
is uniform and equal to $u=-0.3$ and the thermal diffusivity is equal
to $\kappa=0.7$. Parameters of the phase-field are $W_{0}=0.0125$,
$\tau_{0}=1.5625\times10^{-4}$, $\lambda=10$ and $\varepsilon_{s}=0.05$.

\begin{figure*}
\begin{centering}
\begin{tabular}{ccc}
{\small (a) Phase-field $\phi=0$ ($t=t'\times10^{3}\delta t$)} &  & {\small (b) Temperature iso-values}\tabularnewline
\includegraphics[angle=-90,scale=0.4]{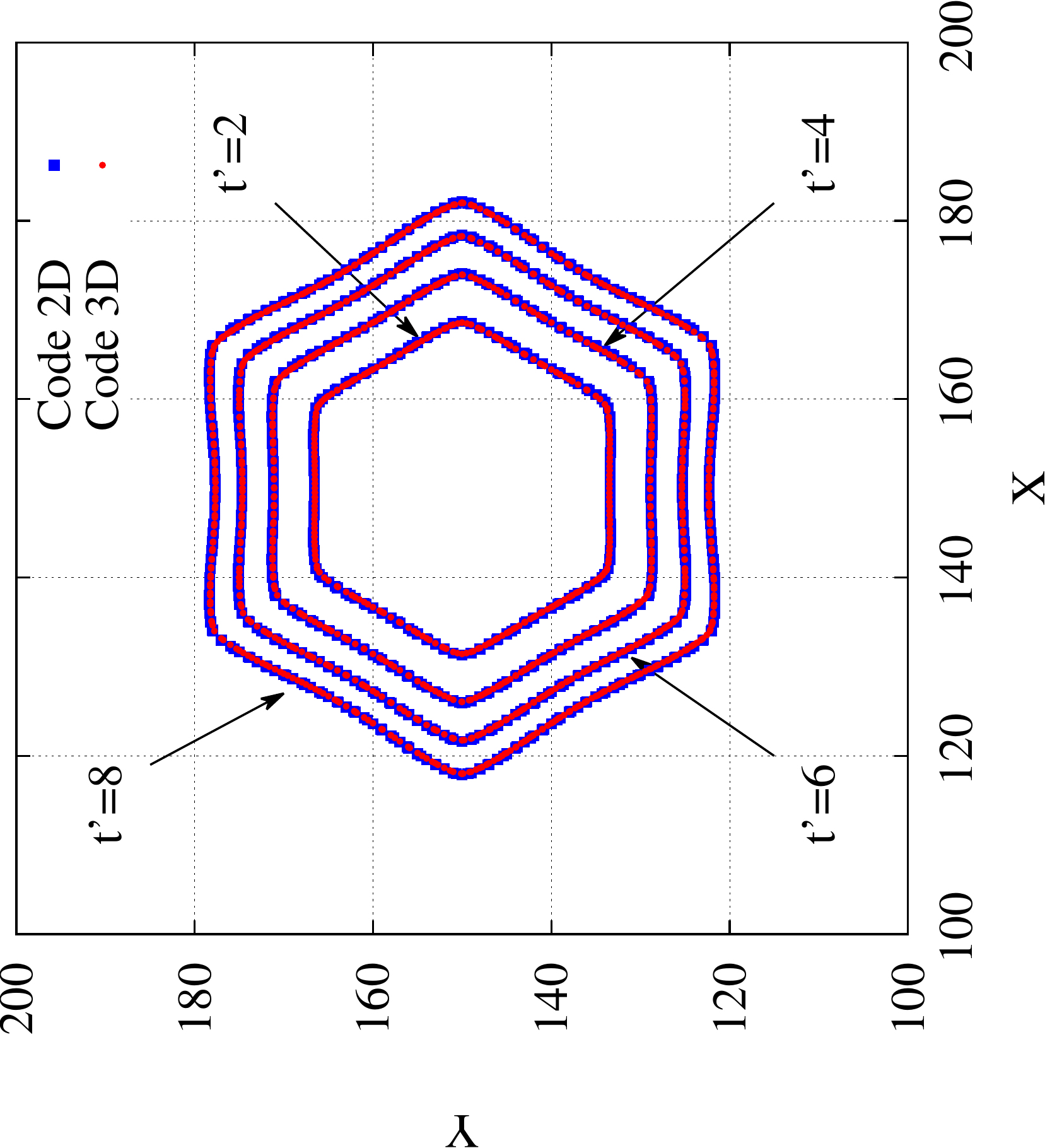} &  & \includegraphics[angle=-90,scale=0.4]{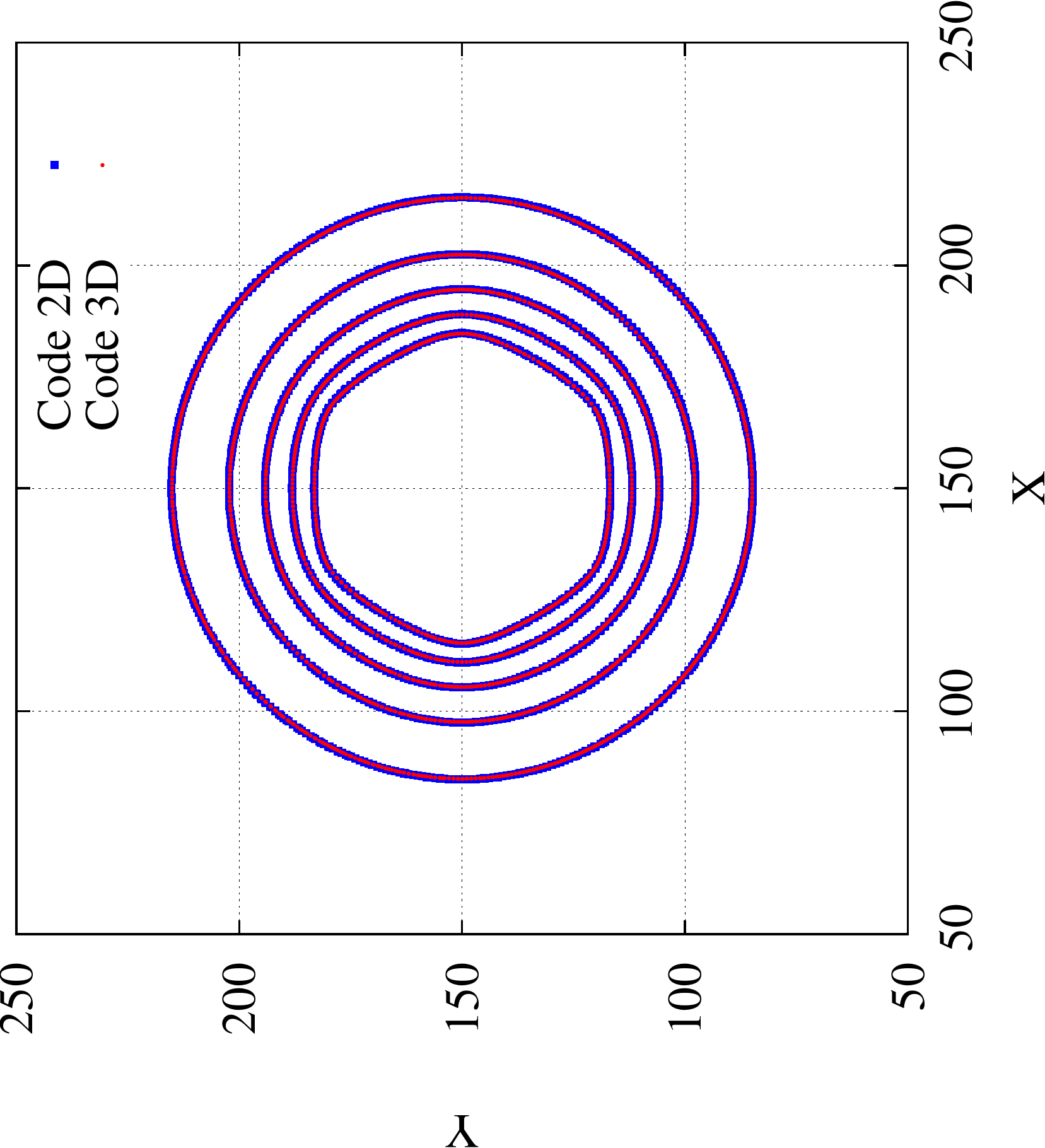}\tabularnewline
\end{tabular}
\par\end{centering}

~

\caption{\label{fig:Superposition-des-iso-valeurs}(a) Iso-values $\phi=0$
for Eq. (\ref{eq:As_Theta}) with $q=6$ and $\varphi_{0}=0$ (blue
squares; 2D code) and Eq. (\ref{eq:As_Y66}) (red dots; 3D code) at
$t=3\times10^{3}\delta t$ and $t=8\times10^{3}\delta t$. (b) Iso-values
$u=-0.25$, $-0.2$, $-0.15$, $-0.1$ and $-0.05$ at $t=8\times10^{3}\delta t$
in 2D (blue square) and 3D (red dots).}
\end{figure*}

Results are presented on Fig. \ref{fig:Superposition-des-iso-valeurs}.
On this figure, the left plot presents the phase-field $\phi=0$ for
$a_{s}(\varphi)$ in 2D (blue squares) and $a_{s}(\mathbf{n})$ in
3D (red dots) at four times $t=2,\,4,\,6,\,8\times10^{3}\delta t$.
The shape with six tips in the $XY$-plane is well reproduced by using
the anisotropy function defined by Eq. (\ref{eq:As_Y66}) in the 3D
code. For both codes, several iso-values of temperature field at $t=8\times10^{3}\delta t$
are superimposed on the right plot. The 3D implementation of the LB
schemes is validated.

\subsection{\label{sub:Sensitivity-of-parameters}Simulations with non standard
anisotropy functions}

In this section, 3D simulations are carried out by using three anisotropy
functions $a_{s}(\mathbf{n})$. The first anisotropy function is the
standard one, defined by Eq. (\ref{eq:As_Function_Classical}) with
$\varepsilon_{s}=0.05$. That function favors the growth in the $\left\langle 100\right\rangle $-direction
(see Fig. \ref{fig:Representations_As}a). The second one is defined
by Eq. (\ref{eq:As_MatSci}) which favors the crystal development
in the $\left\langle 110\right\rangle $-direction if $\varepsilon_{s}=0$
and $\gamma=-0.02$ (see Fig. \ref{fig:Representations_As}e). The
last one is defined by:

\begin{equation}
a_{s}(\mathbf{n})=66\gamma n_{x}^{2}n_{y}^{2}n_{z}^{2},\label{eq:As_Function_S}
\end{equation}
with $\gamma=0.02$. That function favors the growth in the $\left\langle 111\right\rangle $-direction
(see Fig. \ref{fig:Representations_As}b).

For each simulation, the computational domain is composed of $351^{3}$
nodes and the undercooling is fixed at $\Delta=0.30$. The space step
is $\delta x=0.01$ and the time step is $\delta t=1.5\times10^{-5}$.
The seed is initialized at the center $\mathbf{x}_{c}=(175,\,175,\,175)^{T}$
of the domain. The interface thickness is equal to $W_{0}=0.0125$,
the kinetic time is $\tau_{0}=1.5625\times10^{-4}$, the coupling
coefficient is $\lambda=10$, and the thermal diffusivity is $\kappa=1$.

\begin{figure}
\begin{centering}
\begin{tabular}{ccc}
{\small (a) Dendrite A: growing direction $\left\langle 100\right\rangle $} & {\small (b) Dendrite B: growing direction $\left\langle 110\right\rangle $} & {\small (c) Dendrite C: growing direction $\left\langle 111\right\rangle $}\tabularnewline
\noalign{\vskip2mm}
\includegraphics[scale=0.182]{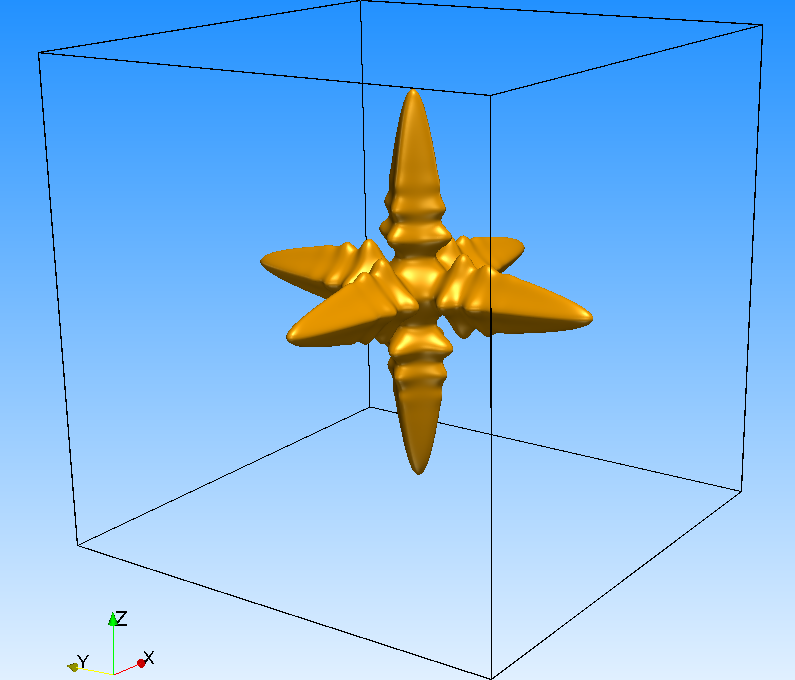} & \includegraphics[scale=0.182]{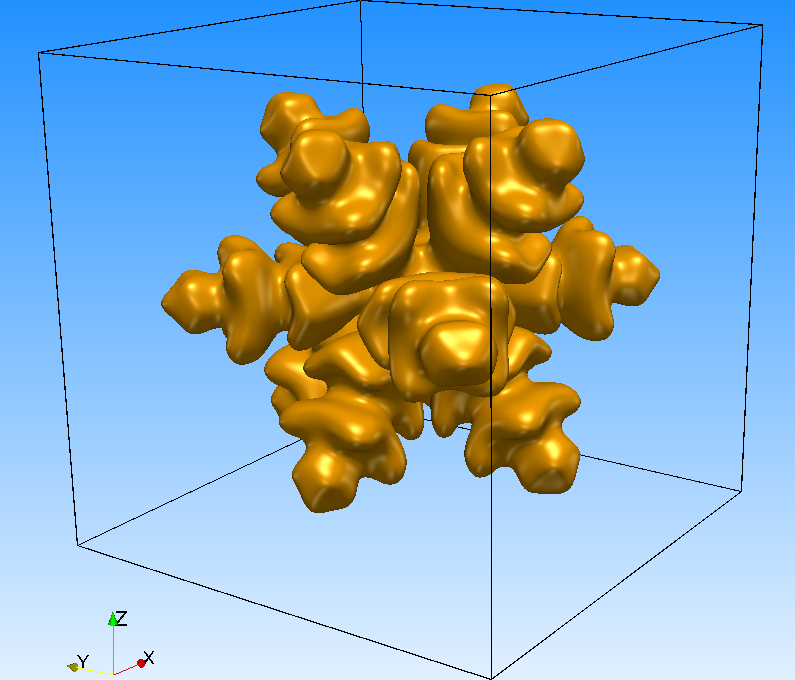} & \includegraphics[scale=0.182]{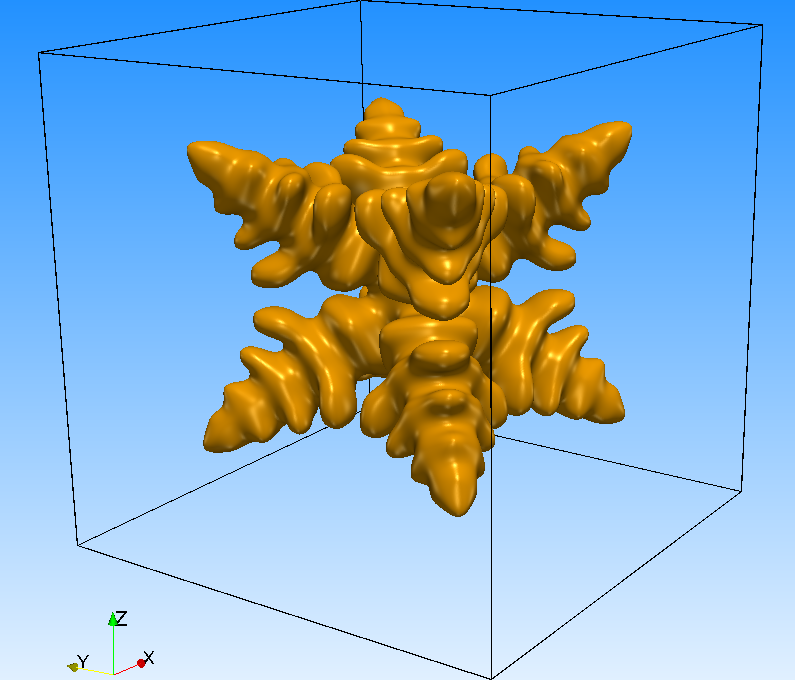}\tabularnewline
\end{tabular}
\par\end{centering}

~

\caption{\label{fig:Comparison_Q-S-K61}Dendritic shapes $\phi=0$ for $a_{s}(\mathbf{n})$
defined by (a) Eq. (\ref{eq:As_Function_Classical}) with $\varepsilon_{s}=0.05$
at $t=5\times10^{3}\delta t$; (b) Eq. (\ref{eq:As_MatSci}) with
$\varepsilon_{s}=0$ and $\gamma=-0.02$ at $t=2.5\times10^{4}\delta t$;
(c) Eq. (\ref{eq:As_Function_S}) with $\gamma=0.02$ at $t=1.5\times10^{4}\delta t$.}
\end{figure}

Dendritic shapes $\phi=0$ of each crystal are presented on Fig. \ref{fig:Comparison_Q-S-K61}
for a same orientation of the coordinate system. Directions of growth
$\left\langle 100\right\rangle $, $\left\langle 110\right\rangle $
and $\left\langle 111\right\rangle $ are presented respectively on
Figs. \ref{fig:Comparison_Q-S-K61}a (dendrite A), \ref{fig:Comparison_Q-S-K61}b
(dendrite B) and \ref{fig:Comparison_Q-S-K61}c (dendrite C). For
dendrite B, as expected in view of Fig. \ref{fig:Representations_As}e,
the crystal shape presents twelve tips: four contained in the $XY$-plane
centered at $\mathbf{x}_{c}$, four other above this plane and four
below. Finally, for dendrite C, we can see four tips above the same
plane and four below, as expected in view of Fig. \ref{fig:Representations_As}b.

\begin{figure}
\begin{centering}
\begin{tabular}{c}
\begin{tabular}{ccc}
{\small (a) Slices for dendrite A; planes} & $\qquad$ & {\small (c) Slices for dendrite C; planes}\tabularnewline
{\small of normal vectors $\mathbf{n}_{1}^{A}$ and $\mathbf{n}_{2}^{A}$} &  & {\small of normal vectors $\mathbf{n}_{1}^{C}$ and $\mathbf{n}_{2}^{C}$}\tabularnewline
\includegraphics[angle=-90,scale=0.38]{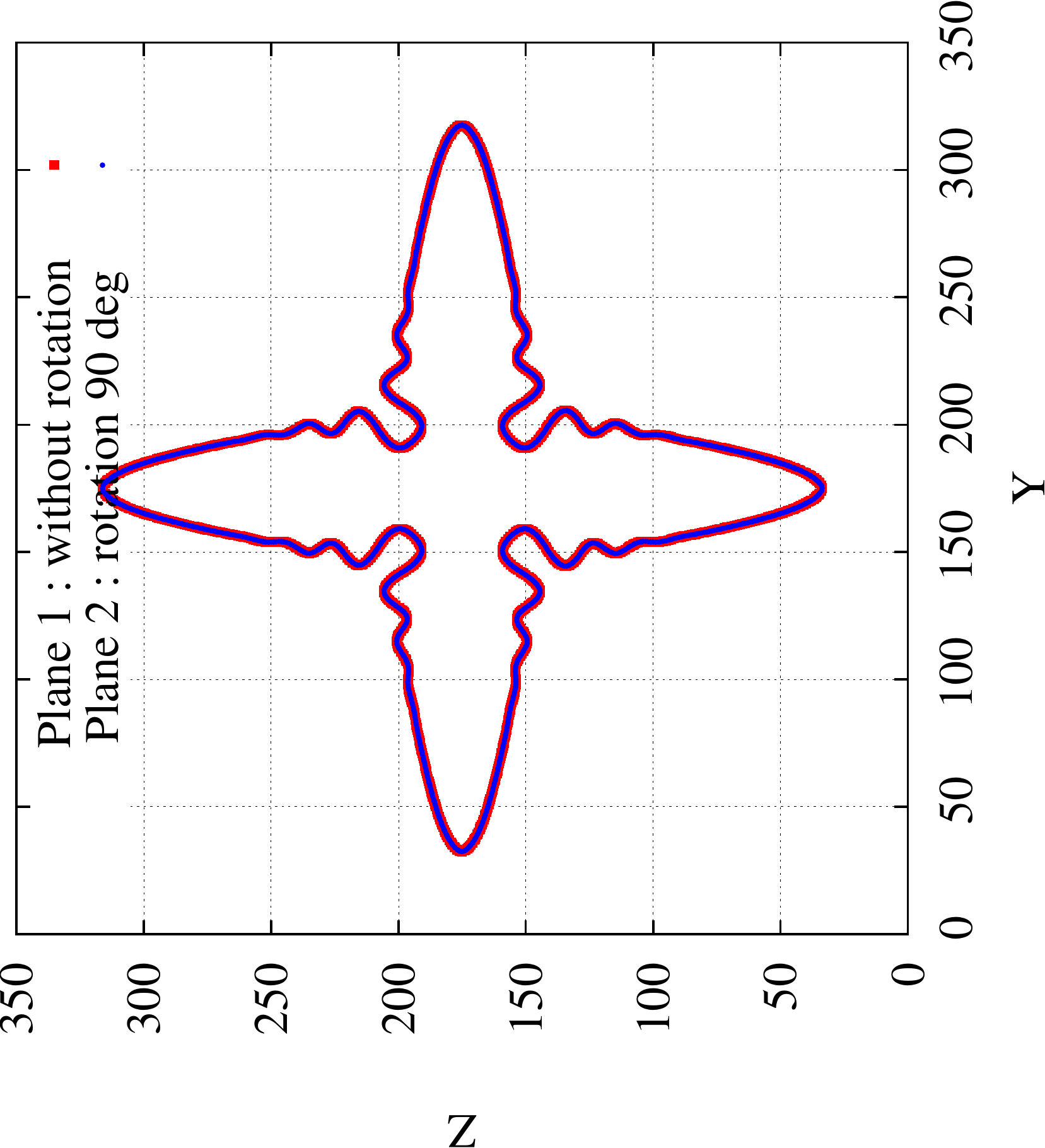} &  & \includegraphics[angle=-90,scale=0.38]{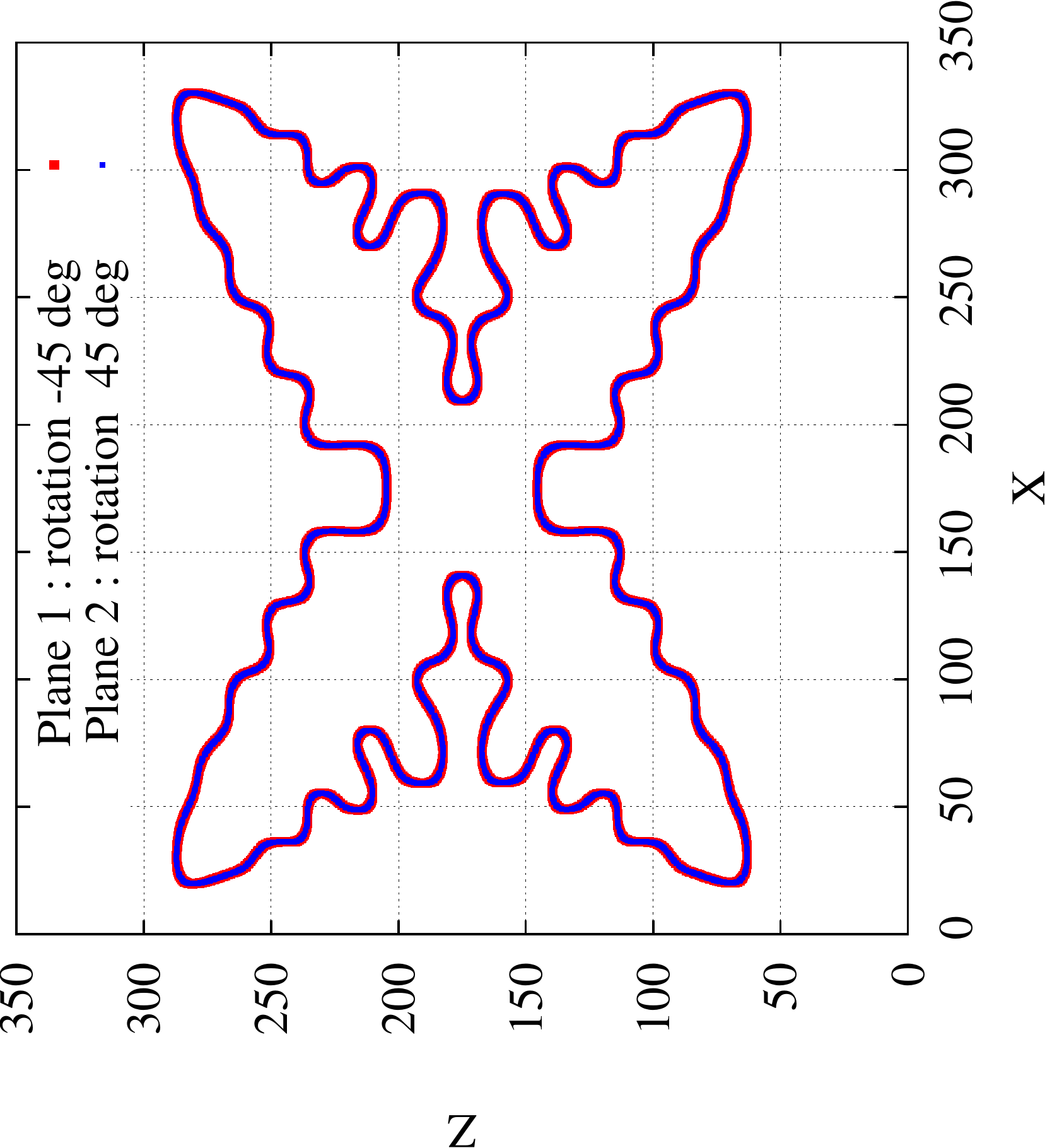}\tabularnewline
\end{tabular}\tabularnewline
\tabularnewline
\tabularnewline
\begin{tabular}{c}
{\small (b) Slices for dendrite B; planes of normal vectors $\mathbf{n}_{1}^{B}$
and $\mathbf{n}_{2}^{B}$}\tabularnewline
\includegraphics[angle=-90,scale=0.38]{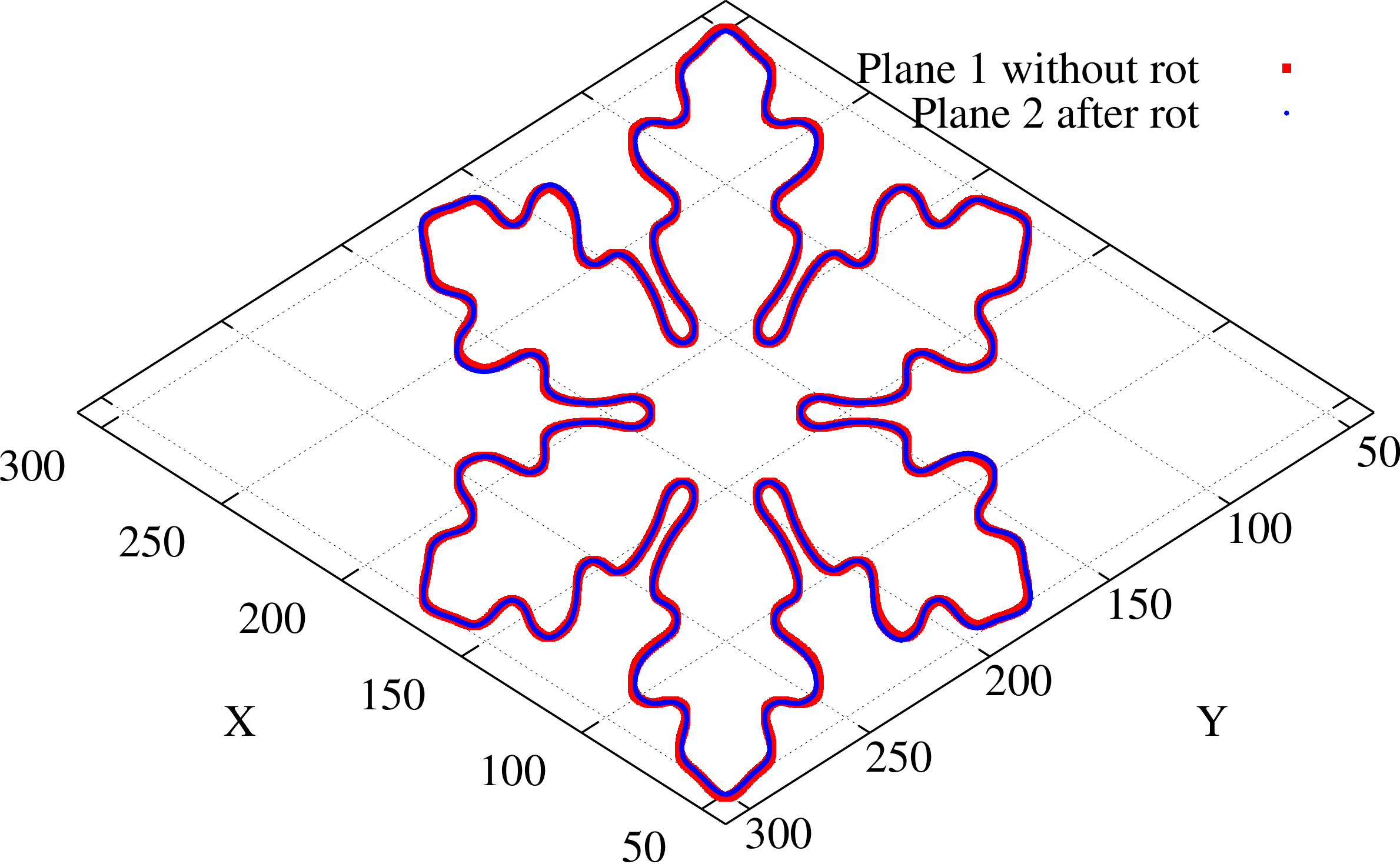}\tabularnewline
\end{tabular}\tabularnewline
\end{tabular}
\par\end{centering}

~

\caption{\label{fig:Slices}All planes are centered at $\mathbf{x}_{c}=(175,\,175,\,175)^{T}$.
(a) Growth in $\left\langle 100\right\rangle $-direction: planes
of normal vectors $\mathbf{n}_{1}^{A}=(1,\,0,\,0)$ and $\mathbf{n}_{2}^{A}=(0,\,1,\,0)$.
Phase-field $\phi=0$ from the first plane (red) matches perfectly
$\phi=0$ from the second plane (blue) after rotation of 90\textdegree{}
around the $z$-axis. (b) Growth in $\left\langle 110\right\rangle $-direction:
planes of normal vector $\mathbf{n}_{1}^{B}=(1,\,1,\,1)$ and $\mathbf{n}_{2}^{B}=(1,\,1,\,-1)$.
Phase-field $\phi=0$ from the first plane (red) matches perfectly
$\phi=0$ from the second plane (blue) after rotation. (c) Growth
in $\left\langle 111\right\rangle $-direction: planes of normal vectors
$\mathbf{n}_{1}^{C}=(1,\,1,\,0)$ and $\mathbf{n}_{2}^{C}=(1,\,-1,\,0)$.
Phase-field from the first plane rotated from -45\textdegree{} around
the $z$-axis (red) matches perfectly $\phi=0$ from the second plane
rotated from 45\textdegree{} (blue).}

\end{figure}

In order to check the isotropy of phase-field $\phi=0$, obtained
with the directional derivatives method, several slices are carried
out for each dendrite. The center of each plane is $\mathbf{x}_{c}$.
For dendrite A, two slices are carried out in the planes of normal
vectors $\mathbf{n}_{1}^{A}=(1,\,0,\,0)$ and $\mathbf{n}_{2}^{A}=(0,\,1,\,0)$.
Solution $\phi=0$ of the $\mathbf{n}_{1}^{A}$-plane ($YZ$-plane)
is compared to the second one ($\mathbf{n}_{2}^{A}$-plane) after
a rotation of 90\textdegree{} around the $z$-axis for the second
plane (see Fig. \ref{fig:Slices}a). For dendrite B, two slices of
normal vectors $\mathbf{n}_{1}^{B}=(1,\,1,\,1)$ and $\mathbf{n}_{2}^{B}=(1,\,1,\,-1)$
are carried out. The solution $\phi=0$ from the first plane is compared
with the second one obtained after rotation (see Fig. \ref{fig:Slices}b).
Finally, for dendrite C, the phase-fields $\phi=0$ from planes of
normal vectors $\mathbf{n}_{1}^{C}=(1,\,1,\,0)$ and $\mathbf{n}_{2}^{C}=(1,\,-1,\,0)$
are compared on Fig. \ref{fig:Slices}c after a rotation of -45\textdegree{}
for the first plane and +45\textdegree{} for the second one. The rotation
is performed around the $z$-axis. For each dendrite, the phase-fields
are superimposed and the directional derivatives method gives satisfying
results for the numerical isotropy of the solution.

On Fig. \ref{fig:Comparison_Q-S-K61}c, which corresponds to the dendrite
C, we can observe a threefold symmetry for the secondary branches
that appear along the main directions of growth $\left\langle 111\right\rangle $.
The threefold symmetry explains why, in the slices of Fig. \ref{fig:Slices}c,
the secondary branches appear only on one side of each branch. The
branches are not symmetric in those planes. For dendrites A and B,
the symmetry is fourfold and the branches are fully symmetric on Figs.
\ref{fig:Slices}a and \ref{fig:Slices}b. For dendrite C, the origin
of the threefold symmetry can be understood by analyzing the $a_{s}(\mathbf{n})$
function and its derivatives with respect to each component $\partial_{x}\phi$,
$\partial_{y}\phi$ and $\partial_{z}\phi$, i.e. the vector $\boldsymbol{\mathcal{N}}'(\mathbf{x},\, t)$:

\begin{equation}
\boldsymbol{\mathcal{N}}'(\mathbf{x},\, t)=\left(\frac{\partial a_{s}(\mathbf{n})}{\partial(\partial_{x}\phi)},\,\frac{\partial a_{s}(\mathbf{n})}{\partial(\partial_{y}\phi)},\,\frac{\partial a_{s}(\mathbf{n})}{\partial(\partial_{z}\phi)}\right)^{T}.\label{eq:VectNprime}
\end{equation}

\begin{figure}

\begin{centering}
\begin{tabular}{ccc}
{\small (a) $a_{s}(\mathbf{n})$ and streamlines} & {\small (b) $\phi=0$ at $t=1.5\times10^{4}\delta t$} & {\small (c) $a_{s}(\mathbf{n})$ in the plane $(\mathbf{n}_{d},\,\mathbf{x}_{p})$}\tabularnewline
{\small of $\boldsymbol{\mathcal{N}}'$ at $t=0$} &  & \tabularnewline
\includegraphics[scale=0.24]{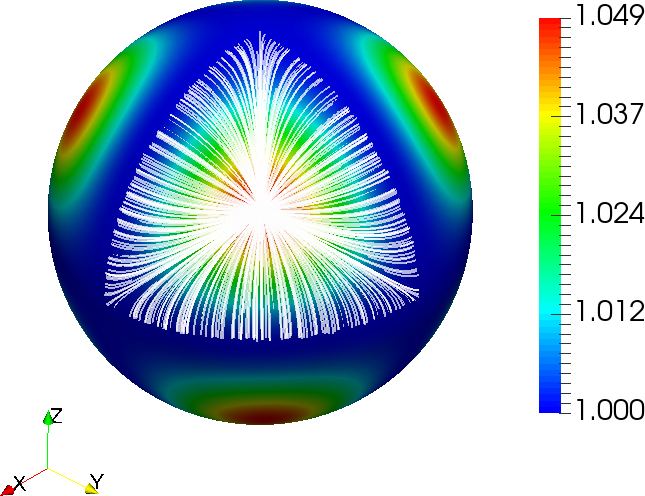} & \includegraphics[scale=0.26]{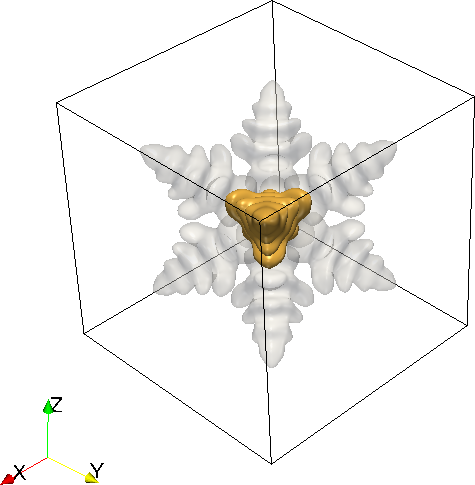} & \includegraphics[scale=0.26]{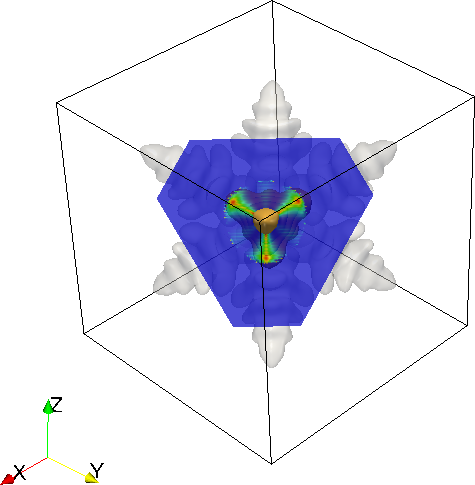}\tabularnewline
\end{tabular}
\par\end{centering}

~

\caption{\label{fig:As-Nprime}(a) Streamlines of $\boldsymbol{\mathcal{N}}'$
(white lines) indicate that growth can occur in three directions which
correspond to the secondary growing directions observed on (b). At
the same time $t=1.5\times10^{5}\delta t$, (c) $a_{s}(\mathbf{n})$
function is plotted in the plane of normal vector $\mathbf{n}_{d}=(1,\,1,\,1)^{T}$
centered at $\mathbf{x}_{p}=(331,\,244,\,244)^{T}$. $a_{s}$ is maximal
in the same three directions.}

\end{figure}

For an anisotropy function defined by Eq. (\ref{eq:As_Function_S}),
the $a_{s}(\mathbf{n})$ function and the streamlines of vectors $\boldsymbol{\mathcal{N}}'(\mathbf{x})$
are plotted on a spherical surface (Fig. \ref{fig:As-Nprime}a). Here,
the <<streamlines>> are the curves that are tangent to vectors $\boldsymbol{\mathcal{N}}'(\mathbf{x})$.
The same word is used by analogy to streamlines that are tangent to
the velocity fields in fluid flow problems. On figure \ref{fig:As-Nprime}a,
the anisotropy function is the colored field and the streamlines are
the white lines. For a better clarity of the figure, the streamlines
are not plotted on the whole sphere, but only in the area around the
specific direction of growth defined by $\mathbf{n}_{d}=(1,\,1,\,1)^{T}$.
The streamlines of $\boldsymbol{\mathcal{N}}'(\mathbf{x})$ indicate
that the growth can also occur in the three directions which correspond
to the secondary growing directions observed on Fig. \ref{fig:As-Nprime}b.
On that figure, the phase-field $\phi=0$ is plotted in the same orientation
of coordinate system of Fig. \ref{fig:As-Nprime}a, at final time
of simulation $t=1.5\times10^{4}\delta t$. The main branch that corresponds
to the direction $\mathbf{n}_{d}$ is highlighted. On Fig. \ref{fig:As-Nprime}c,
the $a_{s}(\mathbf{n})$ function is plotted in a plane of normal
vector $\mathbf{n}_{d}$ and centered at $\mathbf{x}_{p}=(331,\,244,\,244)^{T}$.
That slice is performed at final time, near the tip, where the secondary
branches do not exist yet. The $a_{s}(\mathbf{n})$ function has its
maximal values in the three same directions. Consequently, the three
secondary branches that appear during the calculation are consistent
with the $a_{s}(\mathbf{n})$ function applied for the simulations.

Previous simulations were performed with a seed initialized at the
center of the domain. Nevertheless, with this way to proceed, the
dendrite is influenced by the domain boundaries before its full development.
One first method consists to increase the computational domain by
increasing the number of nodes. However, because 3D simulations require
a lot of time computations, symmetries of problem are considered by
keeping the same mesh. For dendrite B, the seed is initialized at
the origin $(0,\,0,\,0)$ of the domain composed of $351^{3}$ nodes
and the results are post-processed at the end of simulation. For that
simulation, the 3D code calculates in parallel on 50 cores. The global
domain is cut in $z$-direction for both equations. Each core calculates
on a thin part of the global mesh composed of $351\times351\times7$
nodes. For $1.5\times10^{5}$ times steps, the results are obtained
after 98.54 hours ($\sim4$ days). For comparison, results of Figs.
\ref{fig:Comparison_Q-S-K61}(a), (b), (c) are respectively obtained
after 3h, 8.97h and 9.18h for calculations on 100 cores.

\begin{figure*}
\begin{centering}
\begin{tabular}{ccccc}
{\small (a) $t=0$} & $\quad$ & {\small (b) $t=5\times10^{3}\delta t$} & $\quad$ & {\small (c) $t=1.5\times10^{4}\delta t$}\tabularnewline
\noalign{\vskip2mm}
\includegraphics[scale=0.21]{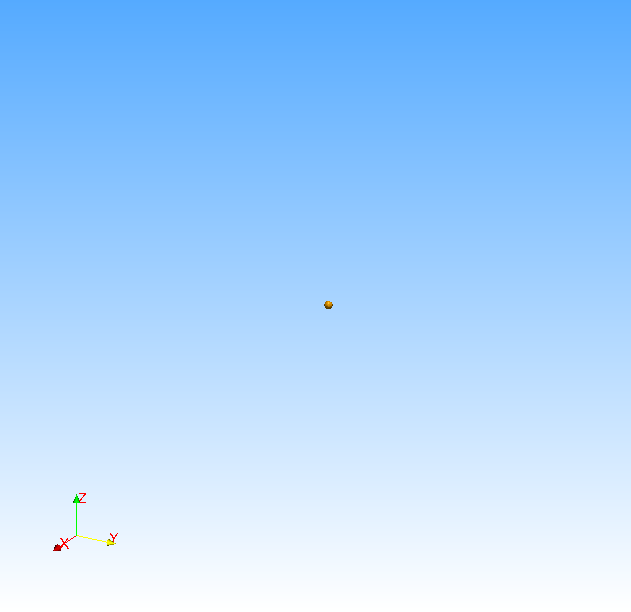} &  & \includegraphics[scale=0.21]{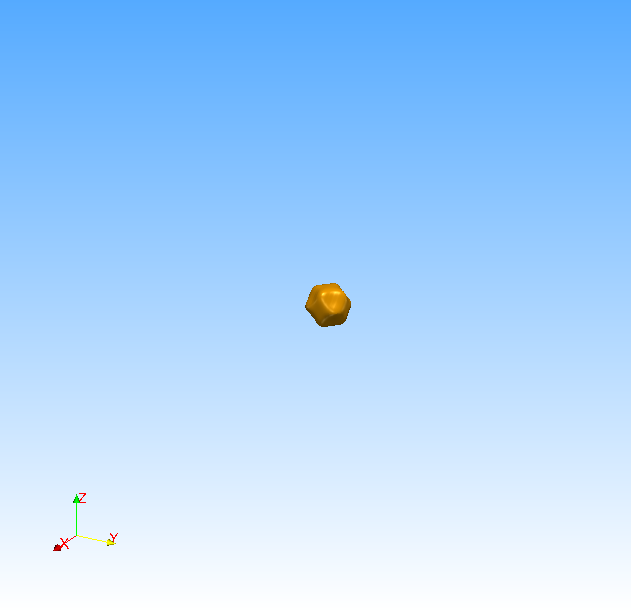} &  & \includegraphics[scale=0.21]{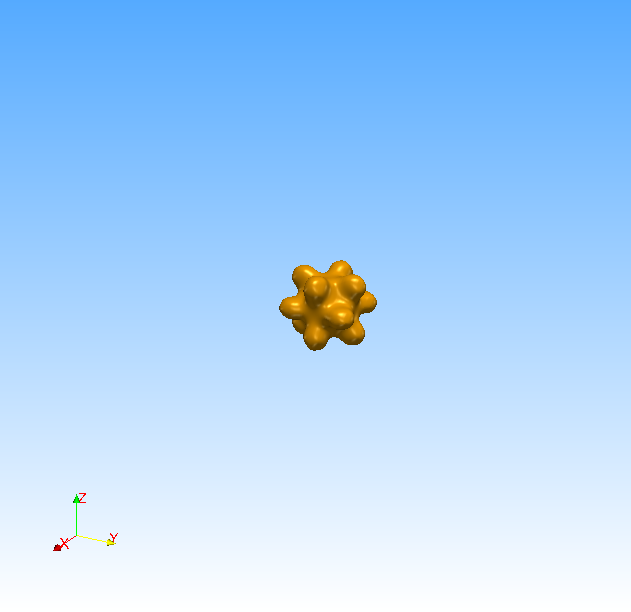}\tabularnewline
 &  &  &  & \tabularnewline
{\small (d) $t=4\times10^{4}\delta t$} &  & {\small (e) $t=10^{5}\delta t$} &  & {\small (f) $t=1.5\times10^{5}\delta t$}\tabularnewline
\noalign{\vskip2mm}
\includegraphics[scale=0.21]{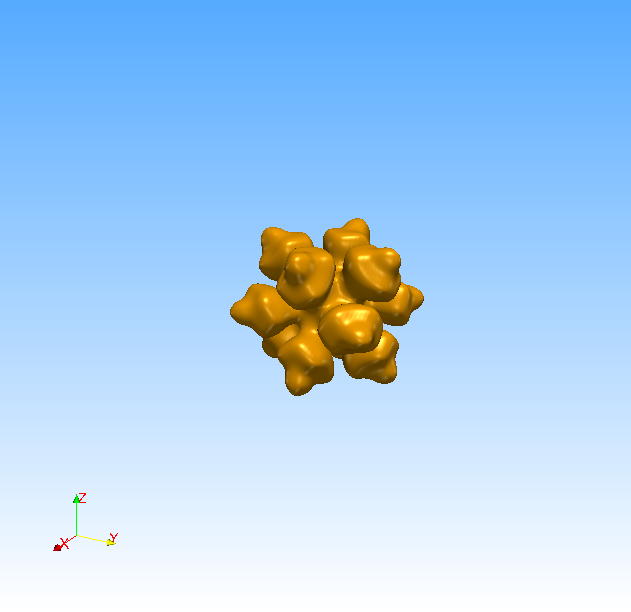} &  & \includegraphics[scale=0.21]{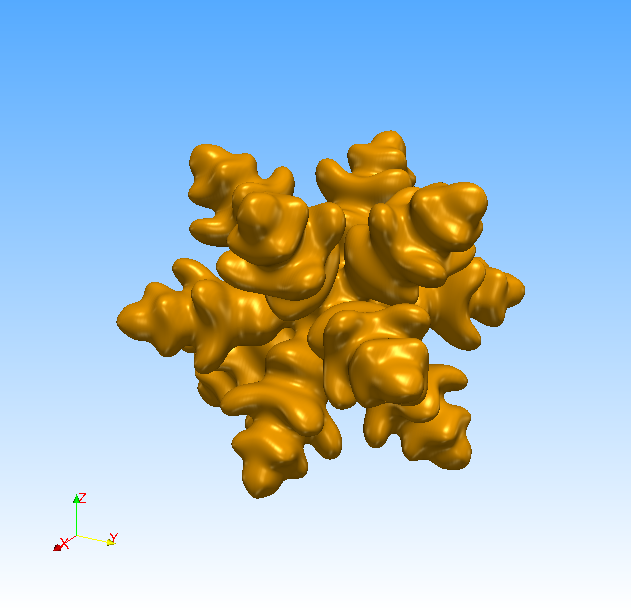} &  & \includegraphics[scale=0.21]{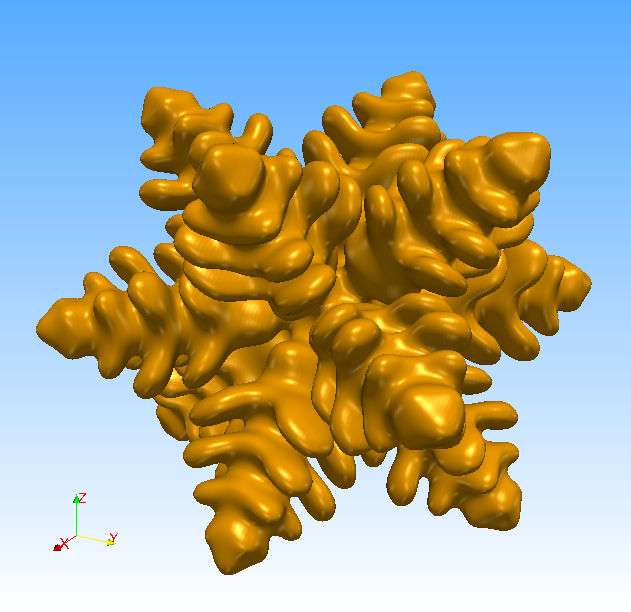}\tabularnewline
\end{tabular}
\par\end{centering}

~

\caption{\label{fig:Simul_K61}Phase-field evolution for $a_{s}(\mathbf{n})$
defined by Eq. (\ref{eq:As_MatSci}) with $\varepsilon_{s}=0$ and
$\gamma=-0.02$. The parameters are $W_{0}=0.0125$, $\tau_{0}=1.5625\times10^{-4}$,
$\lambda=10$, $\Delta=0.25$ and $\kappa=1$.}
\end{figure*}

The full dendrite is obtained by symmetry with respect to the planes
$XY$, $YZ$ and $XZ$. In the simulation, the undercooling is $\Delta=0.25$
and parameters of $a_{s}(\mathbf{n})$ are $\varepsilon_{s}=0$ and
$\gamma=-0.02$. The evolution of the dendritic structure $\phi=0$
is presented on Fig. \ref{fig:Simul_K61} for six different times.
A same orientation of the coordinate system was set for each figure.

\section{\label{sec:Conclusion}Conclusion}

In this paper, the Lattice Boltzmann (LB) method, usually applied
to simulate fluid flows, is applied to simulate crystal growth. The
interface position and the temperature are modeled by a phase-field
model which is numerically solved with a LB method. In this work,
various dendritic shapes in 2D and 3D were simulated through the study
of anisotropy function $a_{s}(\mathbf{n})$. That function is responsible
for the anisotropic growth of crystals and appears in the phase-field
equation.

A special care must be done when calculating the anisotropy function
that involves the normal vector $\mathbf{n}$ of the interface. Indeed,
for a sixfold symmetry in 2D, when a standard method of central finite
differences is used to calculate $\mathbf{n}$, the phase-field $\phi=0$
is not any more isotropic. The solution does not match with itself
after a rotation of 60\textdegree{} because a numerical anisotropy
occurs on $a_{s}(\mathbf{n})$: the pattern of that function is not
periodic. That lattice anisotropy can be decreased by using the Directional
Derivatives (DD) method in order to compute the gradients. The number
of directional derivatives is equal to the number of moving directions
in the LB method. The gradient components are given by their moment
of first order. Those directional derivatives are specific for each
lattice and were used in this work for D2Q9 and D3Q15. The use of
this method increases the accuracy of $a_{s}(\mathbf{n})$ and yields
a pattern that is periodic. The solution is isotropic by rotation.

Next, the whole approach (LB+DD) was applied to simulate simultaneous
growth of three crystals, each of them being defined by its own anisotropy
function. For that purpose, a supplementary field $I(\mathbf{x},\, t)$
was added in the definition of $a_{s}$. That field carries the crystal
number and evolves with the phase-field $\phi$. The method was applied
for simulating in 2D the growth of three crystals with respectively
four, five and six tips. Finally, several runs were performed with
a 3D code for simulating standard and non standard dendrites. The
anisotropy functions that have been used favor the growth in the $\left\langle 100\right\rangle $-,
$\left\langle 110\right\rangle $- and $\left\langle 111\right\rangle $-directions.
For each dendrite, the isotropy of the computational method was checked
by carrying out several slices. After an appropriate rotation, the
phase-fields are perfectly superimposed. For growth in the $\left\langle 111\right\rangle $-direction,
the threefold symmetry of secondary branches is consistent with the
$a_{s}(\mathbf{n})$ function. Indeed, the analysis of that function
and its derivatives show that those secondary branches can appear
in the three directions. This work shows the flexibility of the lattice
Boltzmann method for simulating dendritic shapes defined by various
anisotropy functions in 2D and 3D.

\section*{Acknowledgments}

The authors thank \noun{Mathis Plapp} for his comments about this
paper. The work was supported by the SIVIT project involving AREVA
NC.

\bibliographystyle{elsarticle/elsarticle-num-names}
\bibliography{/media/Elements/REDAC/Biblio-BibTeX/Biblio-PhaseField,/media/Elements/REDAC/Biblio-BibTeX/Biblio-LBM,/media/Elements/REDAC/Biblio-BibTeX/Biblio-Solidification}

\end{document}